\documentclass[reprint,superscriptaddress,preprintnumbers,nofootinbib,amsmath,amssymb,aps,prd,longbibliography]{revtex4-1}

\usepackage{cancel}
\usepackage{accents}
\usepackage{mciteplus,slashed}
\usepackage{amssymb,cancel,amsmath}
\usepackage{dcolumn}
\usepackage{bm}
\usepackage{appendix}
\usepackage{physics,mathtools}
\usepackage{booktabs}
\usepackage{feynmp-auto}
\usepackage[T1]{fontenc}	
\usepackage{csvsimple}
\usepackage[hidelinks]{hyperref}
\usepackage[section]{placeins}
\usepackage[capitalise]{cleveref}
\usepackage{booktabs}
\usepackage{graphicx}
\usepackage{mathrsfs}
\usepackage{enumitem}
\usepackage{capt-of}
\usepackage[nolist,nohyperlinks]{acronym}
\graphicspath{{Figures/}}

\usepackage{float} 
\usepackage{comment}

\def\jcap{{J.\ Cosmology Astropart. Phys.}}

\AtBeginDocument{
\heavyrulewidth=.08em
\lightrulewidth=.05em
\cmidrulewidth=.03em
\belowrulesep=.65ex
\belowbottomsep=0pt
\aboverulesep=.4ex
\abovetopsep=0pt
\cmidrulesep=\doublerulesep
\cmidrulekern=.5em
\defaultaddspace=.5em
}

\usepackage[dvipsnames]{xcolor}
\usepackage[normalem]{ulem}
\usepackage{fontawesome} 

\hypersetup{
    pdfnewwindow=true,      
    colorlinks=true,       
    linkcolor=blue,          
    citecolor=blue,        
    filecolor=blue,      
    urlcolor=blue        
}

\newcommand{\be}{\begin{equation}\begin{aligned}}
\newcommand{\ee}{\end{aligned}\end{equation}}

\newcommand{\orcid}[1]{\begingroup
  \hypersetup{hidelinks}\href{https://orcid.org/#1}{\includegraphics[width=10pt]{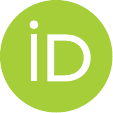}} \endgroup}

\def\bi{\begin{itemize}[noitemsep,leftmargin=*]
\setlength\itemsep{1em}
        }
\def\ei{\end{itemize}}

\def\apjs{{Astrophys.~J.~Supp.}}

\begin{document}

\title{Cross Correlating the Unresolved Gamma-Ray Background with Cosmic Large-Scale Structure from DESI: Implications for Astrophysics and Dark Matter}

\author{Bei Zhou \orcid{0000-0003-1600-8835}}
\email{beizhou@fnal.gov}
\affiliation{Theory division, Fermi National Accelerator Laboratory, Batavia, Illinois 60510, USA}
\affiliation{Kavli Institute for Cosmological Physics, University of Chicago, Chicago, Illinois 60637, USA}
\affiliation{William H. Miller III Department of Physics and Astronomy, Johns Hopkins University, Baltimore, Maryland 21218, USA}

\author{Jos\'e Luis Bernal \orcid{0000-0002-0961-4653}\,}
\email{jlbernal@ifca.unican.es}
\affiliation{Instituto de Física de Cantabria (IFCA), CSIC-Univ. de Cantabria, Avda. de los Castros s/n, E-39005 Santander, Spain}

\author{Elena Pinetti \orcid{0000-0001-7070-0094}\,}
\email{epinetti@fnal.gov}
\affiliation{Theory division, Fermi National Accelerator Laboratory, Batavia, Illinois 60510, USA}
\affiliation{Kavli Institute for Cosmological Physics, University of Chicago, Chicago, Illinois 60637, USA}

\author{Hector Afonso G.\ Cruz \orcid{https://orcid.org/0000-0002-1775-3602}}
\email{hcruz2@jhu.edu}
\affiliation{William H. Miller III Department of Physics and Astronomy, Johns Hopkins University, Baltimore, Maryland 21218, USA}

\author{Marc Kamionkowski \orcid{0000-0001-7018-2055}}
\email{kamion@jhu.edu}
\affiliation{William H. Miller III Department of Physics and Astronomy, Johns Hopkins University, Baltimore, Maryland 21218, USA}

\preprint{FERMILAB-PUB-24-0564-T}

\begin{abstract}
The unresolved gamma-ray background (UGRB) is a diffuse gamma-ray emission arising from numerous extragalactic sources below the detection threshold and is an important component of the gamma-ray sky. Studying the UGRB is crucial for understanding high-energy astrophysical processes in the universe and for probing fundamental physics, such as the nature of dark matter. In this work, we forecast the cross-correlation between the UGRB and galaxy catalogs from the Dark Energy Spectroscopic Instrument (DESI) survey. First, we study the expected astrophysical contributions to the UGRB and their cross-correlation with DESI spectroscopic galaxies. Our calculations show that the cross-correlation signal-to-noise ratio is expected to be significant, with the highest value predicted to be 20.6 for DESI luminous red galaxies due to a higher predicted overlap in the redshift distribution with the UGRB. We consider two science cases that the UGRB-spectroscopic galaxies cross-correlation can be applied to: 1) measuring the UGRB flux as a function of redshift, achieving a precision of 10\% in some redshift bins, and 2) searching for annihilating dark matter potentially up to a mass of about 300~GeV, three times higher than the currently strongest constraints. This work underscores the importance of cross correlating the UGRB with cosmic large-scale structure tracers and highlights the multiwavelength approaches to advancing our understanding of high-energy astrophysical phenomena and fundamental physics.
\end{abstract}

\maketitle 

\begin{acronym}
    \acro{mAGN}{misaligned active galactic nuclei}
    \acro{SF}{galaxies with star-forming activities}
    \acro{BL Lac}{BL Lacertae objects}
    \acro{FSRQ}{flat-spectrum radio quasars}
    \acro{UGRB}{unresolved gamma-ray background}
    \acro{LSS}{large-scale structure}
    \acro{SNR}{signal-to-noise ratio}
    \acro{GLF}{gamma-ray luminosity function}
    \acro{DM}{dark matter}
\end{acronym}

\section{Introduction}

The \ac{UGRB} is a diffuse gamma-ray emission from the extragalactic sources that fall below individual detection thresholds; it is a critical component of the extragalactic gamma-ray sky~\cite{Fermi-LAT:2014ryh}. This background is expected to arise from different populations, such as \ac{mAGN}, \ac{SF}, \ac{BL Lac}, and \ac{FSRQ}~\cite{Ando:2009nk, Xia:2011ax, Harding:2012gk, DiMauro:2013xta, Tamborra:2014xia, DiMauro:2014wha, DiMauro:2015tfa, Xia:2015wka, Linden:2016fdd, Wang:2016zyi, DiMauro:2017ing, Lamastra:2017iyo, Cuoco:2017bpv, Komis:2017jta, Pinetti:2019ztr, Hooper:2016gjy, Stecker:2019ybn, Qu:2019zln, Owen:2021evt, Owen:2021qul, Blanco:2021icw, Korsmeier:2022cwp, Xue:2024tkj, Min:2024bnj, Cholis:2024hmd}. 
In addition, the UGRB may potentially originate from exotic processes, such as the annihilation or decay of \ac{DM} particles, particularly weakly interacting massive particles (WIMPs)~\cite{Ando:2005xg, Fornasa:2012gu, Ando:2013ff, Cholis:2013ena,  Camera:2014rja, Ando:2014aoa, Cuoco:2015rfa, Fermi-LAT:2015qzw, Regis:2015zka, Shirasaki:2015nqp, Regis:2015zka, Feyereisen:2015cea, Liu:2016ngs, Shirasaki:2016kol, Campbell:2017qpa, Ammazzalorso:2018evf, Blanco:2018esa, Pinetti:2019ztr, Arbey:2019vqx, Yang:2020zcu, Liang:2020roo, Zhang:2021mth, Bartlett:2022ztj, Paopiamsap:2023uuo, Delos:2023ipo, Ganjoo:2024hpn, Cholis:2024hmd}.

Understanding the origin of the \ac{UGRB} is crucial for both high-energy astrophysics and fundamental physics. For astrophysics, it informs us of the most extreme environments of the universe. 
While studying the gamma-ray emission from one or a category of resolved sources offers detailed information about individual objects or the population properties (of the brighter ones), studying the unresolved sources that contribute to the \ac{UGRB} will provide us with valuable insights into the population properties of the fainter ones, which may have different properties compared to the brighter ones~\cite{Ando:2009nk, Xia:2011ax, Harding:2012gk, DiMauro:2013xta, Tamborra:2014xia, DiMauro:2014wha, DiMauro:2015tfa, Xia:2015wka, Linden:2016fdd, Wang:2016zyi, DiMauro:2017ing, Lamastra:2017iyo, Cuoco:2017bpv, Komis:2017jta, Pinetti:2019ztr, Hooper:2016gjy, Stecker:2019ybn, Qu:2019zln, Owen:2021evt, Owen:2021qul, Blanco:2021icw, Korsmeier:2022cwp, Xue:2024tkj, Min:2024bnj, Cholis:2024hmd}.
For fundamental physics, understanding the origin of the \ac{UGRB} is also crucial for probing beyond-the-standard-model physics, especially searching for \ac{DM} and constraining its properties~\cite{Ando:2005xg, Fornasa:2012gu, Ando:2013ff, Cholis:2013ena,  Camera:2014rja, Ando:2014aoa, Cuoco:2015rfa, Fermi-LAT:2015qzw, Regis:2015zka, Shirasaki:2015nqp, Regis:2015zka, Feyereisen:2015cea, Liu:2016ngs, Shirasaki:2016kol, Campbell:2017qpa, Ammazzalorso:2018evf, Blanco:2018esa, Pinetti:2019ztr, Arbey:2019vqx, Yang:2020zcu, Liang:2020roo, Zhang:2021mth, Bartlett:2022ztj, Paopiamsap:2023uuo, Delos:2023ipo, Ganjoo:2024hpn, Cholis:2024hmd}. 

The study of the \ac{UGRB} has been approached using various methods. 
One could study the \ac{UGRB} by analyzing the gamma-ray emission from a class of resolved sources and extrapolating the properties to include the unresolved ones~\cite{DiMauro:2013xta, Tamborra:2014xia, DiMauro:2014wha, DiMauro:2015tfa, Linden:2016fdd, Hooper:2016gjy, DiMauro:2017ing, Lamastra:2017iyo, Komis:2017jta, Stecker:2019ybn, Qu:2019zln, Owen:2021evt, Owen:2021qul, Blanco:2021icw, Korsmeier:2022cwp, Xue:2024tkj, Min:2024bnj}. Another approach involves the use of autocorrelation techniques~\cite{Fermi-LAT:2012pez, Harding:2012gk, Fornasa:2016ohl, Fermi-LAT:2018udj}. This method examines the angular power spectrum of the UGRB, allowing us to explore the anisotropies in the UGRB. Last but not least, it is possible to cross correlate the UGRB with different cosmic \ac{LSS} tracers, such as galaxy positions~\cite{Ando:2014aoa, Cuoco:2015rfa, Regis:2015zka, Shirasaki:2015nqp, Cuoco:2017bpv, Ammazzalorso:2018evf, Bartlett:2022ztj}, galaxy clusters~\cite{Branchini:2016glc, Hashimoto:2018ztv, Colavincenzo:2019jtj, DiMauro:2023qat}, cosmic shear~\cite{Camera:2012cj, Shirasaki:2014noa, Shirasaki:2016kol, Troster:2016sgf, Shirasaki:2018dkz, DES:2019ucp},
the integrated Sachs-Wolfe effect from cosmic microwave background (CMB) anisotropies~\cite{Tan:2020fbc}, lensing of CMB~\cite{Fornengo:2014cya}, the cosmic infrared background~\cite{Feng:2016fkl}, the late-time 21cm signal~\cite{Pinetti:2019ztr}, the thermal Sunyaev-Zel’dovich effect~\cite{Shirasaki:2019uls}, and high-energy neutrinos~\cite{Negro:2023kwv}, etc.
The cross-correlation analyses can effectively enhance the \ac{SNR} for detecting weak signatures in the \ac{UGRB} by taking advantage of the high signal-to-noise of other tracers of the cosmic \ac{LSS}.
Moreover, the cross correlation with cosmic \ac{LSS} tracers at different redshifts allows us to measure the redshift-dependence of the \ac{UGRB}, which offers valuable insights into its origin.

The spatial distribution of galaxies is generally the highest signal-to-noise tracer of the matter fluctuations of the Universe and, if redshift information is available, allows for an accurate tomographic reconstruction of the UGRB.
The DESI galaxy survey is one of the most comprehensive surveys mapping the cosmic \ac{LSS} and is conducting the largest galaxy redshift survey to date~\cite{DESI:2016fyo, DESI_web}. 
By measuring tens of millions of galaxies, including luminous red galaxies, emission line galaxies, and quasars, DESI is creating a detailed 3D map of the Universe. DESI's galaxy samples are thus crucial for studying the cross correlation between the \ac{UGRB} and cosmic \ac{LSS}. 

In this paper, we perform the first predictions of the cross correlation between the UGRB and spectroscopic galaxy catalogs from the DESI survey~\cite{DESI:2016fyo, DESI_web}. We derive the astrophysical contributions to the \ac{UGRB},
then calculate their cross correlation with DESI galaxy samples and forecast the \ac{SNR}.
We also forecast the sensitivities for 1) \ac{UGRB} tomography and 2) searching for annihilating \ac{DM}.

The rest of the paper is organized as follows. In Sec.~\ref{sec_cosmo}, we present the framework for angular correlation between the diffuse gamma-ray background and galaxy clustering.
In Sec.~\ref{sec_gamma}, we calculate the gamma-ray emission from various unresolved astrophysical sources.
In Sec.~\ref{sec_galaxies}, we discuss the DESI galaxy catalogs.
In Sec.~\ref{sec_CCresults}, we present our forecasted cross-correlation results, including the angular power spectra, uncertainties, and the \ac{SNR}s.
In Secs.~\ref{sec_UGRBzDep} and \ref{sec_DM}, we forecast the measurement on the redshift dependence of the \ac{UGRB} and the sensitivity reach to \ac{DM} annihilations, respectively, using the cross correlation between the \ac{UGRB} and DESI galaxy samples.
Finally, we present our conclusions in Sec.~\ref{sec_concl}.
Throughout this paper, we use the $\Lambda$CDM cosmological model with cosmological parameters taken from Planck 2018~\cite{Planck:2018vyg}.

\section{Angular correlation between the gamma-ray background and galaxy samples}
\label{sec_cosmo}

In this section, we detail our calculations of the angular correlation between the \ac{UGRB} and galaxy positions.
We employ a two-point angular correlation function to quantify the correlation between the fluctuations in the \ac{UGRB} intensity and the distribution of galaxies (Sec.~\ref{sec_cosmo_APS}). We also quantify the covariance of the angular power spectrum, which is essential for a robust statistical analysis (Sec.~\ref{sec_cosmo_unc}).

\subsection{Angular correlation and power spectrum}
\label{sec_cosmo_APS}

In general, the two-point angular correlation function between two fields, $X$ and $Y$, is given by
\be
\left\langle\delta X\left( \mathbf{n_1} \right) \delta Y\left( \mathbf{n_2} \right)\right\rangle
=
\sum_{\ell} \frac{2 \ell+1}{4 \pi} C_{\ell}^{XY} P_{\ell}[\cos (\theta)]
\label{eq_2pAC} \,,
\ee
where $\mathbf{n_1}$ and $\mathbf{n_2}$ are two angular directions on the sky, 
$\ell$ the angular multipole, 
$C_\ell^{XY}$ the angular power spectrum of the correlation,
$P_\ell$ the Legendre polynomials. The fluctuations $\delta X(\mathbf{n_1})$ and $\delta Y(\mathbf{n_2})$ of the fields $X$ and $Y$ are defined as
\be
\delta X(\mathbf{n_1}) \equiv \frac{X(\mathbf{n_1}) - \langle X \rangle}{\langle X \rangle} \,,
\ee
where $\langle X \rangle$ and $\langle Y \rangle$ are the spatial average of $X$ and $Y$, respectively.

In this work, we are mainly interested in the cross correlation between galaxy overdensity and the \ac{UGRB} fields, for which $X \equiv {\rm g}$ and $Y \equiv \gamma$.
For the gamma rays, we focus on the \ac{UGRB} observed by {\it Fermi}-LAT, detailed in Sec.~\ref{sec_gamma}, and we divide the gamma-ray measurements into different energy bins, indexed by $i$. 
The choice of the binning is the same as Table~III in Ref.~\cite{Fermi-LAT:2018udj}.\footnote{I.e., $E_\gamma=$ 0.5--1.0, 1.0--1.7, 1.7--2.8, 2.8--4.8, 4.8--8.3, 8.3--14.5, 14.5--22.9, 22.9--39.8, 39.8--69.2, 69.2--120.2, 120.2--331.1, 331.1--1000.0 GeV. Twelve bins in total.} 
Details about the galaxy sample can be found in Sec.~\ref {sec_galaxies}.  Therefore, the $C_\ell^{XY}$ term in Eq.~\eqref{eq_2pAC} becomes $C_{i,\ell}^{\gamma {\rm g}}$ and can be calculated by
\be
C_{i, \ell}^{\gamma {\rm g}} 
& =
\frac{2}{\pi}
\int d z_{\rm g} \frac{1}{N_g}\frac{d N_{\rm g}}{d z_{\rm g}} b_{\rm g}(z_{\rm g}) \int d z_\gamma \, \frac{dI^i_\gamma}{dz_\gamma} b_\gamma(z_\gamma) \\
& \times \int d k\, k^2  P(k, z_{\rm g}, z_\gamma)
 j_\ell[k \chi_1(z_{\rm g})] 
j_\ell[k \chi_2(z_\gamma)] \,,
\label{eq_Cl}
\ee
where $dN_{\rm g}/d_{\rm g}$ and $dI^i_\gamma/dz_\gamma$ are the redshift distributions of mean galaxy total number counts and gamma-ray intensity, respectively, $b_{\rm g}(z_{\rm g})$ and $b_\gamma(z_\gamma)$ the bias of the galaxy sample and gamma rays, respectively, $j_\ell[k\chi(z)]$ the spherical Bessel functions, $\chi(z)$ the comoving distance to redshift $z$, $k$ the wavenumber, and $P(k, z_{\rm g}, z_\gamma)$ is the linear power spectrum in the Fourier space. Nonlinear effects, which are important at small scales, are negligible in our calculations, as the cross-correlation \ac{SNR} mainly comes from $\ell$ less than a few hundred (details below), primarily due to the angular resolution of {\it Fermi}-LAT.

Besides the clustering component shown in Eq.~\eqref{eq_Cl}, the cross correlation between the \ac{UGRB} and galaxy clustering involves a shot-noise term proportional to the mean gamma-ray flux coming only from the sources also featured in the galaxy catalog over the total galaxy number density. We anticipate that this term will be subdominant for the cross correlation we are interested in. First, most of the main gamma-ray sources are not targeted by DESI for any of its samples. Second, the number density of DESI galaxies is very large. In addition to this, the shot-noise term would be relevant at small scales, where angular resolution limits suppress the power spectrum measurements. Therefore, we neglect the shot-noise term for the gamma-ray galaxy-clustering cross correlation.

We also need the autocorrelation of the galaxy clustering ($C_\ell^{\rm gg}$) and the gamma-ray intensity ($C_{\ell}^{\gamma\gamma}$) to calculate the uncertainty in the cross-correlation (Eq.~\eqref{eq_cov}) and to validate our \ac{UGRB} model presented below.

Given the angular resolution of \textit{Fermi}-LAT, we can safely assume that gamma-ray emitters are point-like sources. Therefore, we can consider that $C_\ell^{\gamma\gamma}$ is the sum of a clustering component $C_{i, \ell}^{\gamma\gamma, \rm clust}$ and a scale-independent shot-noise component $C_{i, \ell}^{\gamma\gamma, \rm shot}$:\footnote{The $C_{i, \ell}^{\gamma\gamma, \rm clust}$ and $C_{i, \ell}^{\gamma\gamma, \rm shot}$ are equivalent, in the context of a halo model, to the 2-halo and the 1-halo term, respectively, when the intensity profile is assumed to follow a Dirac-delta distribution.}
\be
C_{i, \ell}^{\gamma\gamma} = C_{i, \ell}^{\gamma\gamma, \rm shot} + C_{i, \ell}^{\gamma\gamma, \rm clust} 
\,,
\label{eq_Cl_gmgm}
\ee 
Because the calculation of $C_{i, \ell}^{\gamma \gamma, \rm shot}$ is related to the \ac{UGRB} source properties, we detail it in Sec.~\ref{sec_gamma_general}.

In contrast, the clustering component effectively captures the \ac{LSS} of the universe. It can be calculated by
\be
C_{i, \ell}^{\gamma\gamma, \rm clust} = 
& \frac{2}{\pi} 
\int dz \frac{dI^i_\gamma}{dz} b_\gamma(z) \int dz^{\prime} \frac{dI^i_\gamma}{dz^{\prime}} b_\gamma(z') \\
& \times \int dk \, k^2 j_\ell(k \chi(z)) j_\ell\left(k \chi^{\prime}(z^\prime)\right) P\left(k, z, z^{\prime}\right)
\,.
\ee

Finally, the autocorrelation of the galaxy samples can be calculated by
\be
C_\ell^{{\rm g}{\rm g}}
&=
\frac{2}{\pi} 
\int dz \frac{1}{N_{\rm g}}\frac{dN_{\rm g}}{dz} b_{\rm g}(z) \int dz^{\prime} \frac{1}{N_{\rm g}}\frac{dN_{\rm g}}{dz^\prime} b_{\rm g}(z') \\
&\times \int dk \, k^2 j_\ell(k \chi(z)) j_\ell\left(k \chi^{\prime}(z^\prime)\right) P\left(k, z, z^{\prime}\right) .
\label{eq_Cl_gg}
\ee
In this work, we assume that both shot-noise components in the galaxy clustering and \ac{UGRB} are Poissonian. Since, for the former, this only depends on the density of galaxies, we consider it only as a noise contribution to the covariance rather than part of the signal.

We compute the clustering components of the power spectra using \texttt{Multi\_CLASS}~\cite{Bellomo:2020pnw, Bernal:2020pwq},\footnote{\href{https://github.com/nbellomo/Multi_CLASS}{https://github.com/nbellomo/Multi\_CLASS}} a public extension of the Boltzmann code {\tt CLASS}~\cite{2011JCAP...07..034B}\footnote{\href{https://github.com/lesgourg/class_public}{https://github.com/lesgourg/class\_public}} that allows computing angular power spectra of clustering statistics for different tracers of the \ac{LSS}.

\subsection{Covariance of the power spectrum}
\label{sec_cosmo_unc}
We assume a Gaussian covariance, hence diagonal, for which the variance for each multipole of the angular power spectrum is given by
\begin{equation}
\begin{aligned}
\left(\Delta C_{i, \ell}^{\gamma {\rm g}}\right)^2
&= \frac{1}{(2 \ell+1) f_{i, \mathrm{sky}}} 
\left\{ \left( C_{i, \ell}^{\gamma {\rm g}} \right)^2 \right. \\
& \quad + \left. \left[ C_{i, \ell}^{\gamma\gamma} + \frac{C_{N, i}^\gamma}{\left(B_{i, \ell}^\gamma\right)^2} \right] \right. 
\left. \left[ C_\ell^{{\rm g}{\rm g}} + \frac{C_N^{\rm g}}{\left( B_{i, \ell}^{\rm g} \right)^2} \right] 
\right\} \,,
\label{eq_cov}
\end{aligned}
\end{equation}
where $f_{i, \mathrm{sky}}$ is the fraction of the sky involved in the calculation, which is the overlap of the sky coverage of the DESI survey (about 14,000 $\rm deg^2$~\cite{DESI:2016fyo, DESI_web}) and {\it Fermi}-LAT (i.e., regions outside masking the galactic plane and other bright sources) that can be found in Table~III of Ref.~\cite{Fermi-LAT:2018udj}.
$C^{\gamma\gamma}_{i,\ell}$ and $C_\ell^{{\rm g}{\rm g}}$ are the autocorrelations given in Eqs.~\eqref{eq_Cl_gmgm} and \eqref{eq_Cl_gg}, respectively.
The $C_{N, i}^\gamma$ and $C_N^{\rm {\rm g}}$ are the noise terms for the gamma rays and galaxies, respectively, and $B_{i, \ell}^\gamma$ and $B_{i, \ell}^{\rm g}$ are the beam functions that describe the angular resolution of the gamma-ray and galaxy-survey telescopes, respectively. The details of the noise terms and beam functions are given below.

We understand the noise term for galaxy clustering as the (Poissonain) shot noise,  given by
\begin{equation}
C_N^{\rm g} 
= \frac{4 \pi f_{\rm sky}^{\rm DESI}}{N_{\rm g}} 
= \frac{\Omega_{\rm sky}^{\rm DESI}}{N_{\rm g}}
\, ,
\end{equation}
where $f_{\rm sky}^{\rm DESI} \simeq 14,000\ {\rm deg^2}/41252.96\ {\rm deg^2} \simeq 0.339 $ is the sky fraction of the DESI survey
and $N_{\rm g}$ is the number of galaxies in a survey sample, given in Sec.~\ref{sec_galaxies}. 

We consider perfect angular resolution for the galaxy survey, so that 
\begin{equation}
B_\ell^{\rm g} = 1\,.
\label{eq_Bl}
\end{equation}
This assumption does not introduce any error since DESI angular resolution is much better than the smallest angular scales considered in our study.

The noise term of the gamma rays, for each energy bin $i$, is given by
\begin{equation}
C_{N, i}^\gamma 
=
\frac{N_i^\gamma}{\epsilon_i^2 \Omega_i^{\rm Fermi}},
\end{equation}
where $N_i^\gamma$ is the number of photons. The $\epsilon_i$ represents the exposure of the {\it Fermi}-LAT, specifically for the ``P8R3\_ULTRACLEAN\_V3'' event class, which is used in diffuse analyses. 
We calculate the exposure using {\tt  gtexpcube2}\footnote{\href{https://fermi.gsfc.nasa.gov/ssc/data/analysis/scitools/overview.html}{https://fermi.gsfc.nasa.gov/ssc/data/analysis/scitools/overview.html}} of the {\tt Fermitools} for a 16-year period (i.e., from about June 2008, when {\it Fermi}-LAT was launched, to June 2024) and rescale it to a 20-year period, roughly aligning with the completion of the DESI survey.
The $\Omega_i^{\rm Fermi}$ is the solid angle of {\it Fermi}-LAT sky coverage, $4\pi f_{i, \rm sky}^{\rm Fermi}$ for each energy bin, which is given in Table~III of Ref.~\cite{Fermi-LAT:2018udj}.

The energy-dependent gamma-ray beam function, $B_\ell^\gamma(E_\gamma)$, of {\it Fermi}-LAT can be found in Refs.~\cite{Fermi-LAT:2018udj, Pinetti:2019ztr},
\begin{equation}
B_\ell^\gamma(E_\gamma)
=
\exp \left[-\frac{\sigma_{\mathrm{b}}(\ell, E_\gamma)^2\, \ell^2}{2}\right] \, ,
\end{equation}
with
\begin{equation}
\sigma_{\mathrm{b}}(\ell, E_\gamma)
=
\sigma_0^{\text{Fermi}}(E_\gamma) \left[1+0.25 \sigma_0^{\text {Fermi}}(E_\gamma) \, \ell\right]^{-1} \, ,
\end{equation}
where $\sigma_0^{\text{Fermi}}(E_\gamma)$ is the 68\% containment angle of {\it Fermi}-LAT at energy
$E_\gamma$ and can be parameterized by
\begin{equation}
\sigma_0^{\text {Fermi}}(E_\gamma)
=
\sigma_0^{\text {Fermi}}\left( E_\gamma^{\text{ref}} \right) \times
\left( \frac{E_\gamma}{E_\gamma^{{\rm ref}}}\right)^{-0.95}+0.05\ \mathrm{deg} \, ,
\label{eq_sigma0_Fermi}
\end{equation}
with $E_\gamma^{\text{ref}} = 0.5$ GeV and $\sigma_0^{\text {Fermi}} \left(E_\gamma^{\text{ref}}\right) = 1.20$ deg. 
Eq.~\eqref{eq_sigma0_Fermi} follows a power-law behavior and then gradually flattens at around 0.05 degrees for higher energies, consistent with the ``PSF2'' response function specifications of {\it Fermi}-LAT. This empirical relation reproduces the beam function used in Ref.~\cite{Fermi-LAT:2018udj}, which was derived from Fermi-LAT data.

\section{Gamma-ray emission from unresolved astrophysical sources}
\label{sec_gamma}

In this section, we calculate the \ac{UGRB} emission from different astrophysical components, including misaligned active galactic nuclei (mAGN), galaxies with star-forming activities (SF), BL Lacertae objects (BL Lac), and flat-spectrum radio quasars (FSRQ).
We adopt the framework used by many previous work (e.g., Refs.~\cite{DiMauro:2013xta, DiMauro:2015tfa, Linden:2016fdd, Hooper:2016gjy, Blanco:2021icw, Cholis:2024hmd}).
In addition, we also show the bias, $b(z)$, of the different astrophysical sources.

\subsection{\ac{UGRB} spectrum, redshift distribution, and the shot-noise contribution} 
\label{sec_gamma_general}

The intensity of the \ac{UGRB} contributed by a certain type of source can be calculated by
\be
\frac{d^2 I}{d E_{\gamma} d z }
&=
\frac{d^2 V}{d z d \Omega} 
\int_{L_\gamma^{\min}}^{L_\gamma^{\max}} d \ln{L_\gamma} \times \frac{d F_\gamma}{d E_{\gamma}} \, \Phi_\gamma \left(L_\gamma, z\right) \\
& \times \left\{1-\omega\left(F_\gamma\left(L_\gamma, z\right)\right)\right\} 
\exp \left(-\tau_{\gamma\gamma}(E_{\gamma}, z)\right) \,,
\label{eq_d2IdEgmdz}
\ee
where ${d^2 V}/{d z d \Omega}$ is the comoving volume element, $L_\gamma$ the gamma-ray luminosity of a source, 
$F_\gamma$ the gamma-ray flux for a source with $L_\gamma$ at redshift $z$, given in Eq.~\eqref{eq_Fgamma}, 
and $\Phi_\gamma$ is the \ac{GLF}.
The $\omega(F_\gamma)$ is the detection efficiency of $Fermi$-LAT, and $1-\omega$ accounts for the fact that the sources resolvable by $Fermi$-LAT do not contribute to the \ac{UGRB}. We employ a modified Fermi detection efficiency function as presented in Ref.~\cite{Hooper:2016gjy}. 
The $\exp \left(-\tau_{\gamma\gamma}\right)$ term accounts for the gamma-ray absorption during propagation, through scattering off the extragalactic background light, for which we use the model from Ref.~\cite{Finke:2009xi}.
Note that $d^2I/dE_\gamma dz$ is in units of [$\rm cm^{-2}\, s^{-1}\, sr^{-1}\, GeV^{-1}$].

The $d F_\gamma / d E_\gamma$ can be derived by integrating the power-law spectrum, $E_\gamma (E_\gamma/E_{\min})^{-\Gamma}$, and equating it to the gamma-ray luminosity $L_\gamma$, i.e.,
\begin{equation}
\frac{d F_\gamma}{d E_\gamma}
=
\frac{(1+z)^{2-\Gamma}}{4 \pi d_L(z)^2} \frac{(2-\Gamma)}{\left[\left(\frac{E_{\rm max}}{E_{\rm min}}\right)^{2-\Gamma}-1\right]}\left(\frac{E_\gamma}{E_{\rm min}}\right)^{-\Gamma} \frac{L_\gamma}{E_{\rm min}^2} \,,
\label{eq_Fgamma}
\end{equation}
where $d_L$ is the luminosity distance and $\Gamma$ is the spectral index of the sources. We choose $E_{\rm min} = 0.1$~GeV and $E_{\rm max} = 10^3$~GeV to match {\it Fermi}-LAT observations.

\bigskip
On the other hand, we compute the shot-noise contribution following
Refs.~\cite{Pinetti:2019ztr, Pinetti:2021jjs}\footnote{See Appendix D in Ref.~\cite{Pinetti:2021jjs} for a full derivation.}.
\be
C_{\ell}^{\gamma\gamma, \rm shot}
& =
\Omega_{\rm sky} \int \frac{\mathrm{d} \chi}{\chi^2} \frac{dI_\gamma}{d\chi} \frac{dI_\gamma}{d\chi} P_{\gamma\gamma}^{\rm shot} \left(k=\frac{\ell}{\chi}, \chi\right) 
\\
& =
\Omega_{\rm sky} \int dz \frac{1}{\chi(z)^2} \frac{H(z)}{c}  \frac{dI_\gamma}{dz} \frac{dI_\gamma}{dz}
P_{\gamma\gamma}^{\rm shot}\left(z\right)
\,,
\label{eq_Cl_gmgm1h}
\ee
where $P_{\gamma\gamma}^{\rm shot}$ is the shot-noise contribution to the Fourier power spectrum, and we drop its variable $k$ because the shot-noise contribution is independent of $k$ as all the sources are point sources under {\it Fermi}-LAT angular resolution. As a result, $C_{\ell}^{\gamma\gamma, \rm shot}$ is also $\ell$-independent.

The $P_{\gamma\gamma}^{\rm shot}$ can be calculated by
\be
P_{\gamma\gamma}^{\mathrm{shot}}(z)
&=
\int_{L_\gamma^{\min}}^{L_\gamma^{\max}} 
d\ln L_\gamma 
\left(\frac{L_\gamma}{\langle L_\gamma \rangle(z)}\right)^2 \\
& \quad \times \Phi_{\gamma}(L_\gamma, z) 
\left\{1-\omega\left(F_\gamma\left(L_\gamma, z\right)\right)\right\}  \,,
\ee
where $\langle L_\gamma \rangle (z)$ is the mean luminosity produced by the \ac{UGRB} sources (below {\it Fermi}-LAT sensitivity) at redshift $z$, given by
\be
\langle L_\gamma \rangle (z)  =
\int_{L_\gamma^{\min}}^{L_\gamma^{\max}} d \ln{L_\gamma} \, L_\gamma
\Phi_\gamma \left(L_\gamma, z\right)
\left\{1-\omega\left(F_\gamma\left(L_\gamma, z\right)\right)\right\}  .
\label{eq_dIdEgm}
\ee

\medskip
In the following subsections, we discuss the technical details entering the calculation of each component for Eq.~\eqref{eq_d2IdEgmdz}, notably the minimum and maximum luminosities $L_\gamma^{\min}$ and $L_\gamma^{\max}$, the spectral index $\Gamma$, the \ac{GLF} and the bias of each population. 
Specifically, we discuss \ac{mAGN} in Sec.~\ref{sec_gamma_mAGN}, \ac{SF} in Sec.~\ref{sec_gamma_SF},
\ac{BL Lac} in Sec.~\ref{sec_gamma_BLLac}, and \ac{FSRQ} in Sec.~\ref{sec_gamma_FSRQ}. 
Finally, in Sec.~\ref{sec_gamma_results}, we present our gamma-ray results and {\it Fermi}-LAT measurements, including energy spectra, autocorrelation angular power spectra, and redshift distribution.

\begin{figure*}[t]
\begin{minipage}[t]{0.49\textwidth}
    \centering
    \includegraphics[width=\textwidth]{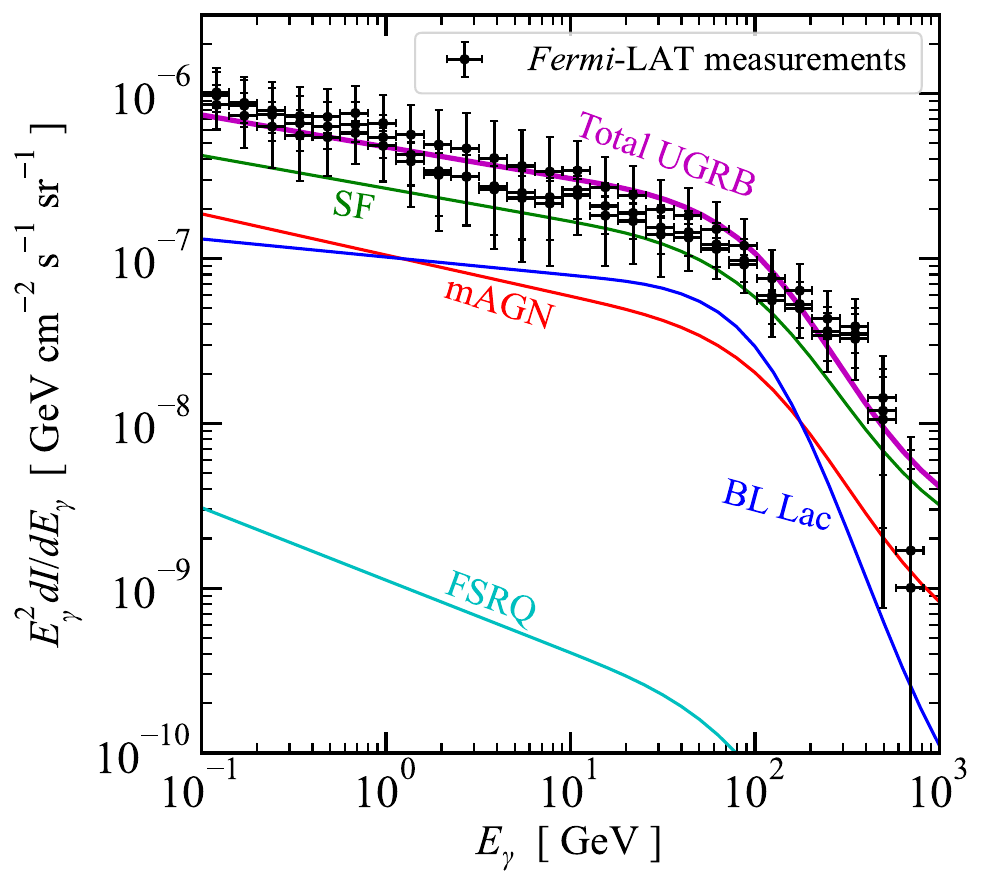}
    \caption{Our calculated gamma-ray energy spectra from different astrophysical sources (as labeled) and the total signal (purple). The {\it Fermi}-LAT measurements~\cite{Fermi-LAT:2014ryh}  are shown in black, with the three datasets corresponding to different choices of the galactic foreground models. Note that our results are in good agreement with the gamma-ray observations.}
    \label{fig_gamma_spec}
\end{minipage}\hfill
\begin{minipage}[t]{0.474\textwidth}
    \centering
    \includegraphics[width=\textwidth]{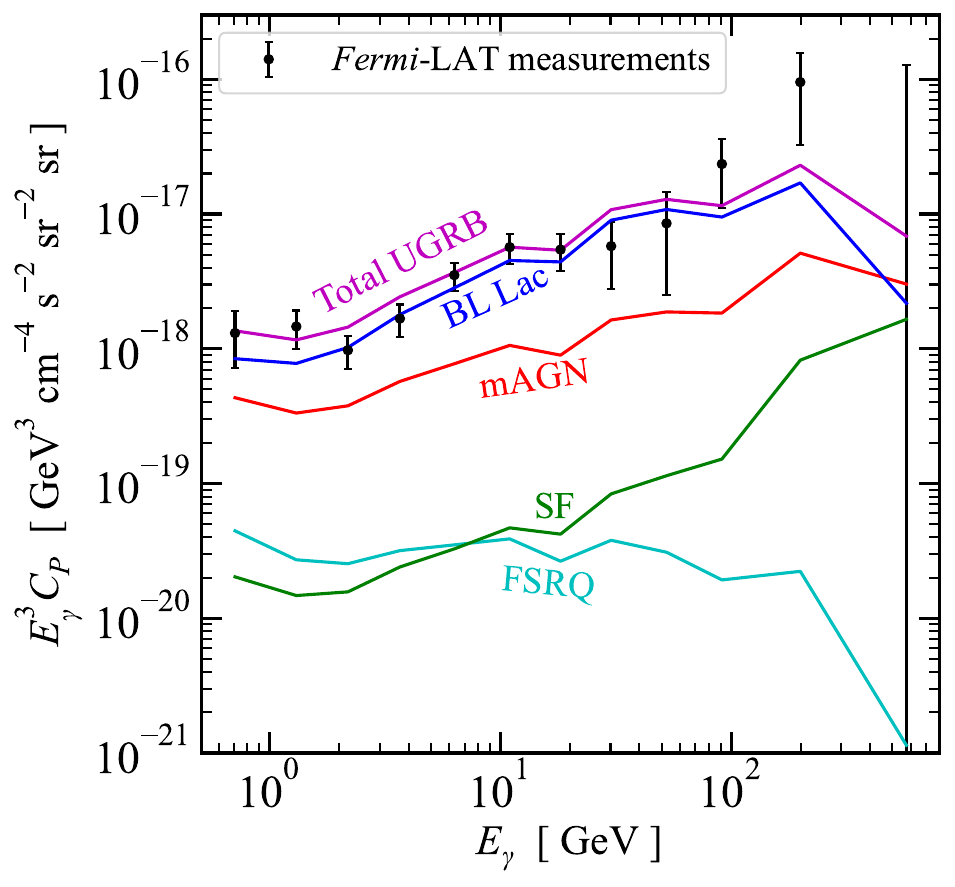}
    \caption{Our calculated gamma-ray auto-correlation angular power-spectrum amplitude $C_P$ from different astrophysical sources (as labeled) and the total signal, which is in good agreement with the {\it Fermi}-LAT measurements~\cite{Fermi-LAT:2018udj}. 
The $C_P$ in the y-axis label is defined as $(\sum_{\ell_{\min}}^{\ell_{\max}} C_\ell)/(\ell_{\max} - \ell_{\min})$, with $\ell_{\min}$ and $\ell_{\max}$ listed in Table~1 of Ref.~\cite{Fermi-LAT:2018udj}.
We multiply $C_P$ by $E_\gamma^3$ for visibility.}
    \label{fig_gamma_Cp}
\end{minipage}
\end{figure*}

\subsection{Astrophysical \ac{UGRB} sources}

\subsubsection{Misaligned AGN}
\label{sec_gamma_mAGN}

As most of the \ac{mAGN} and \ac{SF} fall well below the detection threshold of {\it Fermi}-LAT, it would be impossible to directly measure their \ac{GLF} from gamma-ray observations. Instead, we derive their \ac{GLF} from the luminosity function (LF) in other wavelengths.

The \ac{GLF} of \ac{mAGN} are calculated from the 5-GHz core LF $\Phi_{\rm r}\left(L_{\rm r}, z\right)$ [Eq.~\eqref{eq_Phi_r}], which is well measured, using the empirical scaling relation between $L_\gamma$ and $L_{\rm r}$, which is the 5-GHz core luminosity~\cite{DiMauro:2013xta, Hooper:2016gjy, Blanco:2021icw, Cholis:2024hmd}, i.e.,
\be
\Phi_\gamma\left(L_{\gamma}, z \right)
=
\frac{d \ln L_{\rm r}}{d \ln L_{\gamma}} \int \Phi_{\rm r}\left(L_{\rm r}, z\right) P\left( L_{\gamma}, L_{\rm r} \right) d \ln  L_{\rm r}
  \,,
\label{eq_Phi_gm_mAGN}
\ee
where $P\left( L_{\gamma}, L_{\rm r} \right)$ is the dispersion of the $L_{\gamma}$-$L_r$ relation for \ac{mAGN}, given by
\be
P(L_{\gamma}, L_{\rm r}) 
= 
\frac{1}{\sqrt{2\pi} \sigma_{\rm m}} 
\exp\left( 
\frac{\left[ \log_{10}{\left( \frac{L_{\gamma}/{\rm (erg/s)}}{ [L_r/({\rm 10^{40}erg/s})]^b } \right)} -d \right]^2}
{-2 \sigma^2_{\rm m}} \right), 
\ee
and the $L_{\gamma}$-$L_r$  relation is 
\begin{equation}
\log_{10}\left(\frac{L_{\gamma}}{\mathrm{erg} / \mathrm{s}}\right)=b \log_{10}\left(\frac{L_{\rm r}}{10^{40} \mathrm{erg} / \mathrm{s}}\right)+d \,,
\end{equation}
where the parameters $b=0.78$, $d=40.78$, and $\sigma_{\rm m}=0.88$ are from Table~IV of Ref.~\cite{Blanco:2021icw}.\footnote{Note that their $\sigma_{\rm mAGN}$ is referred to as $\sigma_{\rm m}$ here.}

The 5-GHz core LF of \ac{mAGN} is given by~\cite{Yuan2018LFs}, 
\be
\Phi_{\rm r}\left(L_{\rm r}, z\right) & d \ln L_{\rm r}
= 
e_1(z) \phi_1 \times \\
& \left[\left(\frac{L_{\rm r}}{L^\ast \, e_2(z)}\right)^\beta+\left(\frac{L_{\rm r}}{L^\ast \, e_2(z)}\right)^\gamma\right]^{-1}  d \ln L_{\rm r} \; ,
 \label{eq_Phi_r}
\ee
where the values of the parameters are from Ref.~\cite{Yuan2018LFs}.

Finally, for the other parameters in Eq.~\eqref{eq_d2IdEgmdz}, we take $\Gamma = 2.25$, $L_\gamma^{\min} =10^{40}$ erg/s and $L_\gamma^{\max} = 10^{50}$ erg/s~\cite{Hooper:2016gjy, Blanco:2021icw}. Note that Eq.~\eqref{eq_d2IdEgmdz} converges within the range of $L_\gamma^{\min}$ and $L_\gamma^{\max}$.
For the bias of \ac{mAGN}, we use the same as in Ref.~\cite{Bernal:2018myq}.

\subsubsection{Galaxies with star-forming activities (SF)}
\label{sec_gamma_SF}

The derivation of the \ac{GLF} of \ac{SF} activities is similar to that of \ac{mAGN}~\cite{Blanco:2021icw}, but from the infrared LF~\cite{Gruppioni:2013jna}, i.e., 
\begin{equation}
\Phi_\gamma\left(L_{\gamma}, z \right)
=
\frac{d \ln L_{\rm IR}}{d \ln L_{\gamma}} \int \Phi_{\rm IR}\left(L_{\rm IR}, z\right) P\left( L_{\gamma}, L_{\rm IR} \right) d \ln L_{\rm IR} \, ,
\label{eq_Phi_gm_SF}
\end{equation}
where $L_{\rm IR}$ is the infrared luminosity and $P\left( L_{\gamma}, L_{\rm IR} \right)$ is the dispersion of the $L_{\gamma}$-$L_{\rm IR}$ relation, given by
\be
P\left(L_{\gamma}, L_{\rm IR}\right)
& =
\frac{1}{\sqrt{2 \pi} \sigma_{\rm SF}} \times \\
& \quad \exp \left( \frac{\left(\log_{10}\left(\frac{L_{\gamma}/(\rm erg/s)}{[L_{\rm IR}/(10^{45} {\rm erg/s})]^a}\right)-g\right)^2}{- 2 \sigma_{\rm SF}^2} \right) \; ,
\ee
and the $L_{\gamma}$-$L_{\rm IR}$  relation is 
\begin{equation}
\log_{10}\left(\frac{L_{\gamma}}{\mathrm{erg} / \mathrm{s}}\right)
=
a \log_{10}\left(\frac{L_{\rm IR}}{10^{45} \mathrm{erg} / \mathrm{s}}\right)+g \, ,
\end{equation}
where the parameters $a=1.09$, $g=40.8$, and $\sigma_{\rm SF}=0.20$ are from Table~III of Ref.~\cite{Blanco:2021icw}.

The IR LF is given in Ref.~\cite{Gruppioni:2013jna},
\be
\Phi_{\rm IR}(L_{\rm IR}, z) & \mathrm{d} \ln L_{\rm IR}
=
\Phi_{\rm IR}^*\left(\frac{L_{\rm IR}}{L_{\rm IR}^*}\right)^{1-\alpha}  \\
&\times \exp \left[-\frac{1}{2 \sigma^2} \log_{10}^2\left(1+\frac{L_{\rm IR}}{L_{\rm IR}^*}\right)\right] \mathrm{d} \ln L_{\rm IR}.
\label{eq_LF_IR}
\ee 

The total IR LF in Eq.~\eqref{eq_LF_IR} is the sum of three components: 
1) normal galaxies, for which $\log_{10}L^*=9.46$ and $\log_{10}\Phi^*=-2.08$,
2) starburst galaxies, for which $\log_{10}L^*=11.02$ and $\log_{10}\Phi^*=-4.74$, and
3) star-forming AGN, for which $\log_{10}L^*=10.57$ and $\log_{10}\Phi^*=-3.25$. 

Finally, for the other parameters in Eq.~\eqref{eq_d2IdEgmdz}, we take $\Gamma = 2.2 $~\cite{Ambrosone:2020evo, Blanco:2021icw}, $L_\gamma^{\min} = 10^{36}$ erg/s and $L_\gamma^{\max} = 10^{44}$ erg/s~\cite{Gruppioni:2013jna, Blanco:2021icw}. Note that Eq.~\eqref{eq_d2IdEgmdz} converges within the range of $L_\gamma^{\min}$ and $L_\gamma^{\max}$.
For the bias of \ac{SF}, we follow Ref.~\cite{Bernal:2018myq}.

\subsubsection{BL Lac objects}
\label{sec_gamma_BLLac}

For the \ac{GLF} of the \ac{BL Lac}, we use the LDDE1 (luminosity-dependent density evolution) model 
derived by Ref.~\cite{Ajello:2013lka}, which has the relevant equations and the parameters (in the first line of its Table 3).

For the other parameters in Eq.~\eqref{eq_d2IdEgmdz}, we take $\Gamma = 2.11 $~\cite{Pinetti:2019ztr, Pinetti:2021jjs}, $L_\gamma^{\min} = 7\times10^{43}$ erg/s and $L_\gamma^{\max} = 10^{52}$ erg/s~\cite{Ajello:2013lka, Pinetti:2019ztr, Pinetti:2021jjs}.
For the bias, we use the one derived in Ref.~\cite{Pinetti:2019ztr, Pinetti:2021jjs}.

\begin{figure*}[t!]
\includegraphics[width=0.49\textwidth]{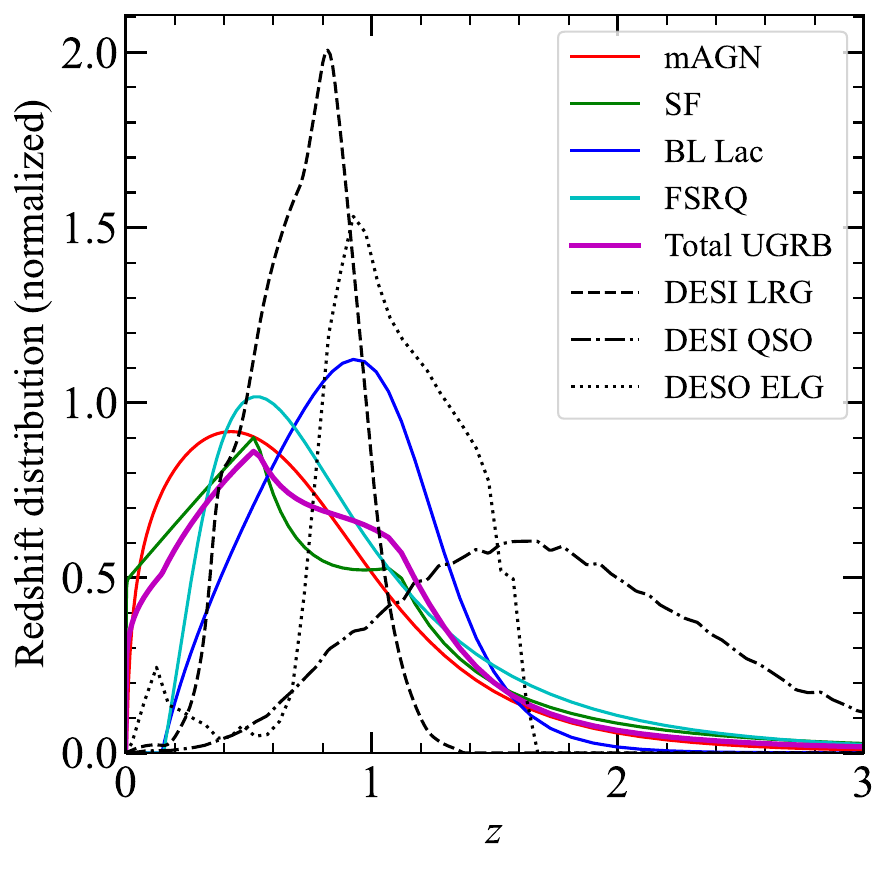}
\includegraphics[width=0.47\textwidth]{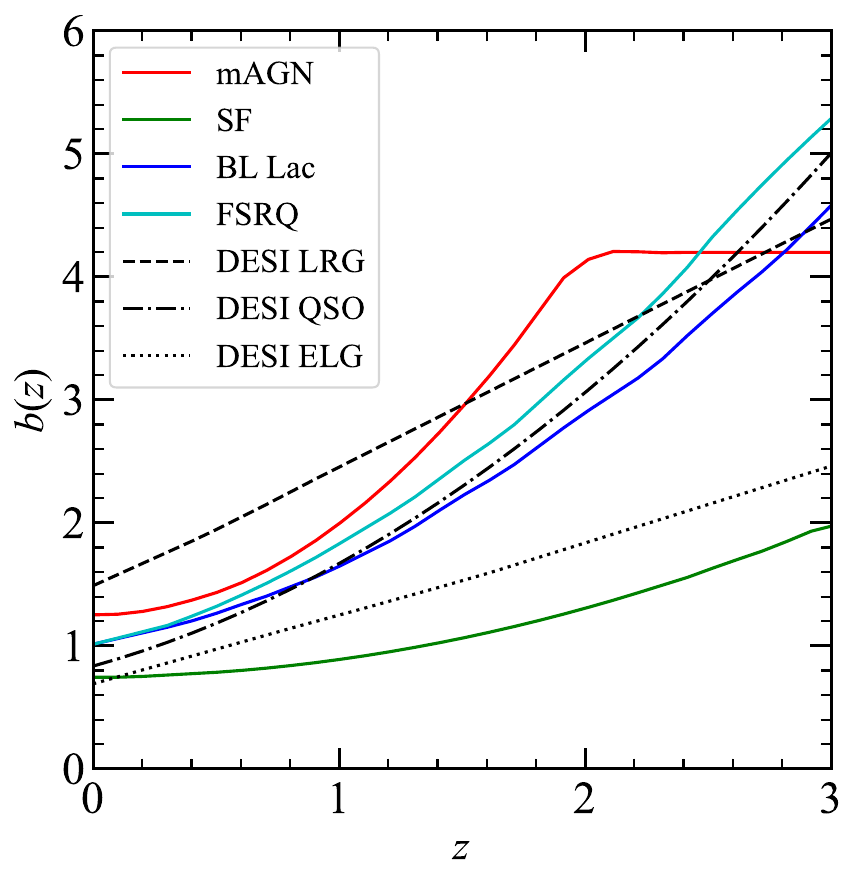}
\caption{\textbf{Left:}
Normalized redshift distribution of the \ac{UGRB} components in the first energy bin ($\int_{\rm 0.5\, GeV}^{\rm 1.0\, GeV} dE_\gamma (d^2I/dzdE_\gamma)$) and DESI samples ($dN/dz$).
\textbf{Right:} Bias of the \ac{UGRB} sources and DESI samples.  
}
\label{fig_zdist_bias}
\end{figure*}

\subsubsection{Flat-spectrum radio quasars (FSRQ)}
\label{sec_gamma_FSRQ}
For the \ac{GLF} of \ac{FSRQ}, we use the LDDE (luminosity-dependent density evolution) model 
derived by Ref.~\cite{Ajello:2011zi}, which has the relevant equations and the parameters (in the first line of its Table 3).

For the other parameters in Eq.~\eqref{eq_d2IdEgmdz}, we take $\Gamma = 2.7 $~\cite{Pinetti:2019ztr, Pinetti:2021jjs}, $L_\gamma^{\min} = 10^{44}$ erg/s and $L_\gamma^{\max} = 10^{52}$ erg/s~\cite{Ajello:2011zi, Pinetti:2019ztr, Pinetti:2021jjs}. For the bias, we use the one derived in Ref.~\cite{Pinetti:2019ztr, Pinetti:2021jjs}.

\subsection{Gamma-ray results and model validation}
\label{sec_gamma_results}

Fig.~\ref{fig_gamma_spec} and Fig.~\ref{fig_gamma_Cp} show our calculated gamma-ray energy spectra and angular power-spectrum amplitude, respectively, from different astrophysical components and their sum.
Also shown in both figures are the {\it Fermi}-LAT measurements~\cite{Fermi-LAT:2014ryh, Fermi-LAT:2018udj}. Our predictions match the data very well for both energy and angular power spectra.

The left panel of Fig.~\ref{fig_zdist_bias} shows the redshift distributions of the gamma-ray intensity from different sources and the total. 
Overall, \ac{mAGN}, \ac{SF}, and \ac{FSRQ} have similar redshift distributions that peak around $z \simeq 0.5$, whereas \ac{BL Lac} peak at higher redshifts.
The right panel of Fig.~\ref{fig_zdist_bias} shows the bias as a function of redshift.

\section{DESI galaxy samples}
\label{sec_galaxies}

For the galaxy catalogs, we focus on the Dark Energy Spectroscopic Instrument (DESI) galaxy survey, which is one of the most comprehensive surveys aimed at mapping the \ac{LSS} of the universe and is conducting the largest galaxy redshift survey to date~\cite{DESI:2016fyo, DESI_web}. DESI is designed to measure tens of millions of galaxies, providing a detailed 3D map of the universe up to unprecedentedly high redshift. This extensive dataset is invaluable for studying the correlation between the UGRB and cosmic \ac{LSS}. DESI targets several types of galaxies and quasars, including luminous red galaxies (LRGs), emission line galaxies (ELGs), and quasars (a.k.a. quasi-stellar objects, or QSOs)~\cite{DESI:2016fyo, DESI_web}.

LRGs are an important type of galaxies for cosmic \ac{LSS} studies because of two main advantages~\cite{DESI:2022gle}: (1) their brightness and prominent 4000 \AA\  break in spectra allow for easy target selection and redshift measurements; and (2) they are highly biased tracers of the cosmic \ac{LSS}, which yields a higher \ac{SNR} per object for the baryon acoustic oscillations (BAO) measurement than typical galaxies. We take the LRG redshift distribution from  Ref.~\cite{DESI:2022gle}, which is consistent with, e.g., Refs.~\cite{Yuan:2023ezi, Prada:2023lmw}. We assume the bias used in  Refs.~\cite{Zhou:2020nwq, Yuan:2023ezi} for $z<0.95$ ($1.5/D(z)$, where $D(z)$ is the growth factor) and linearly extrapolate it to $z>0.95$. For the number of LRGs, we use $N_{\rm g} = 12\times10^6$ from Ref.~\cite{Rezaie:2023lvi}, which corresponds to a target density of $\simeq 800$ $\rm deg^{-2}$. Note that the density is slightly higher than Refs.~\cite{DESI:2022gle, Yuan:2023ezi} because the latter focus on a narrower redshift range ($z=0.4$--0.8).

ELGs, in turn, provide high abundance numbers and have the advantage that their redshift can be reliably measured in a short period of observation time using the emission lines in their spectra --- in particular, the [O II] doublet $\lambda\lambda$ 3726,29 \AA~\cite{Raichoor:2022jab}. For the DESI-ELG redshift distribution, we use the sum of ``ELG\_LOP\_DESI'' and ``ELG\_VLO\_DESI'' from Ref.~\cite{Raichoor:2022jab}, consistent with, e.g., Ref.~\cite{Rocher:2023zyh}. In this case, we take the bias from Ref.~\cite{Rocher:2023zyh}, which can be well-approximated by $0.76/D(z)$. For the total number, we use $N_{\rm g} = 17\times10^6$~\cite{Rocher:2023zyh}.

Finally, QSOs, among the most luminous extragalactic sources, can trace the high-redshift \ac{LSS}. We use Ref.~\cite{Chaussidon:2022pqg} for the DESI-QSO redshift distribution, which is consistent with, e.g., Refs.~\cite{Yuan:2023ezi, Prada:2023lmw, DESI:2023duv}. We use Ref.~\cite{Prada:2023lmw} for the bias, and $N_{\rm g} = 3\times10^6$ for the total number~\cite{Chaussidon:2022pqg, DESI:2023duv}.

The left panel of Fig.~\ref{fig_zdist_bias} shows the redshift distributions of different DESI galaxy samples discussed above. The DESI-LRG ranges from $z \simeq 0.2$ to $\simeq 1.2$, and DESI-ELG from $z \simeq 0.6$ to $\simeq 1.6$, and DESI-QSO from $z \simeq 0.5$ to $>3$ thanks to their brightness. The right panel of Fig.~\ref{fig_zdist_bias} shows the bias as discussed above.

\section{Cross-correlation results}
\label{sec_CCresults}

\subsection{Cross-correlation signals and uncertainties}

\begin{figure*}[t!]
\includegraphics[width=0.48\textwidth]{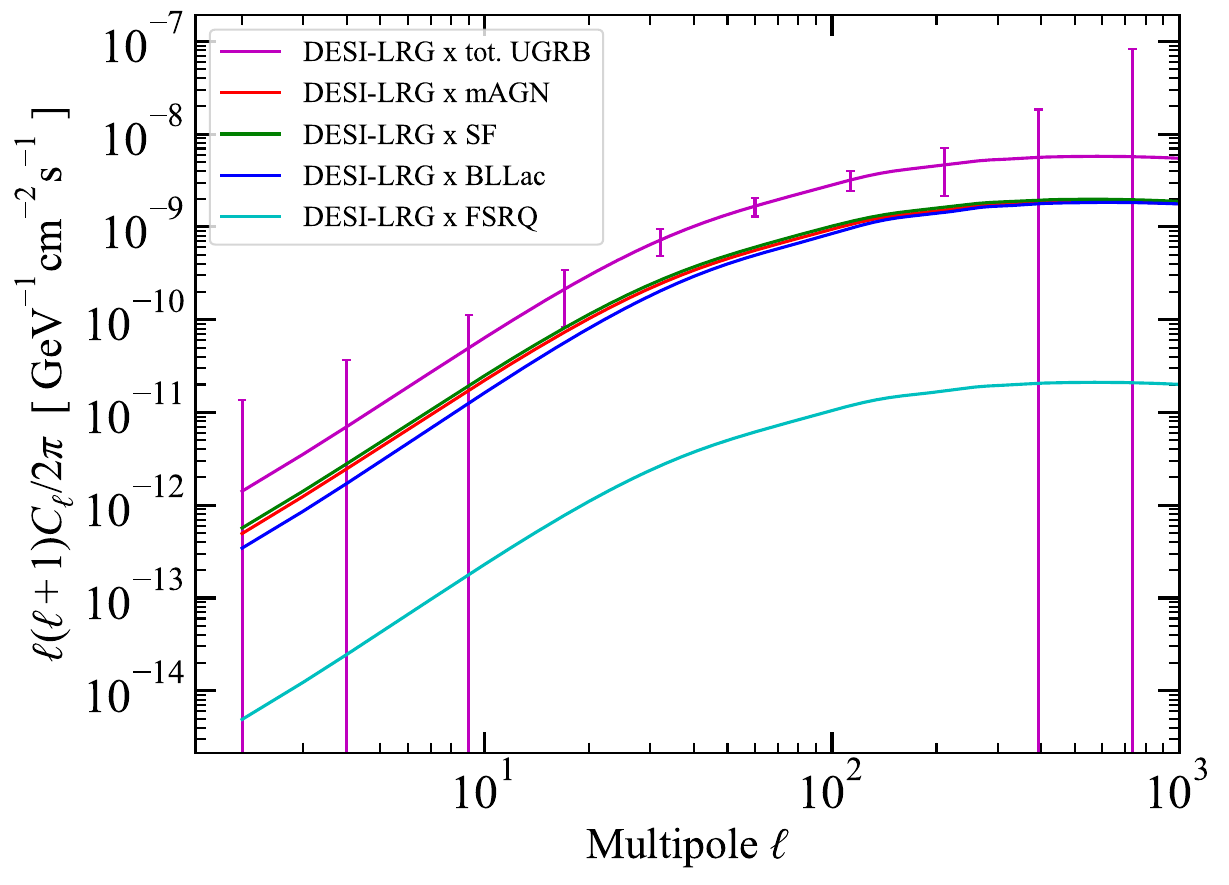}
\includegraphics[width=0.48\textwidth]{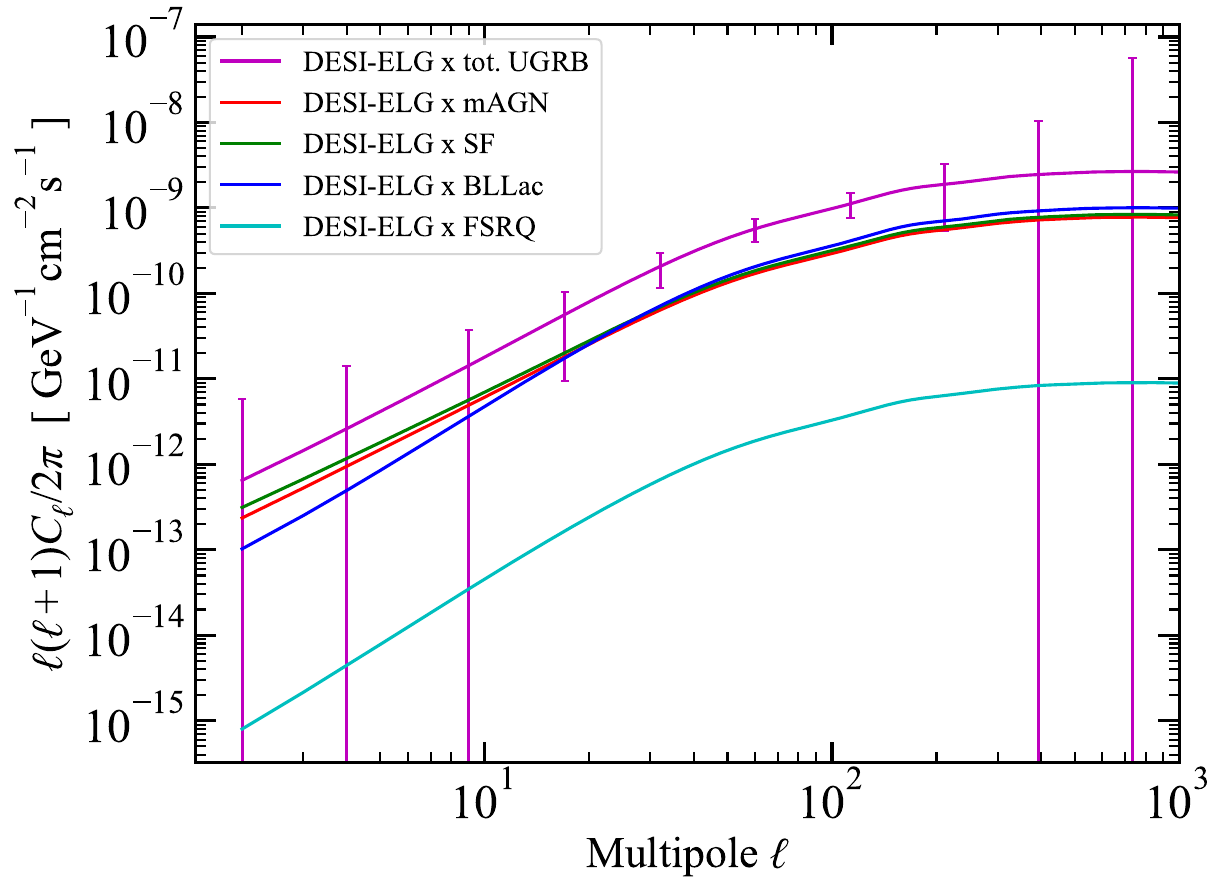}
\includegraphics[width=0.48\textwidth]{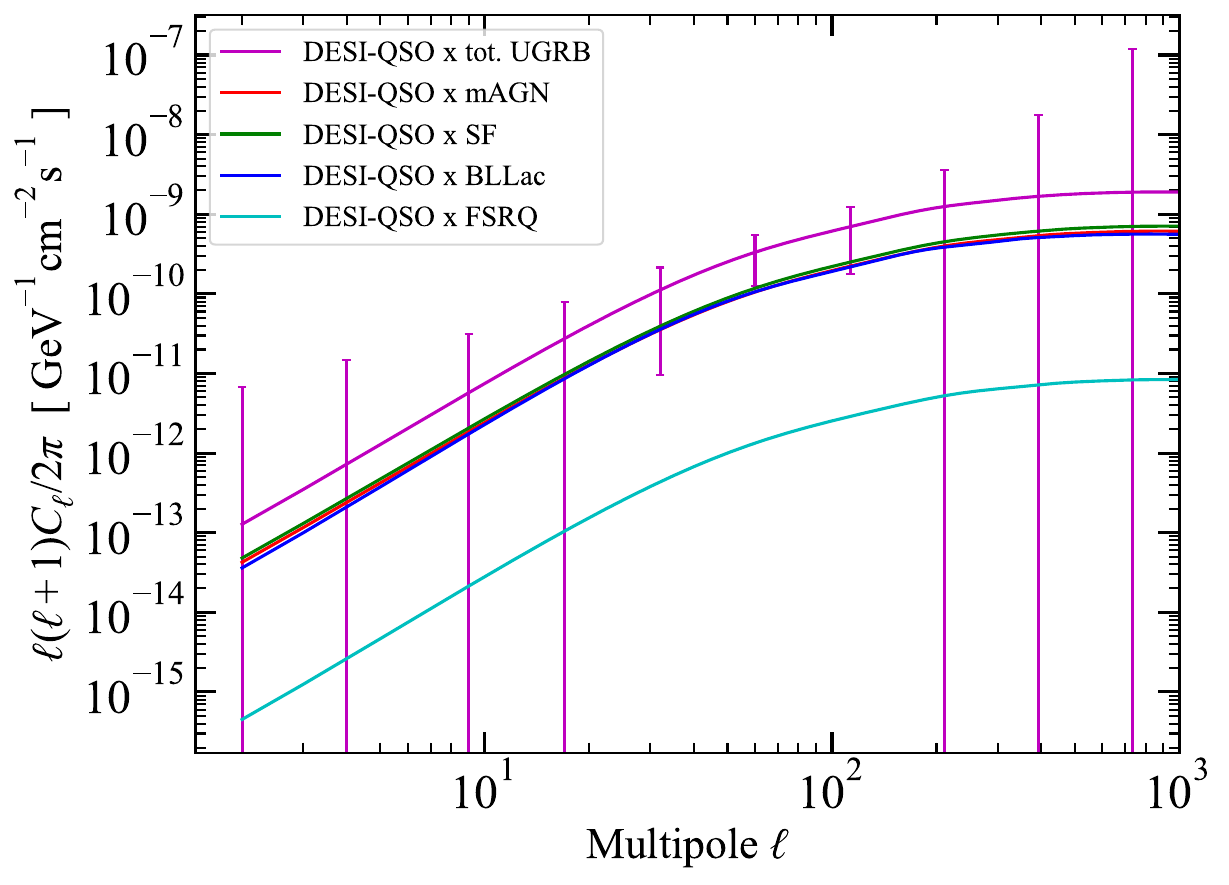}
\caption{Our predicted angular power spectra from cross correlating {\it Fermi}-LAT \ac{UGRB} and DESI galaxy samples. 
Also shown are the expected uncertainty on the total \ac{UGRB} calculated using Eq.~\eqref{eq_cov} and combining the different multipoles into ten logarithmic bins.
The results are shown for the first energy bin, $E_\gamma = 0.5$--1.0 GeV.
{\bf Upper left:} with DESI-LRG.
{\bf Upper right:} with DESI-QSO.
{\bf Bottom:} with DESI-ELG.
}
\label{fig_CC}
\end{figure*}

Fig.~\ref{fig_CC} shows our predicted cross-correlation angular power spectrum between the {\it Fermi}-LAT \ac{UGRB} and DESI galaxy samples, calculated using Eq.~\eqref{eq_Cl}. We also show the expected uncertainty on the total \ac{UGRB} calculated using Eq.~\eqref{eq_cov} and combining the different multipoles into ten logarithmic bins.
The relative heights of the different curves are determined by the gamma-ray intensity (Fig.~\ref{fig_gamma_spec}), bias, and overlap in the redshift distributions (Fig.~\ref{fig_zdist_bias}), and the number of galaxies.
\ac{BL Lac} and \ac{mAGN}, despite their lower fluxes, have cross correlation as high as \ac{SF} because of their larger bias or more overlap in the redshift distribution with the galaxy samples. 
For the cross correlation with DESI-ELG, \ac{SF} has a higher correlation at larger scales, and \ac{BL Lac} has a higher correlation at smaller scales. This is because \ac{SF} overlaps with DESI-ELG at higher redshifts and \ac{BL Lac} overlaps at lower redshifts.

\begin{figure*}
\includegraphics[width=0.49\textwidth]{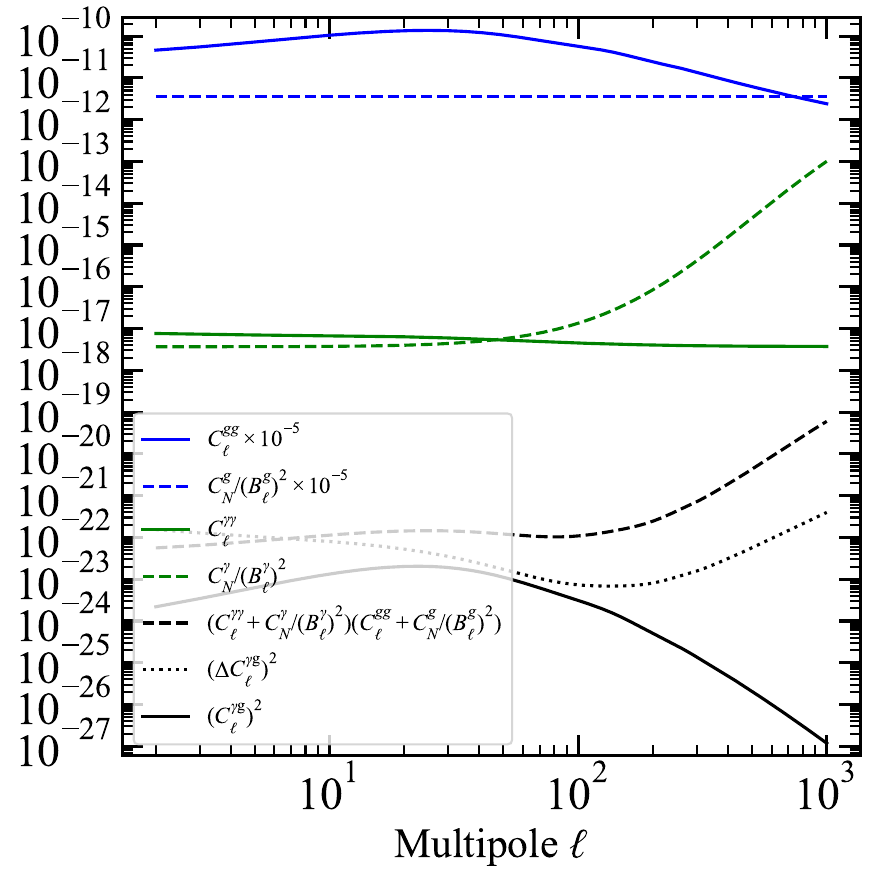}
\includegraphics[width=0.49\textwidth]{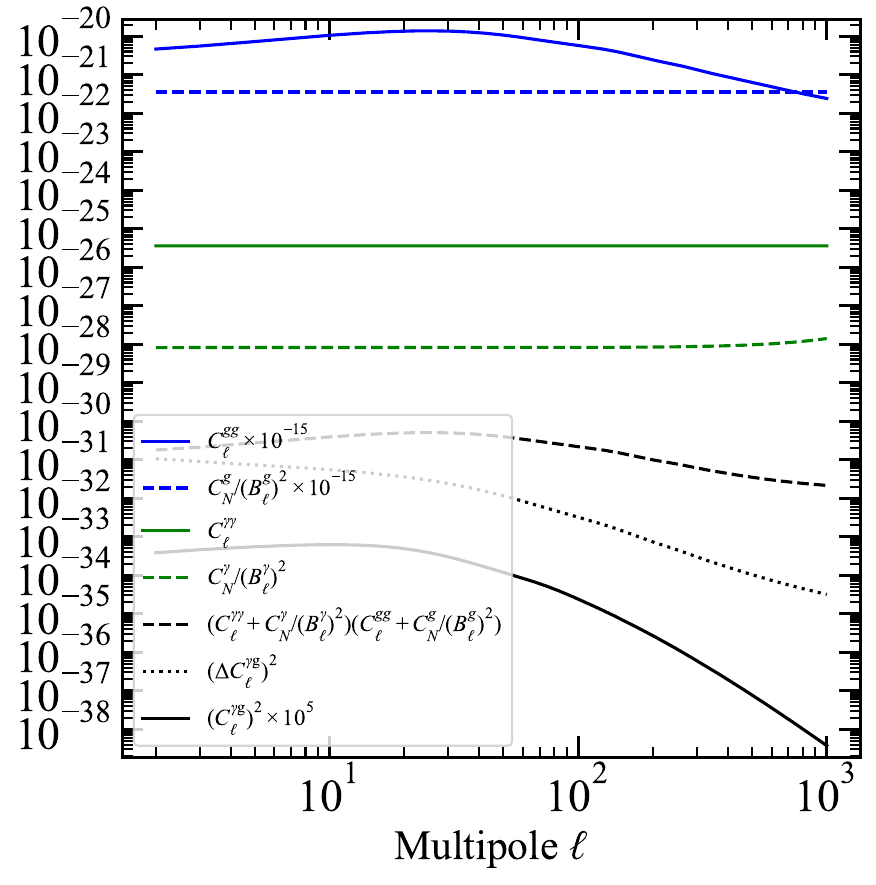}
\caption{The total variance $(\Delta C_\ell^{\gamma {\rm g}})^2$ and each term in Eq.~\eqref{eq_cov} of the angular power spectrum from cross correlating the {\it Fermi}-LAT \ac{UGRB} and DESI-LRG samples. The curves in the same color have the same units.
{\bf Left:} First energy bin, 0.5--1 GeV. 
{\bf Right:} Last energy bin, 331--1000 GeV.
}
\label{fig_var_terms}
\end{figure*}

Fig.~\ref{fig_var_terms} shows different terms in the variance $(\Delta C_\ell^{\gamma {\rm g}})^2$ (as given by Eq.~\eqref{eq_cov}) of the angular power spectrum from cross correlating {\it Fermi}-LAT \ac{UGRB} and DESI-LRG samples, for the first and last energy bins. This helps us understand the \ac{SNR}s better. For the galaxy sample terms (blue curves), the shot noise term (blue dashed) is $\ell$-independent, as $B^{\rm g}_\ell=1$ (as given by Eq.~\eqref{eq_Bl}), and is (nearly) always subdominant compared to $C_\ell^{{\rm g}{\rm g}}$, thanks to the large number of galaxies to be observed by DESI.

For the gamma-ray terms (green curves), the $C_\ell^{\rm \gamma\gamma}$ is $\ell$-independent because it is dominated by the shot-noise term. The shot noise is comparable to the instrumental noise (green dashed) at lower energies and higher than the instrumental noise at higher energies because there is a smaller number of sources at higher energies with a corresponding higher variance in their distribution. The instrumental noise term is flat at small $\ell$ while increasing at large $\ell$ due to the decrease of the beam term $B_\ell^\gamma$, and the turning point is at larger $\ell$ for higher energies because of their better angular resolution.

\subsection{Signal-to-noise ratio (SNR)}

\begin{table}[h!]
\centering
\begin{tabular}{cccccc}
\hline
         & \ac{mAGN}   & \ac{SF}    & \ac{BL Lac} & \ac{FSRQ}  & Total     \\
\hline
DESI-LRG & 13.4 & 15.1 & 10.6 & 0.7 & 20.6  \\
DESI-ELG & 9.2 & 10.3 & 9.7 & 0.5 &  15.8       \\
DESI-QSO & 4.3  & 4.9 & 3.6& 0.2 & 6.8  \\
\hline
\end{tabular}
\caption{
Our predicted cross-correlation SNR (or significance in units of standard normal deviations, i.e., ``$\sigma$'') between the {\it Fermi}-LAT \ac{UGRB} and DESI galaxy samples, combining all the energy bins and multipoles. 
}
\label{tab_SNR}
\end{table}

The signal-to-noise ratio (SNR) of the cross correlation relates to the signal detectability and how well other measurements could be done using the cross correlation and can be calculated by
\begin{equation}
\mathrm{SNR}^2 = 
\sum_{i, \ell} 
\left(\frac{C_{i, \ell}^{\gamma {\rm g}}}
{\Delta C_{i, \ell}^{\gamma {\rm g}}}\right)^2 \,,
\end{equation}
where the index $i$ refers to the gamma-ray energy bins. 

Table~\ref{tab_SNR} shows the cross-correlation SNR (or significance in the unit of ``$\sigma$'') 
between the {\it Fermi}-LAT \ac{UGRB} and DESI galaxy positions, combining all the energy bins and multipoles. The highest SNR (from DESI-LRG) is much higher than current detections and previous forecasts (e.g., Refs.~\cite{Xia:2015wka, Cuoco:2017bpv, Pinetti:2019ztr, Paopiamsap:2023uuo}), thanks to the large number of galaxies, large sky coverage, and wide redshift coverage from DESI survey, and the reduced levels of instrumental noise assumed for \textit{Fermi}LAT, since we assume expected statistics obtained by the time DESI finishes observing.
The relative significance of the different components reflects the relative heights in Fig.~\ref{fig_CC}.
For example, the \ac{FSRQ} population contributes the least to the total significance due to its lowest intensity, and DESI-QSO has the lowest significance due to its smallest total number of counts and least overlap in redshift distribution with the gamma rays.

\begin{figure*}
\includegraphics[width=0.48\textwidth]{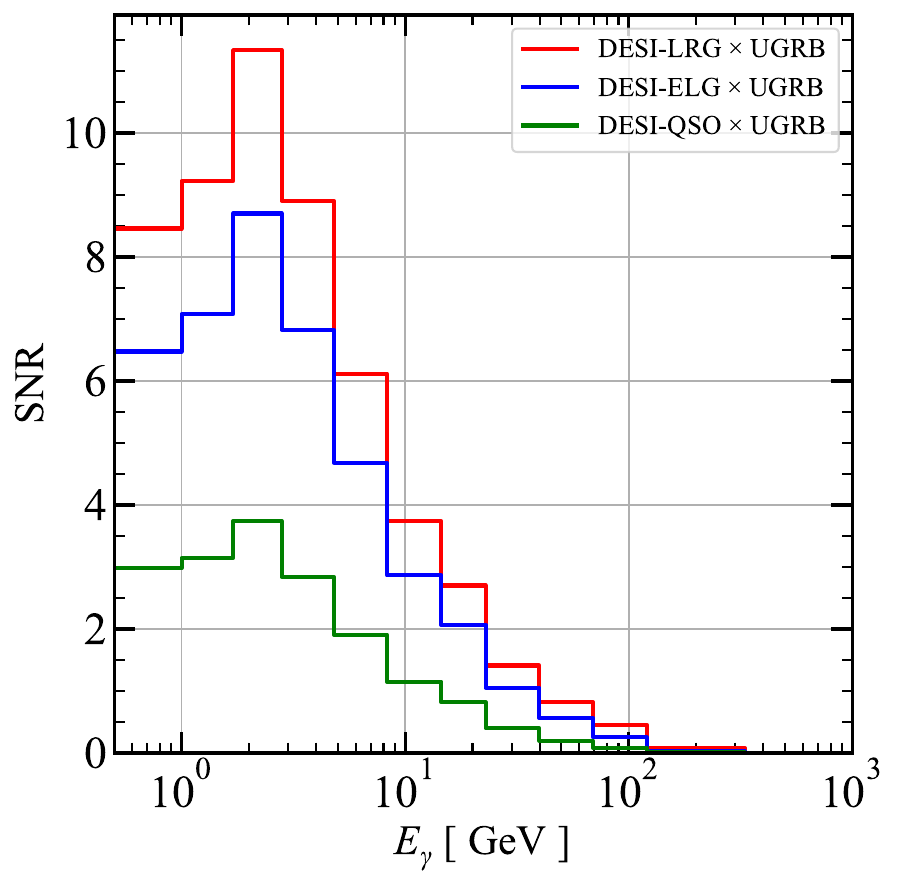}
\includegraphics[width=0.49\textwidth]{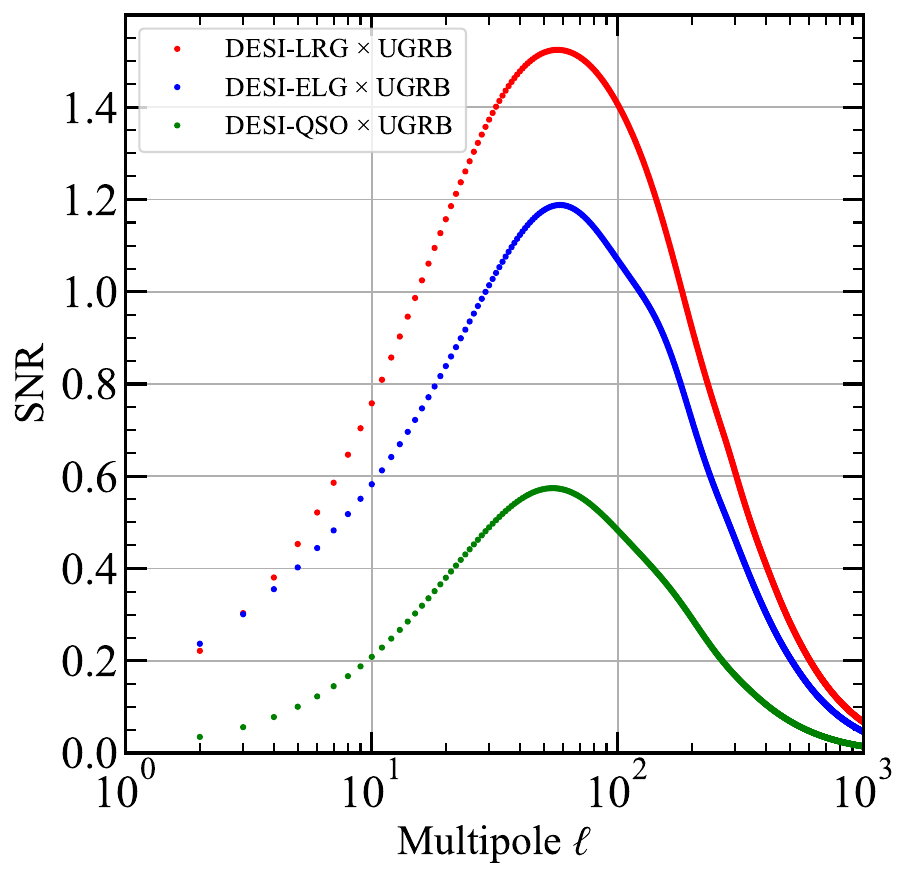}
\caption{Breakdown of the SNRs in Table~\ref{tab_SNR}.
{\bf Left:} The SNRs in the different gamma-ray energy bins.
{\bf Right:} The SNRs at different multipoles $\ell$.
}
\label{fig_SNR_dep}
\end{figure*}

Fig.~\ref{fig_SNR_dep} shows the breakdown of the SNR at different gamma-ray energy bins (left panel) and multipoles $\ell$ (right panel). For the left panel, the SNRs peak at the third energy bin (1.7--2.8 GeV) and decrease at lower and higher energies due to worse angular resolution (which suppresses the power spectrum over a wider range of scales) and lower gamma-ray statistics --- i.e., higher instrumental noise ---, respectively.
For the right panel, the SNRs peak around $\ell=60$ and decrease at smaller and larger $\ell$ due to cosmic variance and the suppression of the power spectra due to the limited angular resolution, respectively.

\section{Measuring the redshift distribution of  the \ac{UGRB}}
\label{sec_UGRBzDep}

Starting from this section, we study how we can learn about the origin of the \ac{UGRB} from the cross-correlation analyses. This section is about measuring the redshift distribution of the UGRB, and the next section is about searching for \ac{DM} origin.

The cross-correlation signal heavily depends on the redshift distribution of the gamma-ray intensity (Sec.~\ref{sec_cosmo_APS}),  suggesting that it can provide \ac{UGRB} redshift tomography. Here, we need to divide the cross correlation into different redshift bins, i.e., redshift tomography, which is feasible for (DESI) spectroscopic surveys. As an illustration, we divide the cross correlation into seven redshift bins: $z=0.0$--0.5, 0.5--1.0, ..., 3.0--3.5.

We use the Fisher information matrix formalism to forecast the constraints on the \ac{UGRB} flux normalization $\lambda$ in each redshift bin (with fiducial values $\lambda=1$ for all redshift bins). The Fisher matrix can be calculated by
\be
F_{\alpha \beta}
& =
\sum_{i, j, \ell}
\frac{1}{ \left(\Delta \hat{C}^{\gamma {\rm g}}_{i, j, \ell}\right)^2 }
\frac{\partial C^{\gamma {\rm g}}_{i, j, \ell}\left(\left\{\bar{\lambda}_\eta\right\}\right)}{\partial \bar{\lambda}_\alpha} 
\frac{\partial C^{\gamma {\rm g}}_{i, j, \ell}\left(\left\{\bar{\lambda}_\eta\right\}\right)}{\partial \bar{\lambda}_\beta} \,,
\label{eq_Fisher_Ebins_zbins}
\ee
where the indices $i$ and $j$ refer to the energy and redshift bins, respectively. Note that here we only split the galaxy samples in different redshift bins, cross correlating in each case with the whole integrated gamma-ray flux. The variance $\left(\Delta \hat{C}^{\gamma {\rm g}}_{i, j, \ell}\right)^2$ at the fiducial values $\bar{\lambda}$ can be calculated using Eq.~\eqref{eq_cov}, and $\{\lambda\} = \lambda_1, \lambda_2, .., \lambda_7$ are the parameters to be constrained, i.e. the flux normalization of the \ac{UGRB} in each redshift bin. For each redshift bin, all energy bins have the same flux normalization parameter $\lambda$. The fiducial values,
$\bar{\lambda}$, correspond to our benchmark \ac{UGRB} model, as presented in Sec.~\ref{sec_gamma}. Note that $\bar{\lambda}$ is degenerate with the bias, $b(z)$, which has uncertainty. Here, we assume the bias is perfectly known, but in a realistic case, their combination, $dI/dz \times b(z)$, is measured. In any case, our reported sensitivities can be straightforwardly interpreted in terms of $dI/dz\times b(z)$.

\begin{figure}
\includegraphics[width=0.46\textwidth]{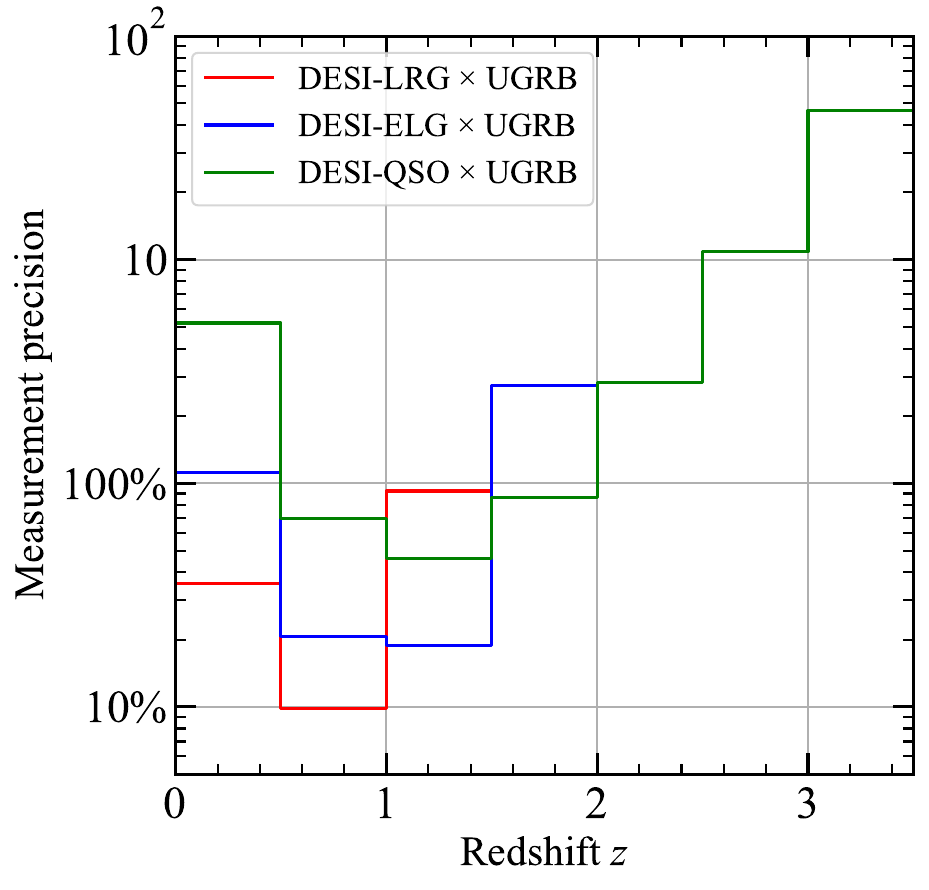}
\caption{Our forecasted precision (i.e., relative error) of measuring the \ac{UGRB} flux in each redshift bin using the cross correlation between the Fermi \ac{UGRB} and DESI galaxy samples.
}
\label{Figfile_z_dep_measurement}
\end{figure}

Fig.~\ref{Figfile_z_dep_measurement} shows our forecasted precision (i.e., relative error) of measuring the UGRB flux in each redshift bin. 
The shape of the lines can be understood from the redshift distributions of the DESI galaxy samples in the left panel of Fig.~\ref{fig_zdist_bias}. Specifically, the more galaxies in a redshift bin, the higher the measurement precision. Overall, the precision can be as high as 10\% for the bin of $0.5<z<1.0$. The DESI-QSO has a lower precision due to a smaller total number of galaxies but extends to higher redshifts. The bins of $z>1.5$ are much less constrained, mainly because of much less gamma-ray emission from our model. If UGRB actually extends to larger redshifts, these bins will be better constrained. 
The precision can be improved if we combine the three cross correlation. 

The reported sensitivities are expected, given the redshift distributions shown in Fig.~\ref{fig_zdist_bias}. One potential way to improve \ac{UGRB} tomography would be to use an adaptive redshift binning in the galaxy samples, rather than uniform as chosen for this work. We leave this study for future work.

The results here are for the total \ac{UGRB}. However, such measurements can provide valuable insights into the origin of the \ac{UGRB}, including both astrophysical and \ac{DM} contributions. For example, BL-Lac-dominated UGRB would favor higher redshifts while \ac{DM}-dominated UGRB would favor lower redshifts and peak around $z=0$.

\section{Dark matter}
\label{sec_DM}
The cross correlation between DESI and {\it Fermi}-LAT \ac{UGRB} is a powerful tool to search for \ac{DM}, which is a possible origin of the \ac{UGRB} in addition to the astrophysical populations~\cite{Ando:2005xg, Fornasa:2012gu, Ando:2013ff, Cholis:2013ena,  Camera:2014rja, Ando:2014aoa, Cuoco:2015rfa, Fermi-LAT:2015qzw, Regis:2015zka, Shirasaki:2015nqp, Regis:2015zka, Feyereisen:2015cea, Liu:2016ngs, Shirasaki:2016kol, Campbell:2017qpa, Ammazzalorso:2018evf, Blanco:2018esa, Pinetti:2019ztr, Arbey:2019vqx, Yang:2020zcu, Liang:2020roo, Zhang:2021mth, Bartlett:2022ztj, Paopiamsap:2023uuo, Delos:2023ipo, Ganjoo:2024hpn, Cholis:2024hmd}. The phenomenology of the signal originating from processes related to \ac{DM} is expected to be different than that from standard astrophysical sources. Therefore, increasing the sensitivity to such features boosts the potential to shed light on the nature of \ac{DM}. 

\begin{figure}
\includegraphics[width=0.49\textwidth]{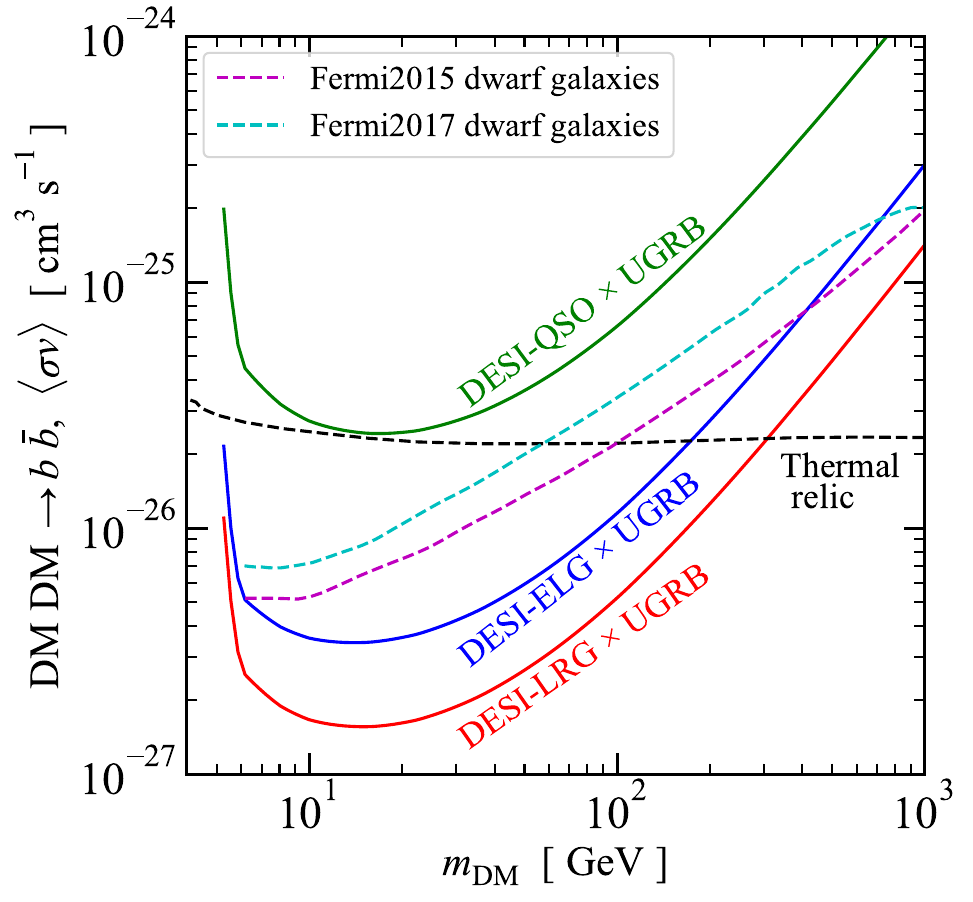}
\caption{Our forecasted sensitivity at $2\sigma$ (95.4\%) confidence level for the WIMP \ac{DM} annihilation parameter space (mass vs. thermal-averaged annihilation cross section times relative velocity) from measuring the cross correlation between the {\it Fermi}-LAT \ac{UGRB} and DESI galaxy samples.
The annihilation channel is DM$\,$DM $\rightarrow b\, \bar{b}$.
For comparison, we also plot the constraints from {\it Fermi}-LAT observations of dwarf spheroidal galaxies~\cite{Fermi-LAT:2015att, Fermi-LAT:2016uux}, which are typically regarded as the strongest current limits in this mass range. 
The black dashed line represents the thermal relic cross section required to produce the observed present \ac{DM} abundance (assuming only $s$-wave annihilation).
}
\label{Fig_DM_Ann}
\end{figure}
We study the annihilation of the weakly-interacting massive particles (WIMPs) \ac{DM} as an example. 
For most of the \ac{DM} calculations, we follow Refs.~\cite{Pinetti:2019ztr, Pinetti:2021jjs}. We summarize the main ingredients below and refer the readers to Refs.~\cite{Pinetti:2019ztr, Pinetti:2021jjs}
\footnote{Mainly in Sec.~4 of Ref.~\cite{Pinetti:2019ztr} and Sec.5.2 and Appendix~E of Ref.~\cite{Pinetti:2021jjs}.}
for further details.

The intensity of the UGRB possibly contributed by WIMP annihilation can be calculated by
\be
\frac{d^2I_{\rm DM}}{dE_\gamma dz}
& = \frac{1}{4 \pi} \frac{c}{H(z)} \, \frac{\langle \sigma v \rangle}{2} \Delta^2(z) \, \left (\frac{\Omega_{\rm DM} \,\rho_{\rm c} }{m_{\rm DM}} \right)^2 \\
& \times (1+z)^3  \frac{d N}{d E_\gamma}[(1+z)E_\gamma]  {\rm e}^{-\tau_{\gamma\gamma} \left[(1+z)E_\gamma,  z \right]},
     \label{eq_Idm}
\ee
where $c$ is the light speed, $H(z)$ the Hubble parameter, $m_{\rm DM}$ the WIMP mass, $\langle \sigma v \rangle$ the thermal-averaged annihilation cross section times the relative velocity of the DM particles, $\Omega_{\rm DM}$ the \ac{DM} density in the Universe today in units of $\rho_{\rm c}$, i.e., the critical density of the Universe, and $d N/d E_\gamma$ is the gamma-ray spectrum per WIMP annihilation.

The clumping factor $\Delta^2(z)$ appears in Eq.~\eqref{eq_Idm} because the DM annihilation signal probes the DM density squared and $\langle\rho^2\rangle \ne \langle\rho\rangle^2$, and it is defined as
\begin{equation}
\Delta^2(z) \,\equiv \,\frac{\langle\rho^2\rangle}{{\langle \rho \rangle}^2}\, =\, \int_{{M_{\min}}}^{{M_{\max}}} d M\, \frac{d N}{d M}(M,z)\int d^3x \;\frac{\rho^2({\bm x}|M)}{{\langle \rho \rangle}^2},
\end{equation}
where $\rho({\bm x}|M)$ is the density profile of a DM halo of mass $M$ and $\langle\rho\rangle$ is its average value. The minimum and maximum of the DM halo mass, $M_{\min}$ and $M_{\max}$, are chosen to be $10^{-6}$ and $10^{18}$ of solar mass, respectively. We describe the DM halo with a Navarro–Frenk–White profile~\cite{Navarro:1996gj} with a concentration parameter related to the halo mass as derived in Ref.~\cite{Correa:2015dva} for the low-redshift regime ($z \le 4$). We also include the presence of substructures within main haloes by replacing $\rho^2 ({\bm x}|M)$ with $[1+B(M, z)] \rho^2 ({\bm x}|M, z)$ where $B(M, z)$ is the boost factor, for which we adopt the parameterization of Ref.~\cite{Moline:2016pbm}.

To obtain the forecasted sensitivity, we use a $\chi^2$ statistic with one free parameter, $\langle \sigma v \rangle$, i.e.,
\be
\Delta \chi^2
=
\sum_{i, j, \ell} \left(\frac{ C_{i, j, \ell}^{\gamma^\prime {\rm g}} }{\Delta C_{i, j, \ell}^{\gamma^\prime {\rm g}}}\right)^2-\sum_{i, j, \ell}\left(\frac{C_{i, j, \ell}^{\gamma {\rm g}}}{\Delta C_{i, j, \ell}^{\gamma {\rm g}}}\right)^2 
\label{eq:deltachi2}
\,,
\ee
where $C_{i, j, \ell}^{\gamma {\rm g}}$ is the cross-correlation signal between DESI galaxy samples and the total astrophysical \ac{UGRB}, and $\Delta C_{i, j, \ell}^{\gamma {\rm g}}$ is its uncertainty.
The $C_{i, j, \ell}^{\gamma^\prime {\rm g}}$ represents the cross correlation between DESI galaxy samples and the total \ac{UGRB}, including both astrophysical and \ac{DM} components, and $\Delta C_{i, j, \ell}^{\gamma^\prime {\rm g}}$ is its uncertainty. 
Therefore, the second term on the right-hand side of Eq.~\eqref{eq:deltachi2} represents the null hypothesis (i.e., only astrophysical contribution to the \ac{UGRB}), and the term on the left to the minus sign represents the alternative hypothesis.
We set the sensitivity on the $\langle \sigma v \rangle$ at $2\sigma$ (95.4\%) confidence level, corresponding to $\Delta \chi^2 = 4$.
The above consideration is for a specific \ac{DM} mass $m_{\rm DM}$, and we repeat the procedure for different masses. 

Fig.~\ref{Fig_DM_Ann} shows our forecasted sensitivity on WIMP \ac{DM} annihilation from measuring the cross correlation between DESI galaxy samples and {\it Fermi}-LAT \ac{UGRB}. 
The strength of the sensitivity is related to the cross-correlation SNR, for which DESI-LRG $>$ DESI-ELG $>$ DESI-QSO. It is also related to the redshift distribution, for which LRG goes to lower redshifts than ELG, which goes to lower redshifts than QSO, and WIMP annihilation favors lower redshifts (peaks at $z=0$).
Stronger sensitivity would be achieved by the combination of all DESI samples, but our current results show that the sensitivity is expected to be completely dominated by LRGs. The sensitivity is the strongest at $m_{\rm DM} \simeq 20$~GeV because its annihilating gamma-ray spectrum peaks around a few GeV, where the cross-correlation SNR peaks (Fig.~\ref{fig_SNR_dep}, left panel). The quick decrease of the sensitivity for \ac{DM} masses below $m_{\rm DM} \simeq 20$~GeV is in part due to the decrease in the cross-correlation significance due to worse angular resolution, but also because our analysis started at $E_\gamma = 0.5$~GeV. Starting the analysis at lower energies (e.g., 0.1~GeV widely used for {\it Fermi}-LAT) will increase the sensitivity for smaller $m_{\rm DM}$.
The DESI-LRG sample gives the most sensitive result, which can probe $m_{\rm DM}$ up to $\simeq 300$ GeV, {\it three times higher than the constraints from {\it Fermi}-LAT observations of dwarf spheroidal galaxies}~\cite{Fermi-LAT:2015att, Fermi-LAT:2016uux} (see also Ref.~\cite{Hu:2023iex} for a more recent analysis) shown in the figure, which generally represents the currently strongest limits in this mass range. This could cover most of the parameter space of the galactic center GeV excess (e.g., Refs.~\cite{Vitale:2009hr, Hooper:2010mq, Abazajian:2010zy, Hooper:2011ti, Hooper:2013rwa, Gordon:2013vta, Abazajian:2014fta, Daylan:2014rsa, Calore:2014xka, Zhou:2014lva, Huang:2015rlu, Fermi-LAT:2015sau, DiMauro:2021raz, Cholis:2021rpp, Calore:2014nla, Agrawal:2014oha, Berlin:2015wwa, Zhong:2024vyi, Song:2024iup}) and the cosmic-ray antiproton excess (e.g., Refs.~\cite{Cui:2016ppb, Cuoco:2016eej, Cholis:2019ejx, Cuoco:2019kuu}), respectively.

Note that the results we report here are optimistic since we do not marginalize the uncertainties in the astrophysical contribution and the booster factor for cosmic \ac{DM} annihilation. 
For the latter, a recent paper~\cite{Paopiamsap:2023uuo} in 2023 found that the largest and smallest boost factors give rise to a factor of $\sim 3$ difference in the $\langle \sigma v \rangle$ sensitivity, which is smaller than an earlier work~\cite{Regis:2015zka} in 2015 that reported a difference of an order of magnitude.
On the other hand, our calculation uses only one redshift bin, and adopting redshift binning would increase the sensitivity. 
Moreover, gaining gamma-ray redshift information, as discussed in Sec.~\ref{sec_UGRBzDep}, may help to discriminate between the astrophysical and \ac{DM}-annihilation contributions.
These aspects should be included in the future data analysis.

\section{Conclusions}
\label{sec_concl}

The unresolved gamma-ray background (\ac{UGRB}), a critical component of the extragalactic gamma-ray sky, is the diffuse gamma-ray emission from numerous unresolved extragalactic sources~\cite{Fermi-LAT:2014ryh}, such as misaligned AGN (mAGN), galaxies with star-forming activities (SF), BL Lacertae objects (BL Lac), and flat-spectrum radio quasars (FSRQ)~\cite{Ando:2009nk, Xia:2011ax, Harding:2012gk, DiMauro:2013xta, Tamborra:2014xia, DiMauro:2014wha, DiMauro:2015tfa, Xia:2015wka, Linden:2016fdd, Wang:2016zyi, DiMauro:2017ing, Lamastra:2017iyo, Cuoco:2017bpv, Komis:2017jta, Pinetti:2019ztr, Hooper:2016gjy, Stecker:2019ybn, Qu:2019zln, Owen:2021evt, Owen:2021qul, Blanco:2021icw, Korsmeier:2022cwp,Xue:2024tkj, Min:2024bnj, Cholis:2024hmd}, as well as from potential contributions by exotic processes such as \ac{DM} annihilation or decay~\cite{Ando:2005xg, Fornasa:2012gu, Ando:2013ff, Cholis:2013ena,  Camera:2014rja, Ando:2014aoa, Cuoco:2015rfa, Fermi-LAT:2015qzw, Regis:2015zka, Shirasaki:2015nqp, Regis:2015zka, Feyereisen:2015cea, Liu:2016ngs, Shirasaki:2016kol, Campbell:2017qpa, Ammazzalorso:2018evf, Blanco:2018esa, Pinetti:2019ztr, Arbey:2019vqx, Yang:2020zcu, Liang:2020roo, Zhang:2021mth, Bartlett:2022ztj, Paopiamsap:2023uuo, Delos:2023ipo, Ganjoo:2024hpn, Cholis:2024hmd}. Understanding the origin of the UGRB is critical for high-energy astrophysics, as it provides insights into some of the most extreme environments in the universe. It also offers valuable clues for probing fundamental physics, particularly in the context of \ac{DM} searches.

In this paper, we study the cross correlation of the UGRB~\cite{Fermi-LAT:2014ryh} with galaxy clustering measurements from DESI~\cite{DESI:2016fyo, DESI_web} and using the cross correlation to study the origin of the UGRB.
DESI is conducting one of the most comprehensive surveys mapping the cosmic \ac{LSS} and the largest galaxy redshift survey to date~\cite{DESI:2016fyo, DESI_web}. 
We start by presenting the calculational framework for the two-point angular correlation between the diﬀuse gamma-ray background and a galaxy sample (Sec.~\ref{sec_cosmo_APS}). We also present the calculation of the measurement uncertainties (Sec.~\ref{sec_cosmo_unc}).

We set up our astrophysical UGRB benchmark model by calculating the astrophysical contributions to the UGRB and comparing with the {\it Fermi}-LAT measurements to validate it (Figs.~\ref{fig_gamma_spec} and \ref{fig_gamma_Cp}). The astrophysical components include mAGN, SF, BL Lac, and FSRQ.
For the galaxy samples from the DESI survey, we use luminous red galaxies (LRGs)~\cite{Zhou:2020nwq, DESI:2022gle, Yuan:2023ezi, Prada:2023lmw, Rezaie:2023lvi}, emission line galaxies (ELGs)~\cite{Raichoor:2022jab, Rocher:2023zyh}, and quasars (QSOs)~\cite{Chaussidon:2022pqg, Yuan:2023ezi, Prada:2023lmw, DESI:2023duv}.

We then compute the cross correlation between the UGRB and DESI galaxies to forecast the angular power spectra (Fig.~\ref{fig_CC}), measurement uncertainties (Fig.~\ref{fig_var_terms}), and the SNRs (Table.~\ref{tab_SNR}). {\it We predict that the cross-correlation SNRs between UGRB and DESI-LRG, DESI-ELG, and DESI-QSO can be about 20.6, 15.8, and 6.8}. The highest SNR (from DESI-LRG) is much higher than the previous forecasts or detections (e.g., Refs.~\cite{Xia:2015wka, Cuoco:2017bpv, Pinetti:2019ztr, Paopiamsap:2023uuo}), thanks to the higher galaxy number density, sky coverage, redshift overlap with the expected distribution of UGRB, and improved gamma-ray statistics.

Finally, we forecast studies of the origin of the UGRB using the cross-correlation analyses. First, we find that by taking advantage of the galaxy redshift survey of DESI, the redshift tomography can inform about the redshift distribution of the UGRB flux. Specifically, the flux of the UGRB in certain redshift bins can be measured with a precision of 10\% (Fig.~\ref{fig_CC}) for $0.5\lesssim z \lesssim 1$. This will offer valuable insights into the origin of the UGRB.
Second, we forecast the sensitivity of searching for WIMP \ac{DM} and find that the $\langle \sigma v \rangle$ (i.e., thermal-averaged cross section times the relative velocity) can be potentially probed up to \ac{DM} mass of $m_{\rm DM} \simeq 300$~GeV at 2$\sigma$ (95.4\%) confidence level, three times higher sensitivities than the currently strongest limits in this mass range from {\it Fermi}-LAT observations of dwarf spheroidal galaxies (Fig.~\ref{Fig_DM_Ann})~\cite{Fermi-LAT:2015att, Fermi-LAT:2016uux}. 
These sensitivities could cover most of the parameter space of the galactic center GeV excess (e.g., Refs.~\cite{Vitale:2009hr, Hooper:2010mq, Abazajian:2010zy, Hooper:2011ti, Hooper:2013rwa, Gordon:2013vta, Abazajian:2014fta, Daylan:2014rsa, Calore:2014xka, Zhou:2014lva, Huang:2015rlu, Fermi-LAT:2015sau, DiMauro:2021raz, Cholis:2021rpp, Calore:2014nla, Agrawal:2014oha, Berlin:2015wwa, Zhong:2024vyi, Song:2024iup}) and the cosmic-ray antiproton excess (e.g., ~\cite{Cui:2016ppb, Cuoco:2016eej, Cholis:2019ejx, Cuoco:2019kuu}), respectively.
We leave a more detailed study that marginalizes the uncertainties in the astrophysical contribution and boost factor, which would decrease the sensitivity, and that includes the redshift tomography, which would increase the sensitivity, to future data analysis.
In addition, more science cases using these cross correlation deserve to be explored in future work.

Our calculation starts from minimum gamma-ray energy of 0.5~GeV to be consistent with Ref.~\cite{Fermi-LAT:2018udj}. Starting from lower energies (e.g., 0.1~GeV widely used for {\it Fermi}-LAT) would increase the SNR and sensitivities to the UGRB redshift-distribution measurements and \ac{DM} searches, especially for lower \ac{DM} masses. Moreover, our results depend more on the total UGRB, which is well-measured, than on the relative contribution of each component.

Our work underscores the importance of cross correlating the UGRB with cosmic large-scale structure tracers and highlights the value of adopting a multiwavelength approach to advance our understanding of high-energy astrophysical phenomena and fundamental physics. As future gamma-ray observations and galaxy surveys continue to improve in precision, this approach will enable more accurate reconstructions of the UGRB's origins, enhance the detection of fainter signals, and provide stronger sensitivity to search for \ac{DM} and other potential new physics.  Finally, analogous cross-correlations can soon be carried out in the redshift ranges of interest for the UGRB, possibly with more volume, with e.g., SPHEREx \cite{SPHEREx:2014bgr}, thus various spectroscopic surveys \cite{Spergel:2015sza, Amendola:2016saw, MUST_web} may open the door to probing the gamma-ray emission at higher redshifts.

\section*{Acknowledgments}
We are grateful for the helpful discussions with Marco Ajello, Carlos Blanco, Alessandro Cuoco, Nicolao Fornengo, Dan Hooper, Xiaoyuan Huang, Hui Kong, Yunfeng Liang, Mehdi Rezaie, and Ben Safdi.
B.Z. and E.P. were supported by the Fermi Research Alliance, LLC, acting under Contract No.\ DE-AC02-07CH11359.
JLB acknowledges funding from the Ramón y Cajal Grant RYC2021-033191-I, financed by MCIN/AEI/10.13039/501100011033 and by
the European Union “NextGenerationEU”/PRTR, as well as the project UC-LIME (PID2022-140670NA-I00), financed by MCIN/AEI/ 10.13039/501100011033/FEDER, UE.
HAC was supported by the National Science Foundation Graduate Research Fellowship under Grant No.\ DGE2139757.
This work was supported at Johns Hopkins by NSF Grant Nos.\ 2112699 and 2412361, the Simons Foundation, and the Templeton Foundation.

\newpage
\twocolumngrid

\bibliography{references}

\begin{thebibliography}{123}%
\makeatletter
\providecommand \@ifxundefined [1]{%
 \@ifx{#1\undefined}
}%
\providecommand \@ifnum [1]{%
 \ifnum #1\expandafter \@firstoftwo
 \else \expandafter \@secondoftwo
 \fi
}%
\providecommand \@ifx [1]{%
 \ifx #1\expandafter \@firstoftwo
 \else \expandafter \@secondoftwo
 \fi
}%
\providecommand \natexlab [1]{#1}%
\providecommand \enquote  [1]{``#1''}%
\providecommand \bibnamefont  [1]{#1}%
\providecommand \bibfnamefont [1]{#1}%
\providecommand \citenamefont [1]{#1}%
\providecommand \href@noop [0]{\@secondoftwo}%
\providecommand \href [0]{\begingroup \@sanitize@url \@href}%
\providecommand \@href[1]{\@@startlink{#1}\@@href}%
\providecommand \@@href[1]{\endgroup#1\@@endlink}%
\providecommand \@sanitize@url [0]{\catcode `\\12\catcode `\$12\catcode
  `\&12\catcode `\#12\catcode `\^12\catcode `\_12\catcode `\%12\relax}%
\providecommand \@@startlink[1]{}%
\providecommand \@@endlink[0]{}%
\providecommand \url  [0]{\begingroup\@sanitize@url \@url }%
\providecommand \@url [1]{\endgroup\@href {#1}{\urlprefix }}%
\providecommand \urlprefix  [0]{URL }%
\providecommand \Eprint [0]{\href }%
\providecommand \doibase [0]{http://dx.doi.org/}%
\providecommand \selectlanguage [0]{\@gobble}%
\providecommand \bibinfo  [0]{\@secondoftwo}%
\providecommand \bibfield  [0]{\@secondoftwo}%
\providecommand \translation [1]{[#1]}%
\providecommand \BibitemOpen [0]{}%
\providecommand \bibitemStop [0]{}%
\providecommand \bibitemNoStop [0]{.\EOS\space}%
\providecommand \EOS [0]{\spacefactor3000\relax}%
\providecommand \BibitemShut  [1]{\csname bibitem#1\endcsname}%
\let\auto@bib@innerbib\@empty
\bibitem [{\citenamefont {Ackermann}\ \emph
  {et~al.}(2015{\natexlab{a}})\citenamefont {Ackermann} \emph
  {et~al.}}]{Fermi-LAT:2014ryh}%
  \BibitemOpen
  \bibfield  {author} {\bibinfo {author} {\bibfnamefont {M.}~\bibnamefont
  {Ackermann}} \emph {et~al.} (\bibinfo {collaboration} {Fermi-LAT}),\
  }\bibfield  {title} {\enquote {\bibinfo {title} {{The spectrum of isotropic
  diffuse gamma-ray emission between 100 MeV and 820 GeV}},}\ }\href {\doibase
  10.1088/0004-637X/799/1/86} {\bibfield  {journal} {\bibinfo  {journal}
  {Astrophys. J.}\ }\textbf {\bibinfo {volume} {799}},\ \bibinfo {pages} {86}
  (\bibinfo {year} {2015}{\natexlab{a}})},\ \Eprint
  {http://arxiv.org/abs/1410.3696} {arXiv:1410.3696 [astro-ph.HE]} \BibitemShut
  {NoStop}%
\bibitem [{\citenamefont {Ando}\ and\ \citenamefont
  {Pavlidou}(2009)}]{Ando:2009nk}%
  \BibitemOpen
  \bibfield  {author} {\bibinfo {author} {\bibfnamefont {Shin'ichiro}\
  \bibnamefont {Ando}}\ and\ \bibinfo {author} {\bibfnamefont {Vasiliki}\
  \bibnamefont {Pavlidou}},\ }\bibfield  {title} {\enquote {\bibinfo {title}
  {{Imprint of Galaxy Clustering in the Cosmic Gamma-Ray Background}},}\ }\href
  {\doibase 10.1111/j.1365-2966.2009.15605.x} {\bibfield  {journal} {\bibinfo
  {journal} {Mon. Not. Roy. Astron. Soc.}\ }\textbf {\bibinfo {volume} {400}},\
  \bibinfo {pages} {2122} (\bibinfo {year} {2009})},\ \Eprint
  {http://arxiv.org/abs/0908.3890} {arXiv:0908.3890 [astro-ph.HE]} \BibitemShut
  {NoStop}%
\bibitem [{\citenamefont {Xia}\ \emph {et~al.}(2011)\citenamefont {Xia},
  \citenamefont {Cuoco}, \citenamefont {Branchini}, \citenamefont {Fornasa},\
  and\ \citenamefont {Viel}}]{Xia:2011ax}%
  \BibitemOpen
  \bibfield  {author} {\bibinfo {author} {\bibfnamefont {Jun-Qing}\
  \bibnamefont {Xia}}, \bibinfo {author} {\bibfnamefont {Alessandro}\
  \bibnamefont {Cuoco}}, \bibinfo {author} {\bibfnamefont {Enzo}\ \bibnamefont
  {Branchini}}, \bibinfo {author} {\bibfnamefont {Mattia}\ \bibnamefont
  {Fornasa}}, \ and\ \bibinfo {author} {\bibfnamefont {Matteo}\ \bibnamefont
  {Viel}},\ }\bibfield  {title} {\enquote {\bibinfo {title} {{A
  cross-correlation study of the Fermi-LAT $\gamma$-ray diffuse extragalactic
  signal}},}\ }\href {\doibase 10.1111/j.1365-2966.2011.19200.x} {\bibfield
  {journal} {\bibinfo  {journal} {Mon. Not. Roy. Astron. Soc.}\ }\textbf
  {\bibinfo {volume} {416}},\ \bibinfo {pages} {2247--2264} (\bibinfo {year}
  {2011})},\ \Eprint {http://arxiv.org/abs/1103.4861} {arXiv:1103.4861
  [astro-ph.CO]} \BibitemShut {NoStop}%
\bibitem [{\citenamefont {Harding}\ and\ \citenamefont
  {Abazajian}(2012)}]{Harding:2012gk}%
  \BibitemOpen
  \bibfield  {author} {\bibinfo {author} {\bibfnamefont {J.~Patrick}\
  \bibnamefont {Harding}}\ and\ \bibinfo {author} {\bibfnamefont {Kevork~N.}\
  \bibnamefont {Abazajian}},\ }\bibfield  {title} {\enquote {\bibinfo {title}
  {{Models of the Contribution of Blazars to the Anisotropy of the
  Extragalactic Diffuse Gamma-ray Background}},}\ }\href {\doibase
  10.1088/1475-7516/2012/11/026} {\bibfield  {journal} {\bibinfo  {journal}
  {JCAP}\ }\textbf {\bibinfo {volume} {11}},\ \bibinfo {pages} {026} (\bibinfo
  {year} {2012})},\ \Eprint {http://arxiv.org/abs/1206.4734} {arXiv:1206.4734
  [astro-ph.HE]} \BibitemShut {NoStop}%
\bibitem [{\citenamefont {Di~Mauro}\ \emph
  {et~al.}(2014{\natexlab{a}})\citenamefont {Di~Mauro}, \citenamefont {Calore},
  \citenamefont {Donato}, \citenamefont {Ajello},\ and\ \citenamefont
  {Latronico}}]{DiMauro:2013xta}%
  \BibitemOpen
  \bibfield  {author} {\bibinfo {author} {\bibfnamefont {M.}~\bibnamefont
  {Di~Mauro}}, \bibinfo {author} {\bibfnamefont {F.}~\bibnamefont {Calore}},
  \bibinfo {author} {\bibfnamefont {F.}~\bibnamefont {Donato}}, \bibinfo
  {author} {\bibfnamefont {M.}~\bibnamefont {Ajello}}, \ and\ \bibinfo {author}
  {\bibfnamefont {L.}~\bibnamefont {Latronico}},\ }\bibfield  {title} {\enquote
  {\bibinfo {title} {{Diffuse $\gamma$-ray emission from misaligned active
  galactic nuclei}},}\ }\href {\doibase 10.1088/0004-637X/780/2/161} {\bibfield
   {journal} {\bibinfo  {journal} {Astrophys. J.}\ }\textbf {\bibinfo {volume}
  {780}},\ \bibinfo {pages} {161} (\bibinfo {year} {2014}{\natexlab{a}})},\
  \Eprint {http://arxiv.org/abs/1304.0908} {arXiv:1304.0908 [astro-ph.HE]}
  \BibitemShut {NoStop}%
\bibitem [{\citenamefont {Tamborra}\ \emph {et~al.}(2014)\citenamefont
  {Tamborra}, \citenamefont {Ando},\ and\ \citenamefont
  {Murase}}]{Tamborra:2014xia}%
  \BibitemOpen
  \bibfield  {author} {\bibinfo {author} {\bibfnamefont {Irene}\ \bibnamefont
  {Tamborra}}, \bibinfo {author} {\bibfnamefont {Shin'ichiro}\ \bibnamefont
  {Ando}}, \ and\ \bibinfo {author} {\bibfnamefont {Kohta}\ \bibnamefont
  {Murase}},\ }\bibfield  {title} {\enquote {\bibinfo {title} {{Star-forming
  galaxies as the origin of diffuse high-energy backgrounds: Gamma-ray and
  neutrino connections, and implications for starburst history}},}\ }\href
  {\doibase 10.1088/1475-7516/2014/09/043} {\bibfield  {journal} {\bibinfo
  {journal} {JCAP}\ }\textbf {\bibinfo {volume} {09}},\ \bibinfo {pages} {043}
  (\bibinfo {year} {2014})},\ \Eprint {http://arxiv.org/abs/1404.1189}
  {arXiv:1404.1189 [astro-ph.HE]} \BibitemShut {NoStop}%
\bibitem [{\citenamefont {Di~Mauro}\ \emph
  {et~al.}(2014{\natexlab{b}})\citenamefont {Di~Mauro}, \citenamefont {Cuoco},
  \citenamefont {Donato},\ and\ \citenamefont
  {Siegal-Gaskins}}]{DiMauro:2014wha}%
  \BibitemOpen
  \bibfield  {author} {\bibinfo {author} {\bibfnamefont {Mattia}\ \bibnamefont
  {Di~Mauro}}, \bibinfo {author} {\bibfnamefont {Alessandro}\ \bibnamefont
  {Cuoco}}, \bibinfo {author} {\bibfnamefont {Fiorenza}\ \bibnamefont
  {Donato}}, \ and\ \bibinfo {author} {\bibfnamefont {Jennifer~M.}\
  \bibnamefont {Siegal-Gaskins}},\ }\bibfield  {title} {\enquote {\bibinfo
  {title} {{Fermi-LAT $/gamma$-ray anisotropy and intensity explained by
  unresolved Radio-Loud Active Galactic Nuclei}},}\ }\href {\doibase
  10.1088/1475-7516/2014/11/021} {\bibfield  {journal} {\bibinfo  {journal}
  {JCAP}\ }\textbf {\bibinfo {volume} {11}},\ \bibinfo {pages} {021} (\bibinfo
  {year} {2014}{\natexlab{b}})},\ \Eprint {http://arxiv.org/abs/1407.3275}
  {arXiv:1407.3275 [astro-ph.HE]} \BibitemShut {NoStop}%
\bibitem [{\citenamefont {Di~Mauro}\ and\ \citenamefont
  {Donato}(2015)}]{DiMauro:2015tfa}%
  \BibitemOpen
  \bibfield  {author} {\bibinfo {author} {\bibfnamefont {Mattia}\ \bibnamefont
  {Di~Mauro}}\ and\ \bibinfo {author} {\bibfnamefont {Fiorenza}\ \bibnamefont
  {Donato}},\ }\bibfield  {title} {\enquote {\bibinfo {title} {{Composition of
  the Fermi-LAT isotropic gamma-ray background intensity: Emission from
  extragalactic point sources and dark matter annihilations}},}\ }\href
  {\doibase 10.1103/PhysRevD.91.123001} {\bibfield  {journal} {\bibinfo
  {journal} {Phys. Rev. D}\ }\textbf {\bibinfo {volume} {91}},\ \bibinfo
  {pages} {123001} (\bibinfo {year} {2015})},\ \Eprint
  {http://arxiv.org/abs/1501.05316} {arXiv:1501.05316 [astro-ph.HE]}
  \BibitemShut {NoStop}%
\bibitem [{\citenamefont {Xia}\ \emph {et~al.}(2015)\citenamefont {Xia},
  \citenamefont {Cuoco}, \citenamefont {Branchini},\ and\ \citenamefont
  {Viel}}]{Xia:2015wka}%
  \BibitemOpen
  \bibfield  {author} {\bibinfo {author} {\bibfnamefont {Jun-Qing}\
  \bibnamefont {Xia}}, \bibinfo {author} {\bibfnamefont {Alessandro}\
  \bibnamefont {Cuoco}}, \bibinfo {author} {\bibfnamefont {Enzo}\ \bibnamefont
  {Branchini}}, \ and\ \bibinfo {author} {\bibfnamefont {Matteo}\ \bibnamefont
  {Viel}},\ }\bibfield  {title} {\enquote {\bibinfo {title} {{Tomography of the
  Fermi-lat $\gamma$-ray Diffuse Extragalactic Signal via Cross Correlations
  With Galaxy Catalogs}},}\ }\href {\doibase 10.1088/0067-0049/217/1/15}
  {\bibfield  {journal} {\bibinfo  {journal} {Astrophys. J. Suppl.}\ }\textbf
  {\bibinfo {volume} {217}},\ \bibinfo {pages} {15} (\bibinfo {year} {2015})},\
  \Eprint {http://arxiv.org/abs/1503.05918} {arXiv:1503.05918 [astro-ph.CO]}
  \BibitemShut {NoStop}%
\bibitem [{\citenamefont {Linden}(2017)}]{Linden:2016fdd}%
  \BibitemOpen
  \bibfield  {author} {\bibinfo {author} {\bibfnamefont {Tim}\ \bibnamefont
  {Linden}},\ }\bibfield  {title} {\enquote {\bibinfo {title} {{Star-Forming
  Galaxies Significantly Contribute to the Isotropic Gamma-Ray Background}},}\
  }\href {\doibase 10.1103/PhysRevD.96.083001} {\bibfield  {journal} {\bibinfo
  {journal} {Phys. Rev. D}\ }\textbf {\bibinfo {volume} {96}},\ \bibinfo
  {pages} {083001} (\bibinfo {year} {2017})},\ \Eprint
  {http://arxiv.org/abs/1612.03175} {arXiv:1612.03175 [astro-ph.HE]}
  \BibitemShut {NoStop}%
\bibitem [{\citenamefont {Wang}\ and\ \citenamefont
  {Loeb}(2016)}]{Wang:2016zyi}%
  \BibitemOpen
  \bibfield  {author} {\bibinfo {author} {\bibfnamefont {Xiawei}\ \bibnamefont
  {Wang}}\ and\ \bibinfo {author} {\bibfnamefont {Abraham}\ \bibnamefont
  {Loeb}},\ }\bibfield  {title} {\enquote {\bibinfo {title} {{Quasar-driven
  outflows account for the missing extragalactic gamma-ray background}},}\
  }\href {\doibase 10.1038/nphys3837} {\  (\bibinfo {year} {2016}),\
  10.1038/nphys3837},\ \Eprint {http://arxiv.org/abs/1607.06472}
  {arXiv:1607.06472 [astro-ph.HE]} \BibitemShut {NoStop}%
\bibitem [{\citenamefont {Di~Mauro}\ \emph {et~al.}(2018)\citenamefont
  {Di~Mauro}, \citenamefont {Manconi}, \citenamefont {Zechlin}, \citenamefont
  {Ajello}, \citenamefont {Charles},\ and\ \citenamefont
  {Donato}}]{DiMauro:2017ing}%
  \BibitemOpen
  \bibfield  {author} {\bibinfo {author} {\bibfnamefont {Mattia}\ \bibnamefont
  {Di~Mauro}}, \bibinfo {author} {\bibfnamefont {Silvia}\ \bibnamefont
  {Manconi}}, \bibinfo {author} {\bibfnamefont {Hannes-S.}\ \bibnamefont
  {Zechlin}}, \bibinfo {author} {\bibfnamefont {Marco}\ \bibnamefont {Ajello}},
  \bibinfo {author} {\bibfnamefont {Eric}\ \bibnamefont {Charles}}, \ and\
  \bibinfo {author} {\bibfnamefont {Fiorenza}\ \bibnamefont {Donato}},\
  }\bibfield  {title} {\enquote {\bibinfo {title} {{Deriving the contribution
  of blazars to the Fermi-LAT Extragalactic $\gamma$-ray background at $E>10$
  GeV with efficiency corrections and photon statistics}},}\ }\href {\doibase
  10.3847/1538-4357/aab3e5} {\bibfield  {journal} {\bibinfo  {journal}
  {Astrophys. J.}\ }\textbf {\bibinfo {volume} {856}},\ \bibinfo {pages} {106}
  (\bibinfo {year} {2018})},\ \Eprint {http://arxiv.org/abs/1711.03111}
  {arXiv:1711.03111 [astro-ph.HE]} \BibitemShut {NoStop}%
\bibitem [{\citenamefont {Lamastra}\ \emph {et~al.}(2017)\citenamefont
  {Lamastra}, \citenamefont {Menci}, \citenamefont {Fiore}, \citenamefont
  {Antonelli}, \citenamefont {Colafrancesco}, \citenamefont {Guetta},\ and\
  \citenamefont {Stamerra}}]{Lamastra:2017iyo}%
  \BibitemOpen
  \bibfield  {author} {\bibinfo {author} {\bibfnamefont {A.}~\bibnamefont
  {Lamastra}}, \bibinfo {author} {\bibfnamefont {N.}~\bibnamefont {Menci}},
  \bibinfo {author} {\bibfnamefont {F.}~\bibnamefont {Fiore}}, \bibinfo
  {author} {\bibfnamefont {L.~A.}\ \bibnamefont {Antonelli}}, \bibinfo {author}
  {\bibfnamefont {S.}~\bibnamefont {Colafrancesco}}, \bibinfo {author}
  {\bibfnamefont {D.}~\bibnamefont {Guetta}}, \ and\ \bibinfo {author}
  {\bibfnamefont {A.}~\bibnamefont {Stamerra}},\ }\bibfield  {title} {\enquote
  {\bibinfo {title} {{Extragalactic gamma-ray background from AGN winds and
  star-forming galaxies in cosmological galaxy formation models}},}\ }\href
  {\doibase 10.1051/0004-6361/201731452} {\bibfield  {journal} {\bibinfo
  {journal} {Astron. Astrophys.}\ }\textbf {\bibinfo {volume} {607}},\ \bibinfo
  {pages} {A18} (\bibinfo {year} {2017})},\ \Eprint
  {http://arxiv.org/abs/1709.03497} {arXiv:1709.03497 [astro-ph.HE]}
  \BibitemShut {NoStop}%
\bibitem [{\citenamefont {Cuoco}\ \emph
  {et~al.}(2017{\natexlab{a}})\citenamefont {Cuoco}, \citenamefont {Bilicki},
  \citenamefont {Xia},\ and\ \citenamefont {Branchini}}]{Cuoco:2017bpv}%
  \BibitemOpen
  \bibfield  {author} {\bibinfo {author} {\bibfnamefont {Alessandro}\
  \bibnamefont {Cuoco}}, \bibinfo {author} {\bibfnamefont {Maciej}\
  \bibnamefont {Bilicki}}, \bibinfo {author} {\bibfnamefont {Jun-Qing}\
  \bibnamefont {Xia}}, \ and\ \bibinfo {author} {\bibfnamefont {Enzo}\
  \bibnamefont {Branchini}},\ }\bibfield  {title} {\enquote {\bibinfo {title}
  {{Tomographic imaging of the Fermi-LAT gamma-ray sky through
  cross-correlations: A wider and deeper look}},}\ }\href {\doibase
  10.3847/1538-4365/aa8553} {\bibfield  {journal} {\bibinfo  {journal}
  {Astrophys. J. Suppl.}\ }\textbf {\bibinfo {volume} {232}},\ \bibinfo {pages}
  {10} (\bibinfo {year} {2017}{\natexlab{a}})},\ \Eprint
  {http://arxiv.org/abs/1709.01940} {arXiv:1709.01940 [astro-ph.HE]}
  \BibitemShut {NoStop}%
\bibitem [{\citenamefont {Komis}\ \emph {et~al.}(2019)\citenamefont {Komis},
  \citenamefont {Pavlidou},\ and\ \citenamefont {Zezas}}]{Komis:2017jta}%
  \BibitemOpen
  \bibfield  {author} {\bibinfo {author} {\bibfnamefont {I.}~\bibnamefont
  {Komis}}, \bibinfo {author} {\bibfnamefont {V.}~\bibnamefont {Pavlidou}}, \
  and\ \bibinfo {author} {\bibfnamefont {A.}~\bibnamefont {Zezas}},\ }\bibfield
   {title} {\enquote {\bibinfo {title} {{Extragalactic Gamma-ray Background
  from Star-forming Galaxies: Will Empirical Scalings Suffice?}}}\ }\href
  {\doibase 10.1093/mnras/sty3354} {\bibfield  {journal} {\bibinfo  {journal}
  {Mon. Not. Roy. Astron. Soc.}\ }\textbf {\bibinfo {volume} {483}},\ \bibinfo
  {pages} {4020--4030} (\bibinfo {year} {2019})},\ \Eprint
  {http://arxiv.org/abs/1711.11046} {arXiv:1711.11046 [astro-ph.HE]}
  \BibitemShut {NoStop}%
\bibitem [{\citenamefont {Pinetti}\ \emph {et~al.}(2020)\citenamefont
  {Pinetti}, \citenamefont {Camera}, \citenamefont {Fornengo},\ and\
  \citenamefont {Regis}}]{Pinetti:2019ztr}%
  \BibitemOpen
  \bibfield  {author} {\bibinfo {author} {\bibfnamefont {Elena}\ \bibnamefont
  {Pinetti}}, \bibinfo {author} {\bibfnamefont {Stefano}\ \bibnamefont
  {Camera}}, \bibinfo {author} {\bibfnamefont {Nicolao}\ \bibnamefont
  {Fornengo}}, \ and\ \bibinfo {author} {\bibfnamefont {Marco}\ \bibnamefont
  {Regis}},\ }\bibfield  {title} {\enquote {\bibinfo {title} {{Synergies across
  the spectrum for particle dark matter indirect detection: how HI intensity
  mapping meets gamma rays}},}\ }\href {\doibase 10.1088/1475-7516/2020/07/044}
  {\bibfield  {journal} {\bibinfo  {journal} {JCAP}\ }\textbf {\bibinfo
  {volume} {07}},\ \bibinfo {pages} {044} (\bibinfo {year} {2020})},\ \Eprint
  {http://arxiv.org/abs/1911.04989} {arXiv:1911.04989 [astro-ph.CO]}
  \BibitemShut {NoStop}%
\bibitem [{\citenamefont {Hooper}\ \emph {et~al.}(2016)\citenamefont {Hooper},
  \citenamefont {Linden},\ and\ \citenamefont {Lopez}}]{Hooper:2016gjy}%
  \BibitemOpen
  \bibfield  {author} {\bibinfo {author} {\bibfnamefont {Dan}\ \bibnamefont
  {Hooper}}, \bibinfo {author} {\bibfnamefont {Tim}\ \bibnamefont {Linden}}, \
  and\ \bibinfo {author} {\bibfnamefont {Alejandro}\ \bibnamefont {Lopez}},\
  }\bibfield  {title} {\enquote {\bibinfo {title} {{Radio Galaxies Dominate the
  High-Energy Diffuse Gamma-Ray Background}},}\ }\href {\doibase
  10.1088/1475-7516/2016/08/019} {\bibfield  {journal} {\bibinfo  {journal}
  {JCAP}\ }\textbf {\bibinfo {volume} {08}},\ \bibinfo {pages} {019} (\bibinfo
  {year} {2016})},\ \Eprint {http://arxiv.org/abs/1604.08505} {arXiv:1604.08505
  [astro-ph.HE]} \BibitemShut {NoStop}%
\bibitem [{\citenamefont {Stecker}\ \emph {et~al.}(2019)\citenamefont
  {Stecker}, \citenamefont {Shrader},\ and\ \citenamefont
  {Malkan}}]{Stecker:2019ybn}%
  \BibitemOpen
  \bibfield  {author} {\bibinfo {author} {\bibfnamefont {Floyd~W.}\
  \bibnamefont {Stecker}}, \bibinfo {author} {\bibfnamefont {Chris~R.}\
  \bibnamefont {Shrader}}, \ and\ \bibinfo {author} {\bibfnamefont
  {Matthew.~A.}\ \bibnamefont {Malkan}},\ }\bibfield  {title} {\enquote
  {\bibinfo {title} {{The Extragalactic Gamma-Ray Background from Core
  Dominated Radio Galaxies}},}\ }\href {\doibase 10.3847/1538-4357/ab23ee}
  {\bibfield  {journal} {\bibinfo  {journal} {Astrophys. J.}\ }\textbf
  {\bibinfo {volume} {879}},\ \bibinfo {pages} {68} (\bibinfo {year} {2019})},\
  \Eprint {http://arxiv.org/abs/1903.06544} {arXiv:1903.06544 [astro-ph.GA]}
  \BibitemShut {NoStop}%
\bibitem [{\citenamefont {Qu}\ \emph {et~al.}(2019)\citenamefont {Qu},
  \citenamefont {Zeng},\ and\ \citenamefont {Yan}}]{Qu:2019zln}%
  \BibitemOpen
  \bibfield  {author} {\bibinfo {author} {\bibfnamefont {Yankun}\ \bibnamefont
  {Qu}}, \bibinfo {author} {\bibfnamefont {Houdun}\ \bibnamefont {Zeng}}, \
  and\ \bibinfo {author} {\bibfnamefont {Dahai}\ \bibnamefont {Yan}},\
  }\bibfield  {title} {\enquote {\bibinfo {title} {{Gamma-ray luminosity
  function of BL Lac objects and contribution to the extragalactic gamma-ray
  background}},}\ }\href {\doibase 10.1093/mnras/stz2651} {\bibfield  {journal}
  {\bibinfo  {journal} {Mon. Not. Roy. Astron. Soc.}\ }\textbf {\bibinfo
  {volume} {490}},\ \bibinfo {pages} {758--765} (\bibinfo {year} {2019})},\
  \Eprint {http://arxiv.org/abs/1909.07542} {arXiv:1909.07542 [astro-ph.HE]}
  \BibitemShut {NoStop}%
\bibitem [{\citenamefont {Owen}\ \emph {et~al.}(2021)\citenamefont {Owen},
  \citenamefont {Lee},\ and\ \citenamefont {Kong}}]{Owen:2021evt}%
  \BibitemOpen
  \bibfield  {author} {\bibinfo {author} {\bibfnamefont {Ellis~R.}\
  \bibnamefont {Owen}}, \bibinfo {author} {\bibfnamefont {Khee-Gan}\
  \bibnamefont {Lee}}, \ and\ \bibinfo {author} {\bibfnamefont {Albert K.~H.}\
  \bibnamefont {Kong}},\ }\bibfield  {title} {\enquote {\bibinfo {title}
  {{Characterizing the signatures of star-forming galaxies in the extragalactic
  \ensuremath{\gamma}-ray background}},}\ }\href {\doibase
  10.1093/mnras/stab1707} {\bibfield  {journal} {\bibinfo  {journal} {Mon. Not.
  Roy. Astron. Soc.}\ }\textbf {\bibinfo {volume} {506}},\ \bibinfo {pages}
  {52--72} (\bibinfo {year} {2021})},\ \Eprint
  {http://arxiv.org/abs/2106.07308} {arXiv:2106.07308 [astro-ph.HE]}
  \BibitemShut {NoStop}%
\bibitem [{\citenamefont {Owen}\ \emph {et~al.}(2022)\citenamefont {Owen},
  \citenamefont {Kong},\ and\ \citenamefont {Lee}}]{Owen:2021qul}%
  \BibitemOpen
  \bibfield  {author} {\bibinfo {author} {\bibfnamefont {Ellis~R.}\
  \bibnamefont {Owen}}, \bibinfo {author} {\bibfnamefont {Albert K.~H.}\
  \bibnamefont {Kong}}, \ and\ \bibinfo {author} {\bibfnamefont {Khee-Gan}\
  \bibnamefont {Lee}},\ }\bibfield  {title} {\enquote {\bibinfo {title} {{The
  extragalactic \ensuremath{\gamma}-ray background: imprints from the physical
  properties and evolution of star-forming galaxy populations}},}\ }\href
  {\doibase 10.1093/mnras/stac1079} {\bibfield  {journal} {\bibinfo  {journal}
  {Mon. Not. Roy. Astron. Soc.}\ }\textbf {\bibinfo {volume} {513}},\ \bibinfo
  {pages} {2335--2348} (\bibinfo {year} {2022})},\ \Eprint
  {http://arxiv.org/abs/2112.09032} {arXiv:2112.09032 [astro-ph.GA]}
  \BibitemShut {NoStop}%
\bibitem [{\citenamefont {Blanco}\ and\ \citenamefont
  {Linden}(2023)}]{Blanco:2021icw}%
  \BibitemOpen
  \bibfield  {author} {\bibinfo {author} {\bibfnamefont {Carlos}\ \bibnamefont
  {Blanco}}\ and\ \bibinfo {author} {\bibfnamefont {Tim}\ \bibnamefont
  {Linden}},\ }\bibfield  {title} {\enquote {\bibinfo {title} {{Star-forming
  galaxies provide a larger contribution to the isotropic gamma-ray background
  than misaligned active galactic nuclei}},}\ }\href {\doibase
  10.1088/1475-7516/2023/02/003} {\bibfield  {journal} {\bibinfo  {journal}
  {JCAP}\ }\textbf {\bibinfo {volume} {02}},\ \bibinfo {pages} {003} (\bibinfo
  {year} {2023})},\ \Eprint {http://arxiv.org/abs/2104.03315} {arXiv:2104.03315
  [astro-ph.HE]} \BibitemShut {NoStop}%
\bibitem [{\citenamefont {Korsmeier}\ \emph {et~al.}(2022)\citenamefont
  {Korsmeier}, \citenamefont {Pinetti}, \citenamefont {Negro}, \citenamefont
  {Regis},\ and\ \citenamefont {Fornengo}}]{Korsmeier:2022cwp}%
  \BibitemOpen
  \bibfield  {author} {\bibinfo {author} {\bibfnamefont {Michael}\ \bibnamefont
  {Korsmeier}}, \bibinfo {author} {\bibfnamefont {Elena}\ \bibnamefont
  {Pinetti}}, \bibinfo {author} {\bibfnamefont {Michela}\ \bibnamefont
  {Negro}}, \bibinfo {author} {\bibfnamefont {Marco}\ \bibnamefont {Regis}}, \
  and\ \bibinfo {author} {\bibfnamefont {Nicolao}\ \bibnamefont {Fornengo}},\
  }\bibfield  {title} {\enquote {\bibinfo {title} {{Flat-spectrum Radio Quasars
  and BL Lacs Dominate the Anisotropy of the Unresolved Gamma-Ray
  Background}},}\ }\href {\doibase 10.3847/1538-4357/ac6c85} {\bibfield
  {journal} {\bibinfo  {journal} {Astrophys. J.}\ }\textbf {\bibinfo {volume}
  {933}},\ \bibinfo {pages} {221} (\bibinfo {year} {2022})},\ \Eprint
  {http://arxiv.org/abs/2201.02634} {arXiv:2201.02634 [astro-ph.HE]}
  \BibitemShut {NoStop}%
\bibitem [{\citenamefont {Xue}\ \emph {et~al.}(2024)\citenamefont {Xue},
  \citenamefont {Wang}, \citenamefont {Joshi},\ and\ \citenamefont
  {Li}}]{Xue:2024tkj}%
  \BibitemOpen
  \bibfield  {author} {\bibinfo {author} {\bibfnamefont {Rui}\ \bibnamefont
  {Xue}}, \bibinfo {author} {\bibfnamefont {Ze-Rui}\ \bibnamefont {Wang}},
  \bibinfo {author} {\bibfnamefont {Jagdish~C.}\ \bibnamefont {Joshi}}, \ and\
  \bibinfo {author} {\bibfnamefont {Wei-Jian}\ \bibnamefont {Li}},\ }\bibfield
  {title} {\enquote {\bibinfo {title} {{Hadronuclear interactions in AGN jets
  as the origin of the diffuse high-energy neutrino background}},}\ }\href@noop
  {} {\  (\bibinfo {year} {2024})},\ \Eprint {http://arxiv.org/abs/2407.04195}
  {arXiv:2407.04195 [astro-ph.HE]} \BibitemShut {NoStop}%
\bibitem [{\citenamefont {Min}\ \emph {et~al.}(2024)\citenamefont {Min},
  \citenamefont {Yao}, \citenamefont {Liu}, \citenamefont {Chen}, \citenamefont
  {Lu},\ and\ \citenamefont {Guo}}]{Min:2024bnj}%
  \BibitemOpen
  \bibfield  {author} {\bibinfo {author} {\bibfnamefont {Fang-Sheng}\
  \bibnamefont {Min}}, \bibinfo {author} {\bibfnamefont {Yu-Hua}\ \bibnamefont
  {Yao}}, \bibinfo {author} {\bibfnamefont {Ruo-Yu}\ \bibnamefont {Liu}},
  \bibinfo {author} {\bibfnamefont {Shi}\ \bibnamefont {Chen}}, \bibinfo
  {author} {\bibfnamefont {Hong}\ \bibnamefont {Lu}}, \ and\ \bibinfo {author}
  {\bibfnamefont {Yi-Qing}\ \bibnamefont {Guo}},\ }\bibfield  {title} {\enquote
  {\bibinfo {title} {{Contribution of $\gamma$-Ray Burst Afterglow Emissions to
  the Isotropic Diffuse $\gamma$-Ray Background}},}\ }\href {\doibase
  10.3847/1538-4357/ad28be} {\bibfield  {journal} {\bibinfo  {journal}
  {Astrophys. J.}\ }\textbf {\bibinfo {volume} {964}},\ \bibinfo {pages} {195}
  (\bibinfo {year} {2024})}\BibitemShut {NoStop}%
\bibitem [{\citenamefont {Cholis}\ and\ \citenamefont
  {Krommydas}(2024)}]{Cholis:2024hmd}%
  \BibitemOpen
  \bibfield  {author} {\bibinfo {author} {\bibfnamefont {Ilias}\ \bibnamefont
  {Cholis}}\ and\ \bibinfo {author} {\bibfnamefont {Iason}\ \bibnamefont
  {Krommydas}},\ }\bibfield  {title} {\enquote {\bibinfo {title} {{Scrutinizing
  the Isotropic Gamma-Ray Background in Search of Dark Matter}},}\ }\href@noop
  {} {\  (\bibinfo {year} {2024})},\ \Eprint {http://arxiv.org/abs/2408.11421}
  {arXiv:2408.11421 [astro-ph.HE]} \BibitemShut {NoStop}%
\bibitem [{\citenamefont {Ando}\ and\ \citenamefont
  {Komatsu}(2006)}]{Ando:2005xg}%
  \BibitemOpen
  \bibfield  {author} {\bibinfo {author} {\bibfnamefont {Shin'ichiro}\
  \bibnamefont {Ando}}\ and\ \bibinfo {author} {\bibfnamefont {Eiichiro}\
  \bibnamefont {Komatsu}},\ }\bibfield  {title} {\enquote {\bibinfo {title}
  {{Anisotropy of the cosmic gamma-ray background from dark matter
  annihilation}},}\ }\href {\doibase 10.1103/PhysRevD.73.023521} {\bibfield
  {journal} {\bibinfo  {journal} {Phys. Rev. D}\ }\textbf {\bibinfo {volume}
  {73}},\ \bibinfo {pages} {023521} (\bibinfo {year} {2006})},\ \Eprint
  {http://arxiv.org/abs/astro-ph/0512217} {arXiv:astro-ph/0512217} \BibitemShut
  {NoStop}%
\bibitem [{\citenamefont {Fornasa}\ \emph {et~al.}(2013)\citenamefont
  {Fornasa}, \citenamefont {Zavala}, \citenamefont {Sanchez-Conde},
  \citenamefont {Siegal-Gaskins}, \citenamefont {Delahaye}, \citenamefont
  {Prada}, \citenamefont {Vogelsberger}, \citenamefont {Zandanel},\ and\
  \citenamefont {Frenk}}]{Fornasa:2012gu}%
  \BibitemOpen
  \bibfield  {author} {\bibinfo {author} {\bibfnamefont {Mattia}\ \bibnamefont
  {Fornasa}}, \bibinfo {author} {\bibfnamefont {Jesus}\ \bibnamefont {Zavala}},
  \bibinfo {author} {\bibfnamefont {Miguel~A.}\ \bibnamefont {Sanchez-Conde}},
  \bibinfo {author} {\bibfnamefont {Jennifer~M.}\ \bibnamefont
  {Siegal-Gaskins}}, \bibinfo {author} {\bibfnamefont {Timur}\ \bibnamefont
  {Delahaye}}, \bibinfo {author} {\bibfnamefont {Francisco}\ \bibnamefont
  {Prada}}, \bibinfo {author} {\bibfnamefont {Mark}\ \bibnamefont
  {Vogelsberger}}, \bibinfo {author} {\bibfnamefont {Fabio}\ \bibnamefont
  {Zandanel}}, \ and\ \bibinfo {author} {\bibfnamefont {Carlos~S.}\
  \bibnamefont {Frenk}},\ }\bibfield  {title} {\enquote {\bibinfo {title}
  {{Characterization of Dark-Matter-induced anisotropies in the diffuse
  gamma-ray background}},}\ }\href {\doibase 10.1093/mnras/sts444} {\bibfield
  {journal} {\bibinfo  {journal} {Mon. Not. Roy. Astron. Soc.}\ }\textbf
  {\bibinfo {volume} {429}},\ \bibinfo {pages} {1529--1553} (\bibinfo {year}
  {2013})},\ \Eprint {http://arxiv.org/abs/1207.0502} {arXiv:1207.0502
  [astro-ph.HE]} \BibitemShut {NoStop}%
\bibitem [{\citenamefont {Ando}\ and\ \citenamefont
  {Komatsu}(2013)}]{Ando:2013ff}%
  \BibitemOpen
  \bibfield  {author} {\bibinfo {author} {\bibfnamefont {Shin'ichiro}\
  \bibnamefont {Ando}}\ and\ \bibinfo {author} {\bibfnamefont {Eiichiro}\
  \bibnamefont {Komatsu}},\ }\bibfield  {title} {\enquote {\bibinfo {title}
  {{Constraints on the annihilation cross section of dark matter particles from
  anisotropies in the diffuse gamma-ray background measured with Fermi-LAT}},}\
  }\href {\doibase 10.1103/PhysRevD.87.123539} {\bibfield  {journal} {\bibinfo
  {journal} {Phys. Rev. D}\ }\textbf {\bibinfo {volume} {87}},\ \bibinfo
  {pages} {123539} (\bibinfo {year} {2013})},\ \Eprint
  {http://arxiv.org/abs/1301.5901} {arXiv:1301.5901 [astro-ph.CO]} \BibitemShut
  {NoStop}%
\bibitem [{\citenamefont {Cholis}\ \emph {et~al.}(2014)\citenamefont {Cholis},
  \citenamefont {Hooper},\ and\ \citenamefont {McDermott}}]{Cholis:2013ena}%
  \BibitemOpen
  \bibfield  {author} {\bibinfo {author} {\bibfnamefont {Ilias}\ \bibnamefont
  {Cholis}}, \bibinfo {author} {\bibfnamefont {Dan}\ \bibnamefont {Hooper}}, \
  and\ \bibinfo {author} {\bibfnamefont {Samuel~D.}\ \bibnamefont
  {McDermott}},\ }\bibfield  {title} {\enquote {\bibinfo {title} {{Dissecting
  the Gamma-Ray Background in Search of Dark Matter}},}\ }\href {\doibase
  10.1088/1475-7516/2014/02/014} {\bibfield  {journal} {\bibinfo  {journal}
  {JCAP}\ }\textbf {\bibinfo {volume} {02}},\ \bibinfo {pages} {014} (\bibinfo
  {year} {2014})},\ \Eprint {http://arxiv.org/abs/1312.0608} {arXiv:1312.0608
  [astro-ph.CO]} \BibitemShut {NoStop}%
\bibitem [{\citenamefont {Camera}\ \emph {et~al.}(2015)\citenamefont {Camera},
  \citenamefont {Fornasa}, \citenamefont {Fornengo},\ and\ \citenamefont
  {Regis}}]{Camera:2014rja}%
  \BibitemOpen
  \bibfield  {author} {\bibinfo {author} {\bibfnamefont {Stefano}\ \bibnamefont
  {Camera}}, \bibinfo {author} {\bibfnamefont {Mattia}\ \bibnamefont
  {Fornasa}}, \bibinfo {author} {\bibfnamefont {Nicolao}\ \bibnamefont
  {Fornengo}}, \ and\ \bibinfo {author} {\bibfnamefont {Marco}\ \bibnamefont
  {Regis}},\ }\bibfield  {title} {\enquote {\bibinfo {title}
  {{Tomographic-spectral approach for dark matter detection in the
  cross-correlation between cosmic shear and diffuse $\gamma$-ray emission}},}\
  }\href {\doibase 10.1088/1475-7516/2015/06/029} {\bibfield  {journal}
  {\bibinfo  {journal} {JCAP}\ }\textbf {\bibinfo {volume} {06}},\ \bibinfo
  {pages} {029} (\bibinfo {year} {2015})},\ \Eprint
  {http://arxiv.org/abs/1411.4651} {arXiv:1411.4651 [astro-ph.CO]} \BibitemShut
  {NoStop}%
\bibitem [{\citenamefont {Ando}(2014)}]{Ando:2014aoa}%
  \BibitemOpen
  \bibfield  {author} {\bibinfo {author} {\bibfnamefont {Shin'ichiro}\
  \bibnamefont {Ando}},\ }\bibfield  {title} {\enquote {\bibinfo {title}
  {{Power spectrum tomography of dark matter annihilation with local galaxy
  distribution}},}\ }\href {\doibase 10.1088/1475-7516/2014/10/061} {\bibfield
  {journal} {\bibinfo  {journal} {JCAP}\ }\textbf {\bibinfo {volume} {10}},\
  \bibinfo {pages} {061} (\bibinfo {year} {2014})},\ \Eprint
  {http://arxiv.org/abs/1407.8502} {arXiv:1407.8502 [astro-ph.CO]} \BibitemShut
  {NoStop}%
\bibitem [{\citenamefont {Cuoco}\ \emph {et~al.}(2015)\citenamefont {Cuoco},
  \citenamefont {Xia}, \citenamefont {Regis}, \citenamefont {Branchini},
  \citenamefont {Fornengo},\ and\ \citenamefont {Viel}}]{Cuoco:2015rfa}%
  \BibitemOpen
  \bibfield  {author} {\bibinfo {author} {\bibfnamefont {Alessandro}\
  \bibnamefont {Cuoco}}, \bibinfo {author} {\bibfnamefont {Jun-Qing}\
  \bibnamefont {Xia}}, \bibinfo {author} {\bibfnamefont {Marco}\ \bibnamefont
  {Regis}}, \bibinfo {author} {\bibfnamefont {Enzo}\ \bibnamefont {Branchini}},
  \bibinfo {author} {\bibfnamefont {Nicolao}\ \bibnamefont {Fornengo}}, \ and\
  \bibinfo {author} {\bibfnamefont {Matteo}\ \bibnamefont {Viel}},\ }\bibfield
  {title} {\enquote {\bibinfo {title} {{Dark Matter Searches in the Gamma-ray
  Extragalactic Background via Cross-correlations With Galaxy Catalogs}},}\
  }\href {\doibase 10.1088/0067-0049/221/2/29} {\bibfield  {journal} {\bibinfo
  {journal} {Astrophys. J. Suppl.}\ }\textbf {\bibinfo {volume} {221}},\
  \bibinfo {pages} {29} (\bibinfo {year} {2015})},\ \Eprint
  {http://arxiv.org/abs/1506.01030} {arXiv:1506.01030 [astro-ph.HE]}
  \BibitemShut {NoStop}%
\bibitem [{\citenamefont {Ackermann}\ \emph
  {et~al.}(2015{\natexlab{b}})\citenamefont {Ackermann} \emph
  {et~al.}}]{Fermi-LAT:2015qzw}%
  \BibitemOpen
  \bibfield  {author} {\bibinfo {author} {\bibfnamefont {M.}~\bibnamefont
  {Ackermann}} \emph {et~al.} (\bibinfo {collaboration} {Fermi-LAT}),\
  }\bibfield  {title} {\enquote {\bibinfo {title} {{Limits on Dark Matter
  Annihilation Signals from the Fermi LAT 4-year Measurement of the Isotropic
  Gamma-Ray Background}},}\ }\href {\doibase 10.1088/1475-7516/2015/09/008}
  {\bibfield  {journal} {\bibinfo  {journal} {JCAP}\ }\textbf {\bibinfo
  {volume} {09}},\ \bibinfo {pages} {008} (\bibinfo {year}
  {2015}{\natexlab{b}})},\ \Eprint {http://arxiv.org/abs/1501.05464}
  {arXiv:1501.05464 [astro-ph.CO]} \BibitemShut {NoStop}%
\bibitem [{\citenamefont {Regis}\ \emph {et~al.}(2015)\citenamefont {Regis},
  \citenamefont {Xia}, \citenamefont {Cuoco}, \citenamefont {Branchini},
  \citenamefont {Fornengo},\ and\ \citenamefont {Viel}}]{Regis:2015zka}%
  \BibitemOpen
  \bibfield  {author} {\bibinfo {author} {\bibfnamefont {Marco}\ \bibnamefont
  {Regis}}, \bibinfo {author} {\bibfnamefont {Jun-Qing}\ \bibnamefont {Xia}},
  \bibinfo {author} {\bibfnamefont {Alessandro}\ \bibnamefont {Cuoco}},
  \bibinfo {author} {\bibfnamefont {Enzo}\ \bibnamefont {Branchini}}, \bibinfo
  {author} {\bibfnamefont {Nicolao}\ \bibnamefont {Fornengo}}, \ and\ \bibinfo
  {author} {\bibfnamefont {Matteo}\ \bibnamefont {Viel}},\ }\bibfield  {title}
  {\enquote {\bibinfo {title} {{Particle dark matter searches outside the Local
  Group}},}\ }\href {\doibase 10.1103/PhysRevLett.114.241301} {\bibfield
  {journal} {\bibinfo  {journal} {Phys. Rev. Lett.}\ }\textbf {\bibinfo
  {volume} {114}},\ \bibinfo {pages} {241301} (\bibinfo {year} {2015})},\
  \Eprint {http://arxiv.org/abs/1503.05922} {arXiv:1503.05922 [astro-ph.CO]}
  \BibitemShut {NoStop}%
\bibitem [{\citenamefont {Shirasaki}\ \emph {et~al.}(2015)\citenamefont
  {Shirasaki}, \citenamefont {Horiuchi},\ and\ \citenamefont
  {Yoshida}}]{Shirasaki:2015nqp}%
  \BibitemOpen
  \bibfield  {author} {\bibinfo {author} {\bibfnamefont {Masato}\ \bibnamefont
  {Shirasaki}}, \bibinfo {author} {\bibfnamefont {Shunsaku}\ \bibnamefont
  {Horiuchi}}, \ and\ \bibinfo {author} {\bibfnamefont {Naoki}\ \bibnamefont
  {Yoshida}},\ }\bibfield  {title} {\enquote {\bibinfo {title}
  {{Cross-Correlation of the Extragalactic Gamma-ray Background with Luminous
  Red Galaxies}},}\ }\href {\doibase 10.1103/PhysRevD.92.123540} {\bibfield
  {journal} {\bibinfo  {journal} {Phys. Rev. D}\ }\textbf {\bibinfo {volume}
  {92}},\ \bibinfo {pages} {123540} (\bibinfo {year} {2015})},\ \Eprint
  {http://arxiv.org/abs/1511.07092} {arXiv:1511.07092 [astro-ph.CO]}
  \BibitemShut {NoStop}%
\bibitem [{\citenamefont {Feyereisen}\ \emph {et~al.}(2015)\citenamefont
  {Feyereisen}, \citenamefont {Ando},\ and\ \citenamefont
  {Lee}}]{Feyereisen:2015cea}%
  \BibitemOpen
  \bibfield  {author} {\bibinfo {author} {\bibfnamefont {Michael~R.}\
  \bibnamefont {Feyereisen}}, \bibinfo {author} {\bibfnamefont {Shin'ichiro}\
  \bibnamefont {Ando}}, \ and\ \bibinfo {author} {\bibfnamefont {Samuel~K.}\
  \bibnamefont {Lee}},\ }\bibfield  {title} {\enquote {\bibinfo {title}
  {{Modelling the flux distribution function of the extragalactic gamma-ray
  background from dark matter annihilation}},}\ }\href {\doibase
  10.1088/1475-7516/2015/9/027} {\bibfield  {journal} {\bibinfo  {journal}
  {JCAP}\ }\textbf {\bibinfo {volume} {09}},\ \bibinfo {pages} {027} (\bibinfo
  {year} {2015})},\ \Eprint {http://arxiv.org/abs/1506.05118} {arXiv:1506.05118
  [astro-ph.CO]} \BibitemShut {NoStop}%
\bibitem [{\citenamefont {Liu}\ \emph {et~al.}(2017)\citenamefont {Liu},
  \citenamefont {Bi}, \citenamefont {Lin},\ and\ \citenamefont
  {Yin}}]{Liu:2016ngs}%
  \BibitemOpen
  \bibfield  {author} {\bibinfo {author} {\bibfnamefont {Wei}\ \bibnamefont
  {Liu}}, \bibinfo {author} {\bibfnamefont {Xiao-Jun}\ \bibnamefont {Bi}},
  \bibinfo {author} {\bibfnamefont {Su-Jie}\ \bibnamefont {Lin}}, \ and\
  \bibinfo {author} {\bibfnamefont {Peng-Fei}\ \bibnamefont {Yin}},\ }\bibfield
   {title} {\enquote {\bibinfo {title} {{Constraints on dark matter
  annihilation and decay from the isotropic gamma-ray background}},}\ }\href
  {\doibase 10.1088/1674-1137/41/4/045104} {\bibfield  {journal} {\bibinfo
  {journal} {Chin. Phys. C}\ }\textbf {\bibinfo {volume} {41}},\ \bibinfo
  {pages} {045104} (\bibinfo {year} {2017})},\ \Eprint
  {http://arxiv.org/abs/1602.01012} {arXiv:1602.01012 [astro-ph.CO]}
  \BibitemShut {NoStop}%
\bibitem [{\citenamefont {Shirasaki}\ \emph {et~al.}(2016)\citenamefont
  {Shirasaki}, \citenamefont {Macias}, \citenamefont {Horiuchi}, \citenamefont
  {Shirai},\ and\ \citenamefont {Yoshida}}]{Shirasaki:2016kol}%
  \BibitemOpen
  \bibfield  {author} {\bibinfo {author} {\bibfnamefont {Masato}\ \bibnamefont
  {Shirasaki}}, \bibinfo {author} {\bibfnamefont {Oscar}\ \bibnamefont
  {Macias}}, \bibinfo {author} {\bibfnamefont {Shunsaku}\ \bibnamefont
  {Horiuchi}}, \bibinfo {author} {\bibfnamefont {Satoshi}\ \bibnamefont
  {Shirai}}, \ and\ \bibinfo {author} {\bibfnamefont {Naoki}\ \bibnamefont
  {Yoshida}},\ }\bibfield  {title} {\enquote {\bibinfo {title} {{Cosmological
  constraints on dark matter annihilation and decay: Cross-correlation analysis
  of the extragalactic $\gamma$-ray background and cosmic shear}},}\ }\href
  {\doibase 10.1103/PhysRevD.94.063522} {\bibfield  {journal} {\bibinfo
  {journal} {Phys. Rev. D}\ }\textbf {\bibinfo {volume} {94}},\ \bibinfo
  {pages} {063522} (\bibinfo {year} {2016})},\ \Eprint
  {http://arxiv.org/abs/1607.02187} {arXiv:1607.02187 [astro-ph.CO]}
  \BibitemShut {NoStop}%
\bibitem [{\citenamefont {Campbell}\ \emph {et~al.}(2018)\citenamefont
  {Campbell}, \citenamefont {Kwa},\ and\ \citenamefont
  {Kaplinghat}}]{Campbell:2017qpa}%
  \BibitemOpen
  \bibfield  {author} {\bibinfo {author} {\bibfnamefont {Sheldon~S.}\
  \bibnamefont {Campbell}}, \bibinfo {author} {\bibfnamefont {Anna}\
  \bibnamefont {Kwa}}, \ and\ \bibinfo {author} {\bibfnamefont {Manoj}\
  \bibnamefont {Kaplinghat}},\ }\bibfield  {title} {\enquote {\bibinfo {title}
  {{The galactic isotropic $\gamma$-ray background and implications for dark
  matter}},}\ }\href {\doibase 10.1093/mnras/sty1483} {\bibfield  {journal}
  {\bibinfo  {journal} {Mon. Not. Roy. Astron. Soc.}\ }\textbf {\bibinfo
  {volume} {479}},\ \bibinfo {pages} {3616--3633} (\bibinfo {year} {2018})},\
  \Eprint {http://arxiv.org/abs/1709.04014} {arXiv:1709.04014 [astro-ph.HE]}
  \BibitemShut {NoStop}%
\bibitem [{\citenamefont {Ammazzalorso}\ \emph {et~al.}(2018)\citenamefont
  {Ammazzalorso}, \citenamefont {Fornengo}, \citenamefont {Horiuchi},\ and\
  \citenamefont {Regis}}]{Ammazzalorso:2018evf}%
  \BibitemOpen
  \bibfield  {author} {\bibinfo {author} {\bibfnamefont {Simone}\ \bibnamefont
  {Ammazzalorso}}, \bibinfo {author} {\bibfnamefont {Nicolao}\ \bibnamefont
  {Fornengo}}, \bibinfo {author} {\bibfnamefont {Shunsaku}\ \bibnamefont
  {Horiuchi}}, \ and\ \bibinfo {author} {\bibfnamefont {Marco}\ \bibnamefont
  {Regis}},\ }\bibfield  {title} {\enquote {\bibinfo {title} {{Characterizing
  the local gamma-ray Universe via angular cross-correlations}},}\ }\href
  {\doibase 10.1103/PhysRevD.98.103007} {\bibfield  {journal} {\bibinfo
  {journal} {Phys. Rev. D}\ }\textbf {\bibinfo {volume} {98}},\ \bibinfo
  {pages} {103007} (\bibinfo {year} {2018})},\ \Eprint
  {http://arxiv.org/abs/1808.09225} {arXiv:1808.09225 [astro-ph.CO]}
  \BibitemShut {NoStop}%
\bibitem [{\citenamefont {Blanco}\ and\ \citenamefont
  {Hooper}(2019)}]{Blanco:2018esa}%
  \BibitemOpen
  \bibfield  {author} {\bibinfo {author} {\bibfnamefont {Carlos}\ \bibnamefont
  {Blanco}}\ and\ \bibinfo {author} {\bibfnamefont {Dan}\ \bibnamefont
  {Hooper}},\ }\bibfield  {title} {\enquote {\bibinfo {title} {{Constraints on
  Decaying Dark Matter from the Isotropic Gamma-Ray Background}},}\ }\href
  {\doibase 10.1088/1475-7516/2019/03/019} {\bibfield  {journal} {\bibinfo
  {journal} {JCAP}\ }\textbf {\bibinfo {volume} {03}},\ \bibinfo {pages} {019}
  (\bibinfo {year} {2019})},\ \Eprint {http://arxiv.org/abs/1811.05988}
  {arXiv:1811.05988 [astro-ph.HE]} \BibitemShut {NoStop}%
\bibitem [{\citenamefont {Arbey}\ \emph {et~al.}(2020)\citenamefont {Arbey},
  \citenamefont {Auffinger},\ and\ \citenamefont {Silk}}]{Arbey:2019vqx}%
  \BibitemOpen
  \bibfield  {author} {\bibinfo {author} {\bibfnamefont {Alexandre}\
  \bibnamefont {Arbey}}, \bibinfo {author} {\bibfnamefont {J\'er\'emy}\
  \bibnamefont {Auffinger}}, \ and\ \bibinfo {author} {\bibfnamefont {Joseph}\
  \bibnamefont {Silk}},\ }\bibfield  {title} {\enquote {\bibinfo {title}
  {{Constraining primordial black hole masses with the isotropic gamma ray
  background}},}\ }\href {\doibase 10.1103/PhysRevD.101.023010} {\bibfield
  {journal} {\bibinfo  {journal} {Phys. Rev. D}\ }\textbf {\bibinfo {volume}
  {101}},\ \bibinfo {pages} {023010} (\bibinfo {year} {2020})},\ \Eprint
  {http://arxiv.org/abs/1906.04750} {arXiv:1906.04750 [astro-ph.CO]}
  \BibitemShut {NoStop}%
\bibitem [{\citenamefont {Yang}(2020)}]{Yang:2020zcu}%
  \BibitemOpen
  \bibfield  {author} {\bibinfo {author} {\bibfnamefont {Yupeng}\ \bibnamefont
  {Yang}},\ }\bibfield  {title} {\enquote {\bibinfo {title} {{The abundance of
  primordial black holes from the global 21cm signal and extragalactic
  gamma-ray background}},}\ }\href {\doibase 10.1140/epjp/s13360-020-00710-3}
  {\bibfield  {journal} {\bibinfo  {journal} {Eur. Phys. J. Plus}\ }\textbf
  {\bibinfo {volume} {135}},\ \bibinfo {pages} {690} (\bibinfo {year}
  {2020})},\ \Eprint {http://arxiv.org/abs/2008.11859} {arXiv:2008.11859
  [astro-ph.CO]} \BibitemShut {NoStop}%
\bibitem [{\citenamefont {Liang}\ \emph {et~al.}(2021)\citenamefont {Liang},
  \citenamefont {Zhang}, \citenamefont {Cheng}, \citenamefont {Zeng},
  \citenamefont {Fan},\ and\ \citenamefont {Liang}}]{Liang:2020roo}%
  \BibitemOpen
  \bibfield  {author} {\bibinfo {author} {\bibfnamefont {Yun-Feng}\
  \bibnamefont {Liang}}, \bibinfo {author} {\bibfnamefont {Xing-Fu}\
  \bibnamefont {Zhang}}, \bibinfo {author} {\bibfnamefont {Ji-Gui}\
  \bibnamefont {Cheng}}, \bibinfo {author} {\bibfnamefont {Hou-Dun}\
  \bibnamefont {Zeng}}, \bibinfo {author} {\bibfnamefont {Yi-Zhong}\
  \bibnamefont {Fan}}, \ and\ \bibinfo {author} {\bibfnamefont {En-Wei}\
  \bibnamefont {Liang}},\ }\bibfield  {title} {\enquote {\bibinfo {title}
  {{Effect of axion-like particles on the spectrum of the extragalactic
  gamma-ray background}},}\ }\href {\doibase 10.1088/1475-7516/2021/11/030}
  {\bibfield  {journal} {\bibinfo  {journal} {JCAP}\ }\textbf {\bibinfo
  {volume} {11}},\ \bibinfo {pages} {030} (\bibinfo {year} {2021})},\ \Eprint
  {http://arxiv.org/abs/2012.15513} {arXiv:2012.15513 [astro-ph.HE]}
  \BibitemShut {NoStop}%
\bibitem [{\citenamefont {Zhang}\ \emph {et~al.}(2022)\citenamefont {Zhang},
  \citenamefont {Cheng}, \citenamefont {Zhu}, \citenamefont {Liu},
  \citenamefont {Liang},\ and\ \citenamefont {Liang}}]{Zhang:2021mth}%
  \BibitemOpen
  \bibfield  {author} {\bibinfo {author} {\bibfnamefont {Xing-Fu}\ \bibnamefont
  {Zhang}}, \bibinfo {author} {\bibfnamefont {Ji-Gui}\ \bibnamefont {Cheng}},
  \bibinfo {author} {\bibfnamefont {Ben-Yang}\ \bibnamefont {Zhu}}, \bibinfo
  {author} {\bibfnamefont {Tian-Ci}\ \bibnamefont {Liu}}, \bibinfo {author}
  {\bibfnamefont {Yun-Feng}\ \bibnamefont {Liang}}, \ and\ \bibinfo {author}
  {\bibfnamefont {En-Wei}\ \bibnamefont {Liang}},\ }\bibfield  {title}
  {\enquote {\bibinfo {title} {{Constraints on ultracompact minihalos from the
  extragalactic gamma-ray background observation}},}\ }\href {\doibase
  10.1103/PhysRevD.105.043011} {\bibfield  {journal} {\bibinfo  {journal}
  {Phys. Rev. D}\ }\textbf {\bibinfo {volume} {105}},\ \bibinfo {pages}
  {043011} (\bibinfo {year} {2022})},\ \Eprint
  {http://arxiv.org/abs/2109.09575} {arXiv:2109.09575 [astro-ph.CO]}
  \BibitemShut {NoStop}%
\bibitem [{\citenamefont {Bartlett}\ \emph {et~al.}(2022)\citenamefont
  {Bartlett}, \citenamefont {Kosti\'c}, \citenamefont {Desmond}, \citenamefont
  {Jasche},\ and\ \citenamefont {Lavaux}}]{Bartlett:2022ztj}%
  \BibitemOpen
  \bibfield  {author} {\bibinfo {author} {\bibfnamefont {Deaglan~J.}\
  \bibnamefont {Bartlett}}, \bibinfo {author} {\bibfnamefont {Andrija}\
  \bibnamefont {Kosti\'c}}, \bibinfo {author} {\bibfnamefont {Harry}\
  \bibnamefont {Desmond}}, \bibinfo {author} {\bibfnamefont {Jens}\
  \bibnamefont {Jasche}}, \ and\ \bibinfo {author} {\bibfnamefont {Guilhem}\
  \bibnamefont {Lavaux}},\ }\bibfield  {title} {\enquote {\bibinfo {title}
  {{Constraints on dark matter annihilation and decay from the large-scale
  structure of the nearby Universe}},}\ }\href {\doibase
  10.1103/PhysRevD.106.103526} {\bibfield  {journal} {\bibinfo  {journal}
  {Phys. Rev. D}\ }\textbf {\bibinfo {volume} {106}},\ \bibinfo {pages}
  {103526} (\bibinfo {year} {2022})},\ \Eprint
  {http://arxiv.org/abs/2205.12916} {arXiv:2205.12916 [astro-ph.CO]}
  \BibitemShut {NoStop}%
\bibitem [{\citenamefont {Paopiamsap}\ \emph {et~al.}(2024)\citenamefont
  {Paopiamsap}, \citenamefont {Alonso}, \citenamefont {Bartlett},\ and\
  \citenamefont {Bilicki}}]{Paopiamsap:2023uuo}%
  \BibitemOpen
  \bibfield  {author} {\bibinfo {author} {\bibfnamefont {Anya}\ \bibnamefont
  {Paopiamsap}}, \bibinfo {author} {\bibfnamefont {David}\ \bibnamefont
  {Alonso}}, \bibinfo {author} {\bibfnamefont {Deaglan~J.}\ \bibnamefont
  {Bartlett}}, \ and\ \bibinfo {author} {\bibfnamefont {Maciej}\ \bibnamefont
  {Bilicki}},\ }\bibfield  {title} {\enquote {\bibinfo {title} {{Constraints on
  dark matter and astrophysics from tomographic \ensuremath{\gamma}-ray
  cross-correlations}},}\ }\href {\doibase 10.1103/PhysRevD.109.103517}
  {\bibfield  {journal} {\bibinfo  {journal} {Phys. Rev. D}\ }\textbf {\bibinfo
  {volume} {109}},\ \bibinfo {pages} {103517} (\bibinfo {year} {2024})},\
  \Eprint {http://arxiv.org/abs/2307.14881} {arXiv:2307.14881 [astro-ph.CO]}
  \BibitemShut {NoStop}%
\bibitem [{\citenamefont {Delos}\ \emph {et~al.}(2024)\citenamefont {Delos},
  \citenamefont {Korsmeier}, \citenamefont {Widmark}, \citenamefont {Blanco},
  \citenamefont {Linden},\ and\ \citenamefont {White}}]{Delos:2023ipo}%
  \BibitemOpen
  \bibfield  {author} {\bibinfo {author} {\bibfnamefont {M.~Sten}\ \bibnamefont
  {Delos}}, \bibinfo {author} {\bibfnamefont {Michael}\ \bibnamefont
  {Korsmeier}}, \bibinfo {author} {\bibfnamefont {Axel}\ \bibnamefont
  {Widmark}}, \bibinfo {author} {\bibfnamefont {Carlos}\ \bibnamefont
  {Blanco}}, \bibinfo {author} {\bibfnamefont {Tim}\ \bibnamefont {Linden}}, \
  and\ \bibinfo {author} {\bibfnamefont {Simon D.~M.}\ \bibnamefont {White}},\
  }\bibfield  {title} {\enquote {\bibinfo {title} {{Limits on dark matter
  annihilation in prompt cusps from the isotropic gamma-ray background}},}\
  }\href {\doibase 10.1103/PhysRevD.109.083512} {\bibfield  {journal} {\bibinfo
   {journal} {Phys. Rev. D}\ }\textbf {\bibinfo {volume} {109}},\ \bibinfo
  {pages} {083512} (\bibinfo {year} {2024})},\ \Eprint
  {http://arxiv.org/abs/2307.13023} {arXiv:2307.13023 [astro-ph.HE]}
  \BibitemShut {NoStop}%
\bibitem [{\citenamefont {Ganjoo}\ and\ \citenamefont
  {Delos}(2024)}]{Ganjoo:2024hpn}%
  \BibitemOpen
  \bibfield  {author} {\bibinfo {author} {\bibfnamefont {Himanish}\
  \bibnamefont {Ganjoo}}\ and\ \bibinfo {author} {\bibfnamefont {M.~Sten}\
  \bibnamefont {Delos}},\ }\bibfield  {title} {\enquote {\bibinfo {title}
  {{Limits on Early Matter Domination from the Isotropic Gamma-Ray
  Background}},}\ }\href@noop {} {\  (\bibinfo {year} {2024})},\ \Eprint
  {http://arxiv.org/abs/2403.18893} {arXiv:2403.18893 [astro-ph.CO]}
  \BibitemShut {NoStop}%
\bibitem [{\citenamefont {Ackermann}\ \emph {et~al.}(2012)\citenamefont
  {Ackermann} \emph {et~al.}}]{Fermi-LAT:2012pez}%
  \BibitemOpen
  \bibfield  {author} {\bibinfo {author} {\bibfnamefont {M.}~\bibnamefont
  {Ackermann}} \emph {et~al.} (\bibinfo {collaboration} {Fermi-LAT}),\
  }\bibfield  {title} {\enquote {\bibinfo {title} {{Anisotropies in the diffuse
  gamma-ray background measured by the Fermi LAT}},}\ }\href {\doibase
  10.1103/PhysRevD.85.083007} {\bibfield  {journal} {\bibinfo  {journal} {Phys.
  Rev. D}\ }\textbf {\bibinfo {volume} {85}},\ \bibinfo {pages} {083007}
  (\bibinfo {year} {2012})},\ \Eprint {http://arxiv.org/abs/1202.2856}
  {arXiv:1202.2856 [astro-ph.HE]} \BibitemShut {NoStop}%
\bibitem [{\citenamefont {Fornasa}\ \emph {et~al.}(2016)\citenamefont {Fornasa}
  \emph {et~al.}}]{Fornasa:2016ohl}%
  \BibitemOpen
  \bibfield  {author} {\bibinfo {author} {\bibfnamefont {Mattia}\ \bibnamefont
  {Fornasa}} \emph {et~al.},\ }\bibfield  {title} {\enquote {\bibinfo {title}
  {{Angular power spectrum of the diffuse gamma-ray emission as measured by the
  Fermi Large Area Telescope and constraints on its dark matter
  interpretation}},}\ }\href {\doibase 10.1103/PhysRevD.94.123005} {\bibfield
  {journal} {\bibinfo  {journal} {Phys. Rev. D}\ }\textbf {\bibinfo {volume}
  {94}},\ \bibinfo {pages} {123005} (\bibinfo {year} {2016})},\ \Eprint
  {http://arxiv.org/abs/1608.07289} {arXiv:1608.07289 [astro-ph.HE]}
  \BibitemShut {NoStop}%
\bibitem [{\citenamefont {Ackermann}\ \emph {et~al.}(2018)\citenamefont
  {Ackermann} \emph {et~al.}}]{Fermi-LAT:2018udj}%
  \BibitemOpen
  \bibfield  {author} {\bibinfo {author} {\bibfnamefont {M.}~\bibnamefont
  {Ackermann}} \emph {et~al.} (\bibinfo {collaboration} {Fermi-LAT}),\
  }\bibfield  {title} {\enquote {\bibinfo {title} {{Unresolved Gamma-Ray Sky
  through its Angular Power Spectrum}},}\ }\href {\doibase
  10.1103/PhysRevLett.121.241101} {\bibfield  {journal} {\bibinfo  {journal}
  {Phys. Rev. Lett.}\ }\textbf {\bibinfo {volume} {121}},\ \bibinfo {pages}
  {241101} (\bibinfo {year} {2018})},\ \Eprint
  {http://arxiv.org/abs/1812.02079} {arXiv:1812.02079 [astro-ph.HE]}
  \BibitemShut {NoStop}%
\bibitem [{\citenamefont {Branchini}\ \emph {et~al.}(2017)\citenamefont
  {Branchini}, \citenamefont {Camera}, \citenamefont {Cuoco}, \citenamefont
  {Fornengo}, \citenamefont {Regis}, \citenamefont {Viel},\ and\ \citenamefont
  {Xia}}]{Branchini:2016glc}%
  \BibitemOpen
  \bibfield  {author} {\bibinfo {author} {\bibfnamefont {Enzo}\ \bibnamefont
  {Branchini}}, \bibinfo {author} {\bibfnamefont {Stefano}\ \bibnamefont
  {Camera}}, \bibinfo {author} {\bibfnamefont {Alessandro}\ \bibnamefont
  {Cuoco}}, \bibinfo {author} {\bibfnamefont {Nicolao}\ \bibnamefont
  {Fornengo}}, \bibinfo {author} {\bibfnamefont {Marco}\ \bibnamefont {Regis}},
  \bibinfo {author} {\bibfnamefont {Matteo}\ \bibnamefont {Viel}}, \ and\
  \bibinfo {author} {\bibfnamefont {Jun-Qing}\ \bibnamefont {Xia}},\ }\bibfield
   {title} {\enquote {\bibinfo {title} {{Cross-correlating the $\gamma$-ray sky
  with Catalogs of Galaxy Clusters}},}\ }\href {\doibase
  10.3847/1538-4365/228/1/8} {\bibfield  {journal} {\bibinfo  {journal}
  {Astrophys. J. Suppl.}\ }\textbf {\bibinfo {volume} {228}},\ \bibinfo {pages}
  {8} (\bibinfo {year} {2017})},\ \Eprint {http://arxiv.org/abs/1612.05788}
  {arXiv:1612.05788 [astro-ph.CO]} \BibitemShut {NoStop}%
\bibitem [{\citenamefont {Hashimoto}\ \emph {et~al.}(2018)\citenamefont
  {Hashimoto}, \citenamefont {Nishizawa}, \citenamefont {Shirasaki},
  \citenamefont {Macias}, \citenamefont {Horiuchi}, \citenamefont {Tashiro},\
  and\ \citenamefont {Oguri}}]{Hashimoto:2018ztv}%
  \BibitemOpen
  \bibfield  {author} {\bibinfo {author} {\bibfnamefont {Daiki}\ \bibnamefont
  {Hashimoto}}, \bibinfo {author} {\bibfnamefont {Atsushi~J.}\ \bibnamefont
  {Nishizawa}}, \bibinfo {author} {\bibfnamefont {Masato}\ \bibnamefont
  {Shirasaki}}, \bibinfo {author} {\bibfnamefont {Oscar}\ \bibnamefont
  {Macias}}, \bibinfo {author} {\bibfnamefont {Shunsaku}\ \bibnamefont
  {Horiuchi}}, \bibinfo {author} {\bibfnamefont {Hiroyuki}\ \bibnamefont
  {Tashiro}}, \ and\ \bibinfo {author} {\bibfnamefont {Masamune}\ \bibnamefont
  {Oguri}},\ }\bibfield  {title} {\enquote {\bibinfo {title} {{Measurement of
  redshift dependent cross correlation of HSC clusters and Fermi $\gamma$
  rays}},}\ }\href {\doibase 10.1093/mnras/stz321} {\  (\bibinfo {year}
  {2018}),\ 10.1093/mnras/stz321},\ \Eprint {http://arxiv.org/abs/1805.08139}
  {arXiv:1805.08139 [astro-ph.CO]} \BibitemShut {NoStop}%
\bibitem [{\citenamefont {Colavincenzo}\ \emph {et~al.}(2020)\citenamefont
  {Colavincenzo}, \citenamefont {Tan}, \citenamefont {Ammazzalorso},
  \citenamefont {Camera}, \citenamefont {Regis}, \citenamefont {Xia},\ and\
  \citenamefont {Fornengo}}]{Colavincenzo:2019jtj}%
  \BibitemOpen
  \bibfield  {author} {\bibinfo {author} {\bibfnamefont {Manuel}\ \bibnamefont
  {Colavincenzo}}, \bibinfo {author} {\bibfnamefont {Xiuhui}\ \bibnamefont
  {Tan}}, \bibinfo {author} {\bibfnamefont {Simone}\ \bibnamefont
  {Ammazzalorso}}, \bibinfo {author} {\bibfnamefont {Stefano}\ \bibnamefont
  {Camera}}, \bibinfo {author} {\bibfnamefont {Marco}\ \bibnamefont {Regis}},
  \bibinfo {author} {\bibfnamefont {Jun-Qing}\ \bibnamefont {Xia}}, \ and\
  \bibinfo {author} {\bibfnamefont {Nicolao}\ \bibnamefont {Fornengo}},\
  }\bibfield  {title} {\enquote {\bibinfo {title} {{Searching for gamma-ray
  emission from galaxy clusters at low redshift}},}\ }\href {\doibase
  10.1093/mnras/stz3263} {\bibfield  {journal} {\bibinfo  {journal} {Mon. Not.
  Roy. Astron. Soc.}\ }\textbf {\bibinfo {volume} {491}},\ \bibinfo {pages}
  {3225--3244} (\bibinfo {year} {2020})},\ \Eprint
  {http://arxiv.org/abs/1907.05264} {arXiv:1907.05264 [astro-ph.CO]}
  \BibitemShut {NoStop}%
\bibitem [{\citenamefont {Di~Mauro}\ \emph {et~al.}(2023)\citenamefont
  {Di~Mauro}, \citenamefont {P\'erez-Romero}, \citenamefont {S\'anchez-Conde},\
  and\ \citenamefont {Fornengo}}]{DiMauro:2023qat}%
  \BibitemOpen
  \bibfield  {author} {\bibinfo {author} {\bibfnamefont {Mattia}\ \bibnamefont
  {Di~Mauro}}, \bibinfo {author} {\bibfnamefont {Judit}\ \bibnamefont
  {P\'erez-Romero}}, \bibinfo {author} {\bibfnamefont {Miguel~A.}\ \bibnamefont
  {S\'anchez-Conde}}, \ and\ \bibinfo {author} {\bibfnamefont {Nicolao}\
  \bibnamefont {Fornengo}},\ }\bibfield  {title} {\enquote {\bibinfo {title}
  {{Constraining the dark matter contribution of \ensuremath{\gamma} rays in
  clusters of galaxies using Fermi-LAT data}},}\ }\href {\doibase
  10.1103/PhysRevD.107.083030} {\bibfield  {journal} {\bibinfo  {journal}
  {Phys. Rev. D}\ }\textbf {\bibinfo {volume} {107}},\ \bibinfo {pages}
  {083030} (\bibinfo {year} {2023})},\ \Eprint
  {http://arxiv.org/abs/2303.16930} {arXiv:2303.16930 [astro-ph.HE]}
  \BibitemShut {NoStop}%
\bibitem [{\citenamefont {Camera}\ \emph {et~al.}(2013)\citenamefont {Camera},
  \citenamefont {Fornasa}, \citenamefont {Fornengo},\ and\ \citenamefont
  {Regis}}]{Camera:2012cj}%
  \BibitemOpen
  \bibfield  {author} {\bibinfo {author} {\bibfnamefont {Stefano}\ \bibnamefont
  {Camera}}, \bibinfo {author} {\bibfnamefont {Mattia}\ \bibnamefont
  {Fornasa}}, \bibinfo {author} {\bibfnamefont {Nicolao}\ \bibnamefont
  {Fornengo}}, \ and\ \bibinfo {author} {\bibfnamefont {Marco}\ \bibnamefont
  {Regis}},\ }\bibfield  {title} {\enquote {\bibinfo {title} {{A Novel Approach
  in the Weakly Interacting Massive Particle Quest: Cross-correlation of
  Gamma-Ray Anisotropies and Cosmic Shear}},}\ }\href {\doibase
  10.1088/2041-8205/771/1/L5} {\bibfield  {journal} {\bibinfo  {journal}
  {Astrophys. J. Lett.}\ }\textbf {\bibinfo {volume} {771}},\ \bibinfo {pages}
  {L5} (\bibinfo {year} {2013})},\ \Eprint {http://arxiv.org/abs/1212.5018}
  {arXiv:1212.5018 [astro-ph.CO]} \BibitemShut {NoStop}%
\bibitem [{\citenamefont {Shirasaki}\ \emph {et~al.}(2014)\citenamefont
  {Shirasaki}, \citenamefont {Horiuchi},\ and\ \citenamefont
  {Yoshida}}]{Shirasaki:2014noa}%
  \BibitemOpen
  \bibfield  {author} {\bibinfo {author} {\bibfnamefont {Masato}\ \bibnamefont
  {Shirasaki}}, \bibinfo {author} {\bibfnamefont {Shunsaku}\ \bibnamefont
  {Horiuchi}}, \ and\ \bibinfo {author} {\bibfnamefont {Naoki}\ \bibnamefont
  {Yoshida}},\ }\bibfield  {title} {\enquote {\bibinfo {title}
  {{Cross-Correlation of Cosmic Shear and Extragalactic Gamma-ray Background:
  Constraints on the Dark Matter Annihilation Cross-Section}},}\ }\href
  {\doibase 10.1103/PhysRevD.90.063502} {\bibfield  {journal} {\bibinfo
  {journal} {Phys. Rev. D}\ }\textbf {\bibinfo {volume} {90}},\ \bibinfo
  {pages} {063502} (\bibinfo {year} {2014})},\ \Eprint
  {http://arxiv.org/abs/1404.5503} {arXiv:1404.5503 [astro-ph.CO]} \BibitemShut
  {NoStop}%
\bibitem [{\citenamefont {Tr\"oster}\ \emph {et~al.}(2017)\citenamefont
  {Tr\"oster} \emph {et~al.}}]{Troster:2016sgf}%
  \BibitemOpen
  \bibfield  {author} {\bibinfo {author} {\bibfnamefont {Tilman}\ \bibnamefont
  {Tr\"oster}} \emph {et~al.},\ }\bibfield  {title} {\enquote {\bibinfo {title}
  {{Cross-correlation of weak lensing and gamma rays: implications for the
  nature of dark matter}},}\ }\href {\doibase 10.1093/mnras/stx365} {\bibfield
  {journal} {\bibinfo  {journal} {Mon. Not. Roy. Astron. Soc.}\ }\textbf
  {\bibinfo {volume} {467}},\ \bibinfo {pages} {2706--2722} (\bibinfo {year}
  {2017})},\ \Eprint {http://arxiv.org/abs/1611.03554} {arXiv:1611.03554
  [astro-ph.CO]} \BibitemShut {NoStop}%
\bibitem [{\citenamefont {Shirasaki}\ \emph {et~al.}(2018)\citenamefont
  {Shirasaki}, \citenamefont {Macias}, \citenamefont {Horiuchi}, \citenamefont
  {Yoshida}, \citenamefont {Lee},\ and\ \citenamefont
  {Nishizawa}}]{Shirasaki:2018dkz}%
  \BibitemOpen
  \bibfield  {author} {\bibinfo {author} {\bibfnamefont {Masato}\ \bibnamefont
  {Shirasaki}}, \bibinfo {author} {\bibfnamefont {Oscar}\ \bibnamefont
  {Macias}}, \bibinfo {author} {\bibfnamefont {Shunsaku}\ \bibnamefont
  {Horiuchi}}, \bibinfo {author} {\bibfnamefont {Naoki}\ \bibnamefont
  {Yoshida}}, \bibinfo {author} {\bibfnamefont {Chien-Hsiu}\ \bibnamefont
  {Lee}}, \ and\ \bibinfo {author} {\bibfnamefont {Atsushi~J.}\ \bibnamefont
  {Nishizawa}},\ }\bibfield  {title} {\enquote {\bibinfo {title} {{Correlation
  of extragalactic \ensuremath{\gamma} rays with cosmic matter density
  distributions from weak gravitational lensing}},}\ }\href {\doibase
  10.1103/PhysRevD.97.123015} {\bibfield  {journal} {\bibinfo  {journal} {Phys.
  Rev. D}\ }\textbf {\bibinfo {volume} {97}},\ \bibinfo {pages} {123015}
  (\bibinfo {year} {2018})},\ \Eprint {http://arxiv.org/abs/1802.10257}
  {arXiv:1802.10257 [astro-ph.CO]} \BibitemShut {NoStop}%
\bibitem [{\citenamefont {Ammazzalorso}\ \emph {et~al.}(2020)\citenamefont
  {Ammazzalorso} \emph {et~al.}}]{DES:2019ucp}%
  \BibitemOpen
  \bibfield  {author} {\bibinfo {author} {\bibfnamefont {S.}~\bibnamefont
  {Ammazzalorso}} \emph {et~al.} (\bibinfo {collaboration} {DES}),\ }\bibfield
  {title} {\enquote {\bibinfo {title} {{Detection of Cross-Correlation between
  Gravitational Lensing and $\gamma$ Rays}},}\ }\href {\doibase
  10.1103/PhysRevLett.124.101102} {\bibfield  {journal} {\bibinfo  {journal}
  {Phys. Rev. Lett.}\ }\textbf {\bibinfo {volume} {124}},\ \bibinfo {pages}
  {101102} (\bibinfo {year} {2020})},\ \Eprint
  {http://arxiv.org/abs/1907.13484} {arXiv:1907.13484 [astro-ph.CO]}
  \BibitemShut {NoStop}%
\bibitem [{\citenamefont {Tan}\ \emph {et~al.}(2020)\citenamefont {Tan},
  \citenamefont {Dai},\ and\ \citenamefont {Xia}}]{Tan:2020fbc}%
  \BibitemOpen
  \bibfield  {author} {\bibinfo {author} {\bibfnamefont {Xiu-Hui}\ \bibnamefont
  {Tan}}, \bibinfo {author} {\bibfnamefont {Ji-Ping}\ \bibnamefont {Dai}}, \
  and\ \bibinfo {author} {\bibfnamefont {Jun-Qing}\ \bibnamefont {Xia}},\
  }\bibfield  {title} {\enquote {\bibinfo {title} {{Searching for Integrated
  Sachs\textendash{}Wolfe Effect from $Fermi$-LAT diffuse $\gamma$-ray map}},}\
  }\href {\doibase 10.1016/j.dark.2020.100585} {\bibfield  {journal} {\bibinfo
  {journal} {Phys. Dark Univ.}\ }\textbf {\bibinfo {volume} {29}},\ \bibinfo
  {pages} {100585} (\bibinfo {year} {2020})},\ \Eprint
  {http://arxiv.org/abs/2005.03833} {arXiv:2005.03833 [astro-ph.CO]}
  \BibitemShut {NoStop}%
\bibitem [{\citenamefont {Fornengo}\ \emph {et~al.}(2015)\citenamefont
  {Fornengo}, \citenamefont {Perotto}, \citenamefont {Regis},\ and\
  \citenamefont {Camera}}]{Fornengo:2014cya}%
  \BibitemOpen
  \bibfield  {author} {\bibinfo {author} {\bibfnamefont {Nicolao}\ \bibnamefont
  {Fornengo}}, \bibinfo {author} {\bibfnamefont {Laurence}\ \bibnamefont
  {Perotto}}, \bibinfo {author} {\bibfnamefont {Marco}\ \bibnamefont {Regis}},
  \ and\ \bibinfo {author} {\bibfnamefont {Stefano}\ \bibnamefont {Camera}},\
  }\bibfield  {title} {\enquote {\bibinfo {title} {{Evidence of
  Cross-correlation between the CMB Lensing and the $\Gamma$-ray sky}},}\
  }\href {\doibase 10.1088/2041-8205/802/1/L1} {\bibfield  {journal} {\bibinfo
  {journal} {Astrophys. J. Lett.}\ }\textbf {\bibinfo {volume} {802}},\
  \bibinfo {pages} {L1} (\bibinfo {year} {2015})},\ \Eprint
  {http://arxiv.org/abs/1410.4997} {arXiv:1410.4997 [astro-ph.CO]} \BibitemShut
  {NoStop}%
\bibitem [{\citenamefont {Feng}\ \emph {et~al.}(2017)\citenamefont {Feng},
  \citenamefont {Cooray},\ and\ \citenamefont {Keating}}]{Feng:2016fkl}%
  \BibitemOpen
  \bibfield  {author} {\bibinfo {author} {\bibfnamefont {Chang}\ \bibnamefont
  {Feng}}, \bibinfo {author} {\bibfnamefont {Asantha}\ \bibnamefont {Cooray}},
  \ and\ \bibinfo {author} {\bibfnamefont {Brian}\ \bibnamefont {Keating}},\
  }\bibfield  {title} {\enquote {\bibinfo {title} {{Planck Lensing and Cosmic
  Infrared Background Cross-Correlation with Fermi-LAT: Tracing Dark Matter
  Signals in the Gamma-Ray Background}},}\ }\href {\doibase
  10.3847/1538-4357/836/1/127} {\bibfield  {journal} {\bibinfo  {journal}
  {Astrophys. J.}\ }\textbf {\bibinfo {volume} {836}},\ \bibinfo {pages} {127}
  (\bibinfo {year} {2017})},\ \Eprint {http://arxiv.org/abs/1608.04351}
  {arXiv:1608.04351 [astro-ph.CO]} \BibitemShut {NoStop}%
\bibitem [{\citenamefont {Shirasaki}\ \emph {et~al.}(2020)\citenamefont
  {Shirasaki}, \citenamefont {Macias}, \citenamefont {Ando}, \citenamefont
  {Horiuchi},\ and\ \citenamefont {Yoshida}}]{Shirasaki:2019uls}%
  \BibitemOpen
  \bibfield  {author} {\bibinfo {author} {\bibfnamefont {Masato}\ \bibnamefont
  {Shirasaki}}, \bibinfo {author} {\bibfnamefont {Oscar}\ \bibnamefont
  {Macias}}, \bibinfo {author} {\bibfnamefont {Shin'ichiro}\ \bibnamefont
  {Ando}}, \bibinfo {author} {\bibfnamefont {Shunsaku}\ \bibnamefont
  {Horiuchi}}, \ and\ \bibinfo {author} {\bibfnamefont {Naoki}\ \bibnamefont
  {Yoshida}},\ }\bibfield  {title} {\enquote {\bibinfo {title}
  {{Cross-correlation of the extragalactic gamma-ray background with the
  thermal Sunyaev-Zel\textquoteright{}dovich effect in the cosmic microwave
  background}},}\ }\href {\doibase 10.1103/PhysRevD.101.103022} {\bibfield
  {journal} {\bibinfo  {journal} {Phys. Rev. D}\ }\textbf {\bibinfo {volume}
  {101}},\ \bibinfo {pages} {103022} (\bibinfo {year} {2020})},\ \Eprint
  {http://arxiv.org/abs/1911.11841} {arXiv:1911.11841 [astro-ph.CO]}
  \BibitemShut {NoStop}%
\bibitem [{\citenamefont {Negro}\ \emph {et~al.}(2023)\citenamefont {Negro},
  \citenamefont {Crnogor\v{c}evi\'c}, \citenamefont {Burns}, \citenamefont
  {Charles}, \citenamefont {Marcotulli},\ and\ \citenamefont
  {Caputo}}]{Negro:2023kwv}%
  \BibitemOpen
  \bibfield  {author} {\bibinfo {author} {\bibfnamefont {Michela}\ \bibnamefont
  {Negro}}, \bibinfo {author} {\bibfnamefont {Milena}\ \bibnamefont
  {Crnogor\v{c}evi\'c}}, \bibinfo {author} {\bibfnamefont {Eric}\ \bibnamefont
  {Burns}}, \bibinfo {author} {\bibfnamefont {Eric}\ \bibnamefont {Charles}},
  \bibinfo {author} {\bibfnamefont {Lea}\ \bibnamefont {Marcotulli}}, \ and\
  \bibinfo {author} {\bibfnamefont {Regina}\ \bibnamefont {Caputo}},\
  }\bibfield  {title} {\enquote {\bibinfo {title} {{A Cross-correlation Study
  between IceCube Neutrino Events and the FERMI Unresolved Gamma-Ray Sky}},}\
  }\href {\doibase 10.3847/1538-4357/acd172} {\bibfield  {journal} {\bibinfo
  {journal} {Astrophys. J.}\ }\textbf {\bibinfo {volume} {951}},\ \bibinfo
  {pages} {83} (\bibinfo {year} {2023})},\ \Eprint
  {http://arxiv.org/abs/2304.10934} {arXiv:2304.10934 [astro-ph.HE]}
  \BibitemShut {NoStop}%
\bibitem [{\citenamefont {Aghamousa}\ \emph {et~al.}(2016)\citenamefont
  {Aghamousa} \emph {et~al.}}]{DESI:2016fyo}%
  \BibitemOpen
  \bibfield  {author} {\bibinfo {author} {\bibfnamefont {Amir}\ \bibnamefont
  {Aghamousa}} \emph {et~al.} (\bibinfo {collaboration} {DESI}),\ }\bibfield
  {title} {\enquote {\bibinfo {title} {{The DESI Experiment Part I:
  Science,Targeting, and Survey Design}},}\ }\href@noop {} {\  (\bibinfo {year}
  {2016})},\ \Eprint {http://arxiv.org/abs/1611.00036} {arXiv:1611.00036
  [astro-ph.IM]} \BibitemShut {NoStop}%
\bibitem [{DES()}]{DESI_web}%
  \BibitemOpen
  \href@noop {} {}\bibinfo {howpublished} {\url{https://www.desi.lbl.gov/
  }}\BibitemShut {NoStop}%
\bibitem [{\citenamefont {Aghanim}\ \emph {et~al.}(2020)\citenamefont {Aghanim}
  \emph {et~al.}}]{Planck:2018vyg}%
  \BibitemOpen
  \bibfield  {author} {\bibinfo {author} {\bibfnamefont {N.}~\bibnamefont
  {Aghanim}} \emph {et~al.} (\bibinfo {collaboration} {Planck}),\ }\bibfield
  {title} {\enquote {\bibinfo {title} {{Planck 2018 results. VI. Cosmological
  parameters}},}\ }\href {\doibase 10.1051/0004-6361/201833910} {\bibfield
  {journal} {\bibinfo  {journal} {Astron. Astrophys.}\ }\textbf {\bibinfo
  {volume} {641}},\ \bibinfo {pages} {A6} (\bibinfo {year} {2020})},\ \bibinfo
  {note} {[Erratum: Astron.Astrophys. 652, C4 (2021)]},\ \Eprint
  {http://arxiv.org/abs/1807.06209} {arXiv:1807.06209 [astro-ph.CO]}
  \BibitemShut {NoStop}%
\bibitem [{\citenamefont {Bellomo}\ \emph {et~al.}(2020)\citenamefont
  {Bellomo}, \citenamefont {Bernal}, \citenamefont {Scelfo}, \citenamefont
  {Raccanelli},\ and\ \citenamefont {Verde}}]{Bellomo:2020pnw}%
  \BibitemOpen
  \bibfield  {author} {\bibinfo {author} {\bibfnamefont {Nicola}\ \bibnamefont
  {Bellomo}}, \bibinfo {author} {\bibfnamefont {Jos\'e~Luis}\ \bibnamefont
  {Bernal}}, \bibinfo {author} {\bibfnamefont {Giulio}\ \bibnamefont {Scelfo}},
  \bibinfo {author} {\bibfnamefont {Alvise}\ \bibnamefont {Raccanelli}}, \ and\
  \bibinfo {author} {\bibfnamefont {Licia}\ \bibnamefont {Verde}},\ }\bibfield
  {title} {\enquote {\bibinfo {title} {{Beware of commonly used approximations.
  Part I. Errors in forecasts}},}\ }\href {\doibase
  10.1088/1475-7516/2020/10/016} {\bibfield  {journal} {\bibinfo  {journal}
  {JCAP}\ }\textbf {\bibinfo {volume} {10}},\ \bibinfo {pages} {016} (\bibinfo
  {year} {2020})},\ \Eprint {http://arxiv.org/abs/2005.10384} {arXiv:2005.10384
  [astro-ph.CO]} \BibitemShut {NoStop}%
\bibitem [{\citenamefont {Bernal}\ \emph {et~al.}(2020)\citenamefont {Bernal},
  \citenamefont {Bellomo}, \citenamefont {Raccanelli},\ and\ \citenamefont
  {Verde}}]{Bernal:2020pwq}%
  \BibitemOpen
  \bibfield  {author} {\bibinfo {author} {\bibfnamefont {Jos\'e~Luis}\
  \bibnamefont {Bernal}}, \bibinfo {author} {\bibfnamefont {Nicola}\
  \bibnamefont {Bellomo}}, \bibinfo {author} {\bibfnamefont {Alvise}\
  \bibnamefont {Raccanelli}}, \ and\ \bibinfo {author} {\bibfnamefont {Licia}\
  \bibnamefont {Verde}},\ }\bibfield  {title} {\enquote {\bibinfo {title}
  {{Beware of commonly used approximations. Part II. Estimating systematic
  biases in the best-fit parameters}},}\ }\href {\doibase
  10.1088/1475-7516/2020/10/017} {\bibfield  {journal} {\bibinfo  {journal}
  {JCAP}\ }\textbf {\bibinfo {volume} {10}},\ \bibinfo {pages} {017} (\bibinfo
  {year} {2020})},\ \Eprint {http://arxiv.org/abs/2005.09666} {arXiv:2005.09666
  [astro-ph.CO]} \BibitemShut {NoStop}%
\bibitem [{\citenamefont {{Blas}}\ \emph {et~al.}(2011)\citenamefont {{Blas}},
  \citenamefont {{Lesgourgues}},\ and\ \citenamefont
  {{Tram}}}]{2011JCAP...07..034B}%
  \BibitemOpen
  \bibfield  {author} {\bibinfo {author} {\bibfnamefont {Diego}\ \bibnamefont
  {{Blas}}}, \bibinfo {author} {\bibfnamefont {Julien}\ \bibnamefont
  {{Lesgourgues}}}, \ and\ \bibinfo {author} {\bibfnamefont {Thomas}\
  \bibnamefont {{Tram}}},\ }\bibfield  {title} {\enquote {\bibinfo {title}
  {{The Cosmic Linear Anisotropy Solving System (CLASS). Part II: Approximation
  schemes}},}\ }\href {\doibase 10.1088/1475-7516/2011/07/034} {\bibfield
  {journal} {\bibinfo  {journal} {\jcap}\ }\textbf {\bibinfo {volume} {2011}},\
  \bibinfo {eid} {034} (\bibinfo {year} {2011})},\ \Eprint
  {http://arxiv.org/abs/1104.2933} {arXiv:1104.2933 [astro-ph.CO]} \BibitemShut
  {NoStop}%
\bibitem [{\citenamefont {Finke}\ \emph {et~al.}(2010)\citenamefont {Finke},
  \citenamefont {Razzaque},\ and\ \citenamefont {Dermer}}]{Finke:2009xi}%
  \BibitemOpen
  \bibfield  {author} {\bibinfo {author} {\bibfnamefont {Justin~D.}\
  \bibnamefont {Finke}}, \bibinfo {author} {\bibfnamefont {Soebur}\
  \bibnamefont {Razzaque}}, \ and\ \bibinfo {author} {\bibfnamefont
  {Charles~D.}\ \bibnamefont {Dermer}},\ }\bibfield  {title} {\enquote
  {\bibinfo {title} {{Modeling the Extragalactic Background Light from Stars
  and Dust}},}\ }\href {\doibase 10.1088/0004-637X/712/1/238} {\bibfield
  {journal} {\bibinfo  {journal} {Astrophys. J.}\ }\textbf {\bibinfo {volume}
  {712}},\ \bibinfo {pages} {238--249} (\bibinfo {year} {2010})},\ \Eprint
  {http://arxiv.org/abs/0905.1115} {arXiv:0905.1115 [astro-ph.HE]} \BibitemShut
  {NoStop}%
\bibitem [{\citenamefont {Pinetti}(2021)}]{Pinetti:2021jjs}%
  \BibitemOpen
  \bibfield  {author} {\bibinfo {author} {\bibfnamefont {Elena}\ \bibnamefont
  {Pinetti}},\ }\emph {\bibinfo {title} {{From gamma rays to radio waves: Dark
  Matter searches across the spectrum}}},\ \href@noop {} {Ph.D. thesis},\
  \bibinfo  {school} {INFN, Turin} (\bibinfo {year} {2021}),\ \Eprint
  {http://arxiv.org/abs/2212.00125} {arXiv:2212.00125 [astro-ph.HE]}
  \BibitemShut {NoStop}%
\bibitem [{\citenamefont {{Yuan}}\ \emph {et~al.}(2018)\citenamefont {{Yuan}},
  \citenamefont {{Wang}}, \citenamefont {{Worrall}}, \citenamefont {{Zhang}},\
  and\ \citenamefont {{Mao}}}]{Yuan2018LFs}%
  \BibitemOpen
  \bibfield  {author} {\bibinfo {author} {\bibfnamefont {Zunli}\ \bibnamefont
  {{Yuan}}}, \bibinfo {author} {\bibfnamefont {Jiancheng}\ \bibnamefont
  {{Wang}}}, \bibinfo {author} {\bibfnamefont {D.~M.}\ \bibnamefont
  {{Worrall}}}, \bibinfo {author} {\bibfnamefont {Bin-Bin}\ \bibnamefont
  {{Zhang}}}, \ and\ \bibinfo {author} {\bibfnamefont {Jirong}\ \bibnamefont
  {{Mao}}},\ }\bibfield  {title} {\enquote {\bibinfo {title} {{Determining the
  Core Radio Luminosity Function of Radio AGNs via Copula}},}\ }\href {\doibase
  10.3847/1538-4365/aaed3b} {\bibfield  {journal} {\bibinfo  {journal} {\apjs}\
  }\textbf {\bibinfo {volume} {239}},\ \bibinfo {eid} {33} (\bibinfo {year}
  {2018})},\ \Eprint {http://arxiv.org/abs/1810.12713} {arXiv:1810.12713
  [astro-ph.GA]} \BibitemShut {NoStop}%
\bibitem [{\citenamefont {Bernal}\ \emph {et~al.}(2019)\citenamefont {Bernal},
  \citenamefont {Raccanelli}, \citenamefont {Kovetz}, \citenamefont
  {Parkinson}, \citenamefont {Norris}, \citenamefont {Danforth},\ and\
  \citenamefont {Schmitt}}]{Bernal:2018myq}%
  \BibitemOpen
  \bibfield  {author} {\bibinfo {author} {\bibfnamefont {Jos\'e~Luis}\
  \bibnamefont {Bernal}}, \bibinfo {author} {\bibfnamefont {Alvise}\
  \bibnamefont {Raccanelli}}, \bibinfo {author} {\bibfnamefont {Ely~D.}\
  \bibnamefont {Kovetz}}, \bibinfo {author} {\bibfnamefont {David}\
  \bibnamefont {Parkinson}}, \bibinfo {author} {\bibfnamefont {Ray~P.}\
  \bibnamefont {Norris}}, \bibinfo {author} {\bibfnamefont {George}\
  \bibnamefont {Danforth}}, \ and\ \bibinfo {author} {\bibfnamefont {Courtney}\
  \bibnamefont {Schmitt}},\ }\bibfield  {title} {\enquote {\bibinfo {title}
  {{Probing $\Lambda$CDM cosmology with the Evolutionary Map of the Universe
  survey}},}\ }\href {\doibase 10.1088/1475-7516/2019/02/030} {\bibfield
  {journal} {\bibinfo  {journal} {JCAP}\ }\textbf {\bibinfo {volume} {02}},\
  \bibinfo {pages} {030} (\bibinfo {year} {2019})},\ \Eprint
  {http://arxiv.org/abs/1810.06672} {arXiv:1810.06672 [astro-ph.CO]}
  \BibitemShut {NoStop}%
\bibitem [{\citenamefont {Gruppioni}\ \emph {et~al.}(2013)\citenamefont
  {Gruppioni} \emph {et~al.}}]{Gruppioni:2013jna}%
  \BibitemOpen
  \bibfield  {author} {\bibinfo {author} {\bibfnamefont {C.}~\bibnamefont
  {Gruppioni}} \emph {et~al.},\ }\bibfield  {title} {\enquote {\bibinfo {title}
  {{The Herschel PEP/HerMES Luminosity Function. I: Probing the Evolution of
  PACS selected Galaxies to z\textasciitilde{}4}},}\ }\href {\doibase
  10.1093/mnras/stt308} {\bibfield  {journal} {\bibinfo  {journal} {Mon. Not.
  Roy. Astron. Soc.}\ }\textbf {\bibinfo {volume} {432}},\ \bibinfo {pages}
  {23} (\bibinfo {year} {2013})},\ \Eprint {http://arxiv.org/abs/1302.5209}
  {arXiv:1302.5209 [astro-ph.CO]} \BibitemShut {NoStop}%
\bibitem [{\citenamefont {Ambrosone}\ \emph {et~al.}(2021)\citenamefont
  {Ambrosone}, \citenamefont {Chianese}, \citenamefont {Fiorillo},
  \citenamefont {Marinelli}, \citenamefont {Miele},\ and\ \citenamefont
  {Pisanti}}]{Ambrosone:2020evo}%
  \BibitemOpen
  \bibfield  {author} {\bibinfo {author} {\bibfnamefont {Antonio}\ \bibnamefont
  {Ambrosone}}, \bibinfo {author} {\bibfnamefont {Marco}\ \bibnamefont
  {Chianese}}, \bibinfo {author} {\bibfnamefont {Damiano F.~G.}\ \bibnamefont
  {Fiorillo}}, \bibinfo {author} {\bibfnamefont {Antonio}\ \bibnamefont
  {Marinelli}}, \bibinfo {author} {\bibfnamefont {Gennaro}\ \bibnamefont
  {Miele}}, \ and\ \bibinfo {author} {\bibfnamefont {Ofelia}\ \bibnamefont
  {Pisanti}},\ }\bibfield  {title} {\enquote {\bibinfo {title} {{Starburst
  galaxies strike back: a multi-messenger analysis with Fermi-LAT and IceCube
  data}},}\ }\href {\doibase 10.1093/mnras/stab659} {\bibfield  {journal}
  {\bibinfo  {journal} {Mon. Not. Roy. Astron. Soc.}\ }\textbf {\bibinfo
  {volume} {503}},\ \bibinfo {pages} {4032--4049} (\bibinfo {year} {2021})},\
  \Eprint {http://arxiv.org/abs/2011.02483} {arXiv:2011.02483 [astro-ph.HE]}
  \BibitemShut {NoStop}%
\bibitem [{\citenamefont {Ajello}\ \emph {et~al.}(2014)\citenamefont {Ajello}
  \emph {et~al.}}]{Ajello:2013lka}%
  \BibitemOpen
  \bibfield  {author} {\bibinfo {author} {\bibfnamefont {M.}~\bibnamefont
  {Ajello}} \emph {et~al.},\ }\bibfield  {title} {\enquote {\bibinfo {title}
  {{The Cosmic Evolution of Fermi BL Lacertae Objects}},}\ }\href {\doibase
  10.1088/0004-637X/780/1/73} {\bibfield  {journal} {\bibinfo  {journal}
  {Astrophys. J.}\ }\textbf {\bibinfo {volume} {780}},\ \bibinfo {pages} {73}
  (\bibinfo {year} {2014})},\ \Eprint {http://arxiv.org/abs/1310.0006}
  {arXiv:1310.0006 [astro-ph.CO]} \BibitemShut {NoStop}%
\bibitem [{\citenamefont {Ajello}\ \emph {et~al.}(2012)\citenamefont {Ajello}
  \emph {et~al.}}]{Ajello:2011zi}%
  \BibitemOpen
  \bibfield  {author} {\bibinfo {author} {\bibfnamefont {M.}~\bibnamefont
  {Ajello}} \emph {et~al.},\ }\bibfield  {title} {\enquote {\bibinfo {title}
  {{The Luminosity Function of Fermi-detected Flat-Spectrum Radio Quasars}},}\
  }\href {\doibase 10.1088/0004-637X/751/2/108} {\bibfield  {journal} {\bibinfo
   {journal} {Astrophys. J.}\ }\textbf {\bibinfo {volume} {751}},\ \bibinfo
  {pages} {108} (\bibinfo {year} {2012})},\ \Eprint
  {http://arxiv.org/abs/1110.3787} {arXiv:1110.3787 [astro-ph.CO]} \BibitemShut
  {NoStop}%
\bibitem [{\citenamefont {Zhou}\ \emph {et~al.}(2023)\citenamefont {Zhou} \emph
  {et~al.}}]{DESI:2022gle}%
  \BibitemOpen
  \bibfield  {author} {\bibinfo {author} {\bibfnamefont {Rongpu}\ \bibnamefont
  {Zhou}} \emph {et~al.} (\bibinfo {collaboration} {DESI}),\ }\bibfield
  {title} {\enquote {\bibinfo {title} {{Target Selection and Validation of DESI
  Luminous Red Galaxies}},}\ }\href {\doibase 10.3847/1538-3881/aca5fb}
  {\bibfield  {journal} {\bibinfo  {journal} {Astron. J.}\ }\textbf {\bibinfo
  {volume} {165}},\ \bibinfo {pages} {58} (\bibinfo {year} {2023})},\ \Eprint
  {http://arxiv.org/abs/2208.08515} {arXiv:2208.08515 [astro-ph.CO]}
  \BibitemShut {NoStop}%
\bibitem [{\citenamefont {Yuan}\ \emph {et~al.}(2024)\citenamefont {Yuan} \emph
  {et~al.}}]{Yuan:2023ezi}%
  \BibitemOpen
  \bibfield  {author} {\bibinfo {author} {\bibfnamefont {Sihan}\ \bibnamefont
  {Yuan}} \emph {et~al.},\ }\bibfield  {title} {\enquote {\bibinfo {title}
  {{The DESI one-per\,cent survey: exploring the halo occupation distribution
  of luminous red galaxies and quasi-stellar objects with AbacusSummit}},}\
  }\href {\doibase 10.1093/mnras/stae359} {\bibfield  {journal} {\bibinfo
  {journal} {Mon. Not. Roy. Astron. Soc.}\ }\textbf {\bibinfo {volume} {530}},\
  \bibinfo {pages} {947--965} (\bibinfo {year} {2024})},\ \Eprint
  {http://arxiv.org/abs/2306.06314} {arXiv:2306.06314 [astro-ph.CO]}
  \BibitemShut {NoStop}%
\bibitem [{\citenamefont {Prada}\ \emph {et~al.}(2023)\citenamefont {Prada}
  \emph {et~al.}}]{Prada:2023lmw}%
  \BibitemOpen
  \bibfield  {author} {\bibinfo {author} {\bibfnamefont {F.}~\bibnamefont
  {Prada}} \emph {et~al.},\ }\bibfield  {title} {\enquote {\bibinfo {title}
  {{The DESI One-Percent Survey: Modelling the clustering and halo occupation
  of all four DESI tracers with Uchuu}},}\ }\href@noop {} {\  (\bibinfo {year}
  {2023})},\ \Eprint {http://arxiv.org/abs/2306.06315} {arXiv:2306.06315
  [astro-ph.CO]} \BibitemShut {NoStop}%
\bibitem [{\citenamefont {Zhou}\ \emph {et~al.}(2021)\citenamefont {Zhou} \emph
  {et~al.}}]{Zhou:2020nwq}%
  \BibitemOpen
  \bibfield  {author} {\bibinfo {author} {\bibfnamefont {Rongpu}\ \bibnamefont
  {Zhou}} \emph {et~al.},\ }\bibfield  {title} {\enquote {\bibinfo {title}
  {{The Clustering of DESI-like Luminous Red Galaxies Using Photometric
  Redshifts}},}\ }\href {\doibase 10.1093/mnras/staa3764} {\bibfield  {journal}
  {\bibinfo  {journal} {Mon. Not. Roy. Astron. Soc.}\ }\textbf {\bibinfo
  {volume} {501}},\ \bibinfo {pages} {3309--3331} (\bibinfo {year} {2021})},\
  \Eprint {http://arxiv.org/abs/2001.06018} {arXiv:2001.06018 [astro-ph.CO]}
  \BibitemShut {NoStop}%
\bibitem [{\citenamefont {Rezaie}\ \emph {et~al.}(2023)\citenamefont {Rezaie}
  \emph {et~al.}}]{Rezaie:2023lvi}%
  \BibitemOpen
  \bibfield  {author} {\bibinfo {author} {\bibfnamefont {Mehdi}\ \bibnamefont
  {Rezaie}} \emph {et~al.},\ }\bibfield  {title} {\enquote {\bibinfo {title}
  {{Local primordial non-Gaussianity from the large-scale clustering of
  photometric DESI luminous red galaxies}},}\ }\href@noop {} {\  (\bibinfo
  {year} {2023})},\ \Eprint {http://arxiv.org/abs/2307.01753} {arXiv:2307.01753
  [astro-ph.CO]} \BibitemShut {NoStop}%
\bibitem [{\citenamefont {Raichoor}\ \emph {et~al.}(2023)\citenamefont
  {Raichoor} \emph {et~al.}}]{Raichoor:2022jab}%
  \BibitemOpen
  \bibfield  {author} {\bibinfo {author} {\bibfnamefont {A.}~\bibnamefont
  {Raichoor}} \emph {et~al.},\ }\bibfield  {title} {\enquote {\bibinfo {title}
  {{Target Selection and Validation of DESI Emission Line Galaxies}},}\ }\href
  {\doibase 10.3847/1538-3881/acb213} {\bibfield  {journal} {\bibinfo
  {journal} {Astron. J.}\ }\textbf {\bibinfo {volume} {165}},\ \bibinfo {pages}
  {126} (\bibinfo {year} {2023})},\ \Eprint {http://arxiv.org/abs/2208.08513}
  {arXiv:2208.08513 [astro-ph.CO]} \BibitemShut {NoStop}%
\bibitem [{\citenamefont {Rocher}\ \emph {et~al.}(2023)\citenamefont {Rocher}
  \emph {et~al.}}]{Rocher:2023zyh}%
  \BibitemOpen
  \bibfield  {author} {\bibinfo {author} {\bibfnamefont {Antoine}\ \bibnamefont
  {Rocher}} \emph {et~al.},\ }\bibfield  {title} {\enquote {\bibinfo {title}
  {{The DESI One-Percent survey: exploring the Halo Occupation Distribution of
  Emission Line Galaxies with AbacusSummit simulations}},}\ }\href {\doibase
  10.1088/1475-7516/2023/10/016} {\bibfield  {journal} {\bibinfo  {journal}
  {JCAP}\ }\textbf {\bibinfo {volume} {10}},\ \bibinfo {pages} {016} (\bibinfo
  {year} {2023})},\ \Eprint {http://arxiv.org/abs/2306.06319} {arXiv:2306.06319
  [astro-ph.CO]} \BibitemShut {NoStop}%
\bibitem [{\citenamefont {Chaussidon}\ \emph {et~al.}(2023)\citenamefont
  {Chaussidon} \emph {et~al.}}]{Chaussidon:2022pqg}%
  \BibitemOpen
  \bibfield  {author} {\bibinfo {author} {\bibfnamefont {Edmond}\ \bibnamefont
  {Chaussidon}} \emph {et~al.},\ }\bibfield  {title} {\enquote {\bibinfo
  {title} {{Target Selection and Validation of DESI Quasars}},}\ }\href
  {\doibase 10.3847/1538-4357/acb3c2} {\bibfield  {journal} {\bibinfo
  {journal} {Astrophys. J.}\ }\textbf {\bibinfo {volume} {944}},\ \bibinfo
  {pages} {107} (\bibinfo {year} {2023})},\ \Eprint
  {http://arxiv.org/abs/2208.08511} {arXiv:2208.08511 [astro-ph.CO]}
  \BibitemShut {NoStop}%
\bibitem [{\citenamefont {Krolewski}\ \emph {et~al.}(2024)\citenamefont
  {Krolewski} \emph {et~al.}}]{DESI:2023duv}%
  \BibitemOpen
  \bibfield  {author} {\bibinfo {author} {\bibfnamefont {Alex}\ \bibnamefont
  {Krolewski}} \emph {et~al.} (\bibinfo {collaboration} {DESI}),\ }\bibfield
  {title} {\enquote {\bibinfo {title} {{Constraining primordial non-Gaussianity
  from DESI quasar targets and Planck CMB lensing}},}\ }\href {\doibase
  10.1088/1475-7516/2024/03/021} {\bibfield  {journal} {\bibinfo  {journal}
  {JCAP}\ }\textbf {\bibinfo {volume} {03}},\ \bibinfo {pages} {021} (\bibinfo
  {year} {2024})},\ \Eprint {http://arxiv.org/abs/2305.07650} {arXiv:2305.07650
  [astro-ph.CO]} \BibitemShut {NoStop}%
\bibitem [{\citenamefont {Ackermann}\ \emph
  {et~al.}(2015{\natexlab{c}})\citenamefont {Ackermann} \emph
  {et~al.}}]{Fermi-LAT:2015att}%
  \BibitemOpen
  \bibfield  {author} {\bibinfo {author} {\bibfnamefont {M.}~\bibnamefont
  {Ackermann}} \emph {et~al.} (\bibinfo {collaboration} {Fermi-LAT}),\
  }\bibfield  {title} {\enquote {\bibinfo {title} {{Searching for Dark Matter
  Annihilation from Milky Way Dwarf Spheroidal Galaxies with Six Years of Fermi
  Large Area Telescope Data}},}\ }\href {\doibase
  10.1103/PhysRevLett.115.231301} {\bibfield  {journal} {\bibinfo  {journal}
  {Phys. Rev. Lett.}\ }\textbf {\bibinfo {volume} {115}},\ \bibinfo {pages}
  {231301} (\bibinfo {year} {2015}{\natexlab{c}})},\ \Eprint
  {http://arxiv.org/abs/1503.02641} {arXiv:1503.02641 [astro-ph.HE]}
  \BibitemShut {NoStop}%
\bibitem [{\citenamefont {Albert}\ \emph {et~al.}(2017)\citenamefont {Albert}
  \emph {et~al.}}]{Fermi-LAT:2016uux}%
  \BibitemOpen
  \bibfield  {author} {\bibinfo {author} {\bibfnamefont {A.}~\bibnamefont
  {Albert}} \emph {et~al.} (\bibinfo {collaboration} {Fermi-LAT, DES}),\
  }\bibfield  {title} {\enquote {\bibinfo {title} {{Searching for Dark Matter
  Annihilation in Recently Discovered Milky Way Satellites with Fermi-LAT}},}\
  }\href {\doibase 10.3847/1538-4357/834/2/110} {\bibfield  {journal} {\bibinfo
   {journal} {Astrophys. J.}\ }\textbf {\bibinfo {volume} {834}},\ \bibinfo
  {pages} {110} (\bibinfo {year} {2017})},\ \Eprint
  {http://arxiv.org/abs/1611.03184} {arXiv:1611.03184 [astro-ph.HE]}
  \BibitemShut {NoStop}%
\bibitem [{\citenamefont {Navarro}\ \emph {et~al.}(1997)\citenamefont
  {Navarro}, \citenamefont {Frenk},\ and\ \citenamefont
  {White}}]{Navarro:1996gj}%
  \BibitemOpen
  \bibfield  {author} {\bibinfo {author} {\bibfnamefont {Julio~F.}\
  \bibnamefont {Navarro}}, \bibinfo {author} {\bibfnamefont {Carlos~S.}\
  \bibnamefont {Frenk}}, \ and\ \bibinfo {author} {\bibfnamefont {Simon D.~M.}\
  \bibnamefont {White}},\ }\bibfield  {title} {\enquote {\bibinfo {title} {{A
  Universal density profile from hierarchical clustering}},}\ }\href {\doibase
  10.1086/304888} {\bibfield  {journal} {\bibinfo  {journal} {Astrophys. J.}\
  }\textbf {\bibinfo {volume} {490}},\ \bibinfo {pages} {493--508} (\bibinfo
  {year} {1997})},\ \Eprint {http://arxiv.org/abs/astro-ph/9611107}
  {arXiv:astro-ph/9611107} \BibitemShut {NoStop}%
\bibitem [{\citenamefont {Correa}\ \emph {et~al.}(2015)\citenamefont {Correa},
  \citenamefont {Wyithe}, \citenamefont {Schaye},\ and\ \citenamefont
  {Duffy}}]{Correa:2015dva}%
  \BibitemOpen
  \bibfield  {author} {\bibinfo {author} {\bibfnamefont {Camila~A.}\
  \bibnamefont {Correa}}, \bibinfo {author} {\bibfnamefont {J.~Stuart~B.}\
  \bibnamefont {Wyithe}}, \bibinfo {author} {\bibfnamefont {Joop}\ \bibnamefont
  {Schaye}}, \ and\ \bibinfo {author} {\bibfnamefont {Alan~R.}\ \bibnamefont
  {Duffy}},\ }\bibfield  {title} {\enquote {\bibinfo {title} {{The accretion
  history of dark matter haloes \textendash{} III. A physical model for the
  concentration\textendash{}mass relation}},}\ }\href {\doibase
  10.1093/mnras/stv1363} {\bibfield  {journal} {\bibinfo  {journal} {Mon. Not.
  Roy. Astron. Soc.}\ }\textbf {\bibinfo {volume} {452}},\ \bibinfo {pages}
  {1217--1232} (\bibinfo {year} {2015})},\ \Eprint
  {http://arxiv.org/abs/1502.00391} {arXiv:1502.00391 [astro-ph.CO]}
  \BibitemShut {NoStop}%
\bibitem [{\citenamefont {Molin\'e}\ \emph {et~al.}(2017)\citenamefont
  {Molin\'e}, \citenamefont {S\'anchez-Conde}, \citenamefont {Palomares-Ruiz},\
  and\ \citenamefont {Prada}}]{Moline:2016pbm}%
  \BibitemOpen
  \bibfield  {author} {\bibinfo {author} {\bibfnamefont {\'Angeles}\
  \bibnamefont {Molin\'e}}, \bibinfo {author} {\bibfnamefont {Miguel~A.}\
  \bibnamefont {S\'anchez-Conde}}, \bibinfo {author} {\bibfnamefont {Sergio}\
  \bibnamefont {Palomares-Ruiz}}, \ and\ \bibinfo {author} {\bibfnamefont
  {Francisco}\ \bibnamefont {Prada}},\ }\bibfield  {title} {\enquote {\bibinfo
  {title} {{Characterization of subhalo structural properties and implications
  for dark matter annihilation signals}},}\ }\href {\doibase
  10.1093/mnras/stx026} {\bibfield  {journal} {\bibinfo  {journal} {Mon. Not.
  Roy. Astron. Soc.}\ }\textbf {\bibinfo {volume} {466}},\ \bibinfo {pages}
  {4974--4990} (\bibinfo {year} {2017})},\ \Eprint
  {http://arxiv.org/abs/1603.04057} {arXiv:1603.04057 [astro-ph.CO]}
  \BibitemShut {NoStop}%
\bibitem [{\citenamefont {Hu}\ \emph {et~al.}(2024)\citenamefont {Hu},
  \citenamefont {Zhu}, \citenamefont {Liu},\ and\ \citenamefont
  {Liang}}]{Hu:2023iex}%
  \BibitemOpen
  \bibfield  {author} {\bibinfo {author} {\bibfnamefont {Xiao-Song}\
  \bibnamefont {Hu}}, \bibinfo {author} {\bibfnamefont {Ben-Yang}\ \bibnamefont
  {Zhu}}, \bibinfo {author} {\bibfnamefont {Tian-Ci}\ \bibnamefont {Liu}}, \
  and\ \bibinfo {author} {\bibfnamefont {Yun-Feng}\ \bibnamefont {Liang}},\
  }\bibfield  {title} {\enquote {\bibinfo {title} {{Constraints on the
  annihilation of heavy dark matter in dwarf spheroidal galaxies with gamma-ray
  observations}},}\ }\href {\doibase 10.1103/PhysRevD.109.063036} {\bibfield
  {journal} {\bibinfo  {journal} {Phys. Rev. D}\ }\textbf {\bibinfo {volume}
  {109}},\ \bibinfo {pages} {063036} (\bibinfo {year} {2024})},\ \Eprint
  {http://arxiv.org/abs/2309.06151} {arXiv:2309.06151 [astro-ph.HE]}
  \BibitemShut {NoStop}%
\bibitem [{\citenamefont {Vitale}\ and\ \citenamefont
  {Morselli}(2009)}]{Vitale:2009hr}%
  \BibitemOpen
  \bibfield  {author} {\bibinfo {author} {\bibfnamefont {Vincenzo}\
  \bibnamefont {Vitale}}\ and\ \bibinfo {author} {\bibfnamefont {Aldo}\
  \bibnamefont {Morselli}} (\bibinfo {collaboration} {Fermi-LAT}),\ }\bibfield
  {title} {\enquote {\bibinfo {title} {{Indirect Search for Dark Matter from
  the center of the Milky Way with the Fermi-Large Area Telescope}},}\ }in\
  \href@noop {} {\emph {\bibinfo {booktitle} {{2009 Fermi Symposium}}}}\
  (\bibinfo {year} {2009})\ \Eprint {http://arxiv.org/abs/0912.3828}
  {arXiv:0912.3828 [astro-ph.HE]} \BibitemShut {NoStop}%
\bibitem [{\citenamefont {Hooper}\ and\ \citenamefont
  {Goodenough}(2011)}]{Hooper:2010mq}%
  \BibitemOpen
  \bibfield  {author} {\bibinfo {author} {\bibfnamefont {Dan}\ \bibnamefont
  {Hooper}}\ and\ \bibinfo {author} {\bibfnamefont {Lisa}\ \bibnamefont
  {Goodenough}},\ }\bibfield  {title} {\enquote {\bibinfo {title} {{Dark Matter
  Annihilation in The Galactic Center As Seen by the Fermi Gamma Ray Space
  Telescope}},}\ }\href {\doibase 10.1016/j.physletb.2011.02.029} {\bibfield
  {journal} {\bibinfo  {journal} {Phys. Lett. B}\ }\textbf {\bibinfo {volume}
  {697}},\ \bibinfo {pages} {412--428} (\bibinfo {year} {2011})},\ \Eprint
  {http://arxiv.org/abs/1010.2752} {arXiv:1010.2752 [hep-ph]} \BibitemShut
  {NoStop}%
\bibitem [{\citenamefont {Abazajian}(2011)}]{Abazajian:2010zy}%
  \BibitemOpen
  \bibfield  {author} {\bibinfo {author} {\bibfnamefont {Kevork~N.}\
  \bibnamefont {Abazajian}},\ }\bibfield  {title} {\enquote {\bibinfo {title}
  {{The Consistency of Fermi-LAT Observations of the Galactic Center with a
  Millisecond Pulsar Population in the Central Stellar Cluster}},}\ }\href
  {\doibase 10.1088/1475-7516/2011/03/010} {\bibfield  {journal} {\bibinfo
  {journal} {JCAP}\ }\textbf {\bibinfo {volume} {03}},\ \bibinfo {pages} {010}
  (\bibinfo {year} {2011})},\ \Eprint {http://arxiv.org/abs/1011.4275}
  {arXiv:1011.4275 [astro-ph.HE]} \BibitemShut {NoStop}%
\bibitem [{\citenamefont {Hooper}\ and\ \citenamefont
  {Linden}(2011)}]{Hooper:2011ti}%
  \BibitemOpen
  \bibfield  {author} {\bibinfo {author} {\bibfnamefont {Dan}\ \bibnamefont
  {Hooper}}\ and\ \bibinfo {author} {\bibfnamefont {Tim}\ \bibnamefont
  {Linden}},\ }\bibfield  {title} {\enquote {\bibinfo {title} {{On The Origin
  Of The Gamma Rays From The Galactic Center}},}\ }\href {\doibase
  10.1103/PhysRevD.84.123005} {\bibfield  {journal} {\bibinfo  {journal} {Phys.
  Rev. D}\ }\textbf {\bibinfo {volume} {84}},\ \bibinfo {pages} {123005}
  (\bibinfo {year} {2011})},\ \Eprint {http://arxiv.org/abs/1110.0006}
  {arXiv:1110.0006 [astro-ph.HE]} \BibitemShut {NoStop}%
\bibitem [{\citenamefont {Hooper}\ and\ \citenamefont
  {Slatyer}(2013)}]{Hooper:2013rwa}%
  \BibitemOpen
  \bibfield  {author} {\bibinfo {author} {\bibfnamefont {Dan}\ \bibnamefont
  {Hooper}}\ and\ \bibinfo {author} {\bibfnamefont {Tracy~R.}\ \bibnamefont
  {Slatyer}},\ }\bibfield  {title} {\enquote {\bibinfo {title} {{Two Emission
  Mechanisms in the Fermi Bubbles: A Possible Signal of Annihilating Dark
  Matter}},}\ }\href {\doibase 10.1016/j.dark.2013.06.003} {\bibfield
  {journal} {\bibinfo  {journal} {Phys. Dark Univ.}\ }\textbf {\bibinfo
  {volume} {2}},\ \bibinfo {pages} {118--138} (\bibinfo {year} {2013})},\
  \Eprint {http://arxiv.org/abs/1302.6589} {arXiv:1302.6589 [astro-ph.HE]}
  \BibitemShut {NoStop}%
\bibitem [{\citenamefont {Gordon}\ and\ \citenamefont
  {Macias}(2013)}]{Gordon:2013vta}%
  \BibitemOpen
  \bibfield  {author} {\bibinfo {author} {\bibfnamefont {Chris}\ \bibnamefont
  {Gordon}}\ and\ \bibinfo {author} {\bibfnamefont {Oscar}\ \bibnamefont
  {Macias}},\ }\bibfield  {title} {\enquote {\bibinfo {title} {{Dark Matter and
  Pulsar Model Constraints from Galactic Center Fermi-LAT Gamma Ray
  Observations}},}\ }\href {\doibase 10.1103/PhysRevD.88.083521} {\bibfield
  {journal} {\bibinfo  {journal} {Phys. Rev. D}\ }\textbf {\bibinfo {volume}
  {88}},\ \bibinfo {pages} {083521} (\bibinfo {year} {2013})},\ \bibinfo {note}
  {[Erratum: Phys.Rev.D 89, 049901 (2014)]},\ \Eprint
  {http://arxiv.org/abs/1306.5725} {arXiv:1306.5725 [astro-ph.HE]} \BibitemShut
  {NoStop}%
\bibitem [{\citenamefont {Abazajian}\ \emph {et~al.}(2014)\citenamefont
  {Abazajian}, \citenamefont {Canac}, \citenamefont {Horiuchi},\ and\
  \citenamefont {Kaplinghat}}]{Abazajian:2014fta}%
  \BibitemOpen
  \bibfield  {author} {\bibinfo {author} {\bibfnamefont {Kevork~N.}\
  \bibnamefont {Abazajian}}, \bibinfo {author} {\bibfnamefont {Nicolas}\
  \bibnamefont {Canac}}, \bibinfo {author} {\bibfnamefont {Shunsaku}\
  \bibnamefont {Horiuchi}}, \ and\ \bibinfo {author} {\bibfnamefont {Manoj}\
  \bibnamefont {Kaplinghat}},\ }\bibfield  {title} {\enquote {\bibinfo {title}
  {{Astrophysical and Dark Matter Interpretations of Extended Gamma-Ray
  Emission from the Galactic Center}},}\ }\href {\doibase
  10.1103/PhysRevD.90.023526} {\bibfield  {journal} {\bibinfo  {journal} {Phys.
  Rev. D}\ }\textbf {\bibinfo {volume} {90}},\ \bibinfo {pages} {023526}
  (\bibinfo {year} {2014})},\ \Eprint {http://arxiv.org/abs/1402.4090}
  {arXiv:1402.4090 [astro-ph.HE]} \BibitemShut {NoStop}%
\bibitem [{\citenamefont {Daylan}\ \emph {et~al.}(2016)\citenamefont {Daylan},
  \citenamefont {Finkbeiner}, \citenamefont {Hooper}, \citenamefont {Linden},
  \citenamefont {Portillo}, \citenamefont {Rodd},\ and\ \citenamefont
  {Slatyer}}]{Daylan:2014rsa}%
  \BibitemOpen
  \bibfield  {author} {\bibinfo {author} {\bibfnamefont {Tansu}\ \bibnamefont
  {Daylan}}, \bibinfo {author} {\bibfnamefont {Douglas~P.}\ \bibnamefont
  {Finkbeiner}}, \bibinfo {author} {\bibfnamefont {Dan}\ \bibnamefont
  {Hooper}}, \bibinfo {author} {\bibfnamefont {Tim}\ \bibnamefont {Linden}},
  \bibinfo {author} {\bibfnamefont {Stephen K.~N.}\ \bibnamefont {Portillo}},
  \bibinfo {author} {\bibfnamefont {Nicholas~L.}\ \bibnamefont {Rodd}}, \ and\
  \bibinfo {author} {\bibfnamefont {Tracy~R.}\ \bibnamefont {Slatyer}},\
  }\bibfield  {title} {\enquote {\bibinfo {title} {{The characterization of the
  gamma-ray signal from the central Milky Way: A case for annihilating dark
  matter}},}\ }\href {\doibase 10.1016/j.dark.2015.12.005} {\bibfield
  {journal} {\bibinfo  {journal} {Phys. Dark Univ.}\ }\textbf {\bibinfo
  {volume} {12}},\ \bibinfo {pages} {1--23} (\bibinfo {year} {2016})},\ \Eprint
  {http://arxiv.org/abs/1402.6703} {arXiv:1402.6703 [astro-ph.HE]} \BibitemShut
  {NoStop}%
\bibitem [{\citenamefont {Calore}\ \emph
  {et~al.}(2015{\natexlab{a}})\citenamefont {Calore}, \citenamefont {Cholis},\
  and\ \citenamefont {Weniger}}]{Calore:2014xka}%
  \BibitemOpen
  \bibfield  {author} {\bibinfo {author} {\bibfnamefont {Francesca}\
  \bibnamefont {Calore}}, \bibinfo {author} {\bibfnamefont {Ilias}\
  \bibnamefont {Cholis}}, \ and\ \bibinfo {author} {\bibfnamefont {Christoph}\
  \bibnamefont {Weniger}},\ }\bibfield  {title} {\enquote {\bibinfo {title}
  {{Background Model Systematics for the Fermi GeV Excess}},}\ }\href {\doibase
  10.1088/1475-7516/2015/03/038} {\bibfield  {journal} {\bibinfo  {journal}
  {JCAP}\ }\textbf {\bibinfo {volume} {03}},\ \bibinfo {pages} {038} (\bibinfo
  {year} {2015}{\natexlab{a}})},\ \Eprint {http://arxiv.org/abs/1409.0042}
  {arXiv:1409.0042 [astro-ph.CO]} \BibitemShut {NoStop}%
\bibitem [{\citenamefont {Zhou}\ \emph {et~al.}(2015)\citenamefont {Zhou},
  \citenamefont {Liang}, \citenamefont {Huang}, \citenamefont {Li},
  \citenamefont {Fan}, \citenamefont {Feng},\ and\ \citenamefont
  {Chang}}]{Zhou:2014lva}%
  \BibitemOpen
  \bibfield  {author} {\bibinfo {author} {\bibfnamefont {Bei}\ \bibnamefont
  {Zhou}}, \bibinfo {author} {\bibfnamefont {Yun-Feng}\ \bibnamefont {Liang}},
  \bibinfo {author} {\bibfnamefont {Xiaoyuan}\ \bibnamefont {Huang}}, \bibinfo
  {author} {\bibfnamefont {Xiang}\ \bibnamefont {Li}}, \bibinfo {author}
  {\bibfnamefont {Yi-Zhong}\ \bibnamefont {Fan}}, \bibinfo {author}
  {\bibfnamefont {Lei}\ \bibnamefont {Feng}}, \ and\ \bibinfo {author}
  {\bibfnamefont {Jin}\ \bibnamefont {Chang}},\ }\bibfield  {title} {\enquote
  {\bibinfo {title} {{GeV excess in the Milky Way: The role of diffuse galactic
  gamma-ray emission templates}},}\ }\href {\doibase
  10.1103/PhysRevD.91.123010} {\bibfield  {journal} {\bibinfo  {journal} {Phys.
  Rev. D}\ }\textbf {\bibinfo {volume} {91}},\ \bibinfo {pages} {123010}
  (\bibinfo {year} {2015})},\ \Eprint {http://arxiv.org/abs/1406.6948}
  {arXiv:1406.6948 [astro-ph.HE]} \BibitemShut {NoStop}%
\bibitem [{\citenamefont {Huang}\ \emph {et~al.}(2016)\citenamefont {Huang},
  \citenamefont {En\ss{}lin},\ and\ \citenamefont {Selig}}]{Huang:2015rlu}%
  \BibitemOpen
  \bibfield  {author} {\bibinfo {author} {\bibfnamefont {Xiaoyuan}\
  \bibnamefont {Huang}}, \bibinfo {author} {\bibfnamefont {Torsten}\
  \bibnamefont {En\ss{}lin}}, \ and\ \bibinfo {author} {\bibfnamefont {Marco}\
  \bibnamefont {Selig}},\ }\bibfield  {title} {\enquote {\bibinfo {title}
  {{Galactic dark matter search via phenomenological astrophysics modeling}},}\
  }\href {\doibase 10.1088/1475-7516/2016/04/030} {\bibfield  {journal}
  {\bibinfo  {journal} {JCAP}\ }\textbf {\bibinfo {volume} {04}},\ \bibinfo
  {pages} {030} (\bibinfo {year} {2016})},\ \Eprint
  {http://arxiv.org/abs/1511.02621} {arXiv:1511.02621 [astro-ph.HE]}
  \BibitemShut {NoStop}%
\bibitem [{\citenamefont {Ajello}\ \emph {et~al.}(2016)\citenamefont {Ajello}
  \emph {et~al.}}]{Fermi-LAT:2015sau}%
  \BibitemOpen
  \bibfield  {author} {\bibinfo {author} {\bibfnamefont {M.}~\bibnamefont
  {Ajello}} \emph {et~al.} (\bibinfo {collaboration} {Fermi-LAT}),\ }\bibfield
  {title} {\enquote {\bibinfo {title} {{Fermi-LAT Observations of High-Energy
  $\gamma$-Ray Emission Toward the Galactic Center}},}\ }\href {\doibase
  10.3847/0004-637X/819/1/44} {\bibfield  {journal} {\bibinfo  {journal}
  {Astrophys. J.}\ }\textbf {\bibinfo {volume} {819}},\ \bibinfo {pages} {44}
  (\bibinfo {year} {2016})},\ \Eprint {http://arxiv.org/abs/1511.02938}
  {arXiv:1511.02938 [astro-ph.HE]} \BibitemShut {NoStop}%
\bibitem [{\citenamefont {Di~Mauro}(2021)}]{DiMauro:2021raz}%
  \BibitemOpen
  \bibfield  {author} {\bibinfo {author} {\bibfnamefont {Mattia}\ \bibnamefont
  {Di~Mauro}},\ }\bibfield  {title} {\enquote {\bibinfo {title}
  {{Characteristics of the Galactic Center excess measured with 11 years of
  $Fermi$-LAT data}},}\ }\href {\doibase 10.1103/PhysRevD.103.063029}
  {\bibfield  {journal} {\bibinfo  {journal} {Phys. Rev. D}\ }\textbf {\bibinfo
  {volume} {103}},\ \bibinfo {pages} {063029} (\bibinfo {year} {2021})},\
  \Eprint {http://arxiv.org/abs/2101.04694} {arXiv:2101.04694 [astro-ph.HE]}
  \BibitemShut {NoStop}%
\bibitem [{\citenamefont {Cholis}\ \emph {et~al.}(2022)\citenamefont {Cholis},
  \citenamefont {Zhong}, \citenamefont {McDermott},\ and\ \citenamefont
  {Surdutovich}}]{Cholis:2021rpp}%
  \BibitemOpen
  \bibfield  {author} {\bibinfo {author} {\bibfnamefont {Ilias}\ \bibnamefont
  {Cholis}}, \bibinfo {author} {\bibfnamefont {Yi-Ming}\ \bibnamefont {Zhong}},
  \bibinfo {author} {\bibfnamefont {Samuel~D.}\ \bibnamefont {McDermott}}, \
  and\ \bibinfo {author} {\bibfnamefont {Joseph~P.}\ \bibnamefont
  {Surdutovich}},\ }\bibfield  {title} {\enquote {\bibinfo {title} {{Return of
  the templates: Revisiting the Galactic Center excess with multimessenger
  observations}},}\ }\href {\doibase 10.1103/PhysRevD.105.103023} {\bibfield
  {journal} {\bibinfo  {journal} {Phys. Rev. D}\ }\textbf {\bibinfo {volume}
  {105}},\ \bibinfo {pages} {103023} (\bibinfo {year} {2022})},\ \Eprint
  {http://arxiv.org/abs/2112.09706} {arXiv:2112.09706 [astro-ph.HE]}
  \BibitemShut {NoStop}%
\bibitem [{\citenamefont {Calore}\ \emph
  {et~al.}(2015{\natexlab{b}})\citenamefont {Calore}, \citenamefont {Cholis},
  \citenamefont {McCabe},\ and\ \citenamefont {Weniger}}]{Calore:2014nla}%
  \BibitemOpen
  \bibfield  {author} {\bibinfo {author} {\bibfnamefont {Francesca}\
  \bibnamefont {Calore}}, \bibinfo {author} {\bibfnamefont {Ilias}\
  \bibnamefont {Cholis}}, \bibinfo {author} {\bibfnamefont {Christopher}\
  \bibnamefont {McCabe}}, \ and\ \bibinfo {author} {\bibfnamefont {Christoph}\
  \bibnamefont {Weniger}},\ }\bibfield  {title} {\enquote {\bibinfo {title} {{A
  Tale of Tails: Dark Matter Interpretations of the Fermi GeV Excess in Light
  of Background Model Systematics}},}\ }\href {\doibase
  10.1103/PhysRevD.91.063003} {\bibfield  {journal} {\bibinfo  {journal} {Phys.
  Rev. D}\ }\textbf {\bibinfo {volume} {91}},\ \bibinfo {pages} {063003}
  (\bibinfo {year} {2015}{\natexlab{b}})},\ \Eprint
  {http://arxiv.org/abs/1411.4647} {arXiv:1411.4647 [hep-ph]} \BibitemShut
  {NoStop}%
\bibitem [{\citenamefont {Agrawal}\ \emph {et~al.}(2015)\citenamefont
  {Agrawal}, \citenamefont {Batell}, \citenamefont {Fox},\ and\ \citenamefont
  {Harnik}}]{Agrawal:2014oha}%
  \BibitemOpen
  \bibfield  {author} {\bibinfo {author} {\bibfnamefont {Prateek}\ \bibnamefont
  {Agrawal}}, \bibinfo {author} {\bibfnamefont {Brian}\ \bibnamefont {Batell}},
  \bibinfo {author} {\bibfnamefont {Patrick~J.}\ \bibnamefont {Fox}}, \ and\
  \bibinfo {author} {\bibfnamefont {Roni}\ \bibnamefont {Harnik}},\ }\bibfield
  {title} {\enquote {\bibinfo {title} {{WIMPs at the Galactic Center}},}\
  }\href {\doibase 10.1088/1475-7516/2015/05/011} {\bibfield  {journal}
  {\bibinfo  {journal} {JCAP}\ }\textbf {\bibinfo {volume} {05}},\ \bibinfo
  {pages} {011} (\bibinfo {year} {2015})},\ \Eprint
  {http://arxiv.org/abs/1411.2592} {arXiv:1411.2592 [hep-ph]} \BibitemShut
  {NoStop}%
\bibitem [{\citenamefont {Berlin}\ \emph {et~al.}(2015)\citenamefont {Berlin},
  \citenamefont {Gori}, \citenamefont {Lin},\ and\ \citenamefont
  {Wang}}]{Berlin:2015wwa}%
  \BibitemOpen
  \bibfield  {author} {\bibinfo {author} {\bibfnamefont {Asher}\ \bibnamefont
  {Berlin}}, \bibinfo {author} {\bibfnamefont {Stefania}\ \bibnamefont {Gori}},
  \bibinfo {author} {\bibfnamefont {Tongyan}\ \bibnamefont {Lin}}, \ and\
  \bibinfo {author} {\bibfnamefont {Lian-Tao}\ \bibnamefont {Wang}},\
  }\bibfield  {title} {\enquote {\bibinfo {title} {{Pseudoscalar Portal Dark
  Matter}},}\ }\href {\doibase 10.1103/PhysRevD.92.015005} {\bibfield
  {journal} {\bibinfo  {journal} {Phys. Rev. D}\ }\textbf {\bibinfo {volume}
  {92}},\ \bibinfo {pages} {015005} (\bibinfo {year} {2015})},\ \Eprint
  {http://arxiv.org/abs/1502.06000} {arXiv:1502.06000 [hep-ph]} \BibitemShut
  {NoStop}%
\bibitem [{\citenamefont {Zhong}\ and\ \citenamefont
  {Cholis}(2024)}]{Zhong:2024vyi}%
  \BibitemOpen
  \bibfield  {author} {\bibinfo {author} {\bibfnamefont {Yi-Ming}\ \bibnamefont
  {Zhong}}\ and\ \bibinfo {author} {\bibfnamefont {Ilias}\ \bibnamefont
  {Cholis}},\ }\bibfield  {title} {\enquote {\bibinfo {title} {{Robustness of
  the Galactic Center excess morphology against masking}},}\ }\href {\doibase
  10.1103/PhysRevD.109.123017} {\bibfield  {journal} {\bibinfo  {journal}
  {Phys. Rev. D}\ }\textbf {\bibinfo {volume} {109}},\ \bibinfo {pages}
  {123017} (\bibinfo {year} {2024})},\ \Eprint
  {http://arxiv.org/abs/2401.02481} {arXiv:2401.02481 [astro-ph.HE]}
  \BibitemShut {NoStop}%
\bibitem [{\citenamefont {Song}\ \emph {et~al.}(2024)\citenamefont {Song},
  \citenamefont {Eckner}, \citenamefont {Gordon}, \citenamefont {Calore},
  \citenamefont {Macias}, \citenamefont {Abazajian}, \citenamefont {Horiuchi},
  \citenamefont {Kaplinghat},\ and\ \citenamefont {Pohl}}]{Song:2024iup}%
  \BibitemOpen
  \bibfield  {author} {\bibinfo {author} {\bibfnamefont {Deheng}\ \bibnamefont
  {Song}}, \bibinfo {author} {\bibfnamefont {Christopher}\ \bibnamefont
  {Eckner}}, \bibinfo {author} {\bibfnamefont {Chris}\ \bibnamefont {Gordon}},
  \bibinfo {author} {\bibfnamefont {Francesca}\ \bibnamefont {Calore}},
  \bibinfo {author} {\bibfnamefont {Oscar}\ \bibnamefont {Macias}}, \bibinfo
  {author} {\bibfnamefont {Kevork~N.}\ \bibnamefont {Abazajian}}, \bibinfo
  {author} {\bibfnamefont {Shunsaku}\ \bibnamefont {Horiuchi}}, \bibinfo
  {author} {\bibfnamefont {Manoj}\ \bibnamefont {Kaplinghat}}, \ and\ \bibinfo
  {author} {\bibfnamefont {Martin}\ \bibnamefont {Pohl}},\ }\bibfield  {title}
  {\enquote {\bibinfo {title} {{Robust inference of the Galactic Centre
  gamma-ray excess spatial properties}},}\ }\href {\doibase
  10.1093/mnras/stae923} {\bibfield  {journal} {\bibinfo  {journal} {Mon. Not.
  Roy. Astron. Soc.}\ }\textbf {\bibinfo {volume} {530}},\ \bibinfo {pages}
  {4395--4411} (\bibinfo {year} {2024})},\ \Eprint
  {http://arxiv.org/abs/2402.05449} {arXiv:2402.05449 [astro-ph.GA]}
  \BibitemShut {NoStop}%
\bibitem [{\citenamefont {Cui}\ \emph {et~al.}(2017)\citenamefont {Cui},
  \citenamefont {Yuan}, \citenamefont {Tsai},\ and\ \citenamefont
  {Fan}}]{Cui:2016ppb}%
  \BibitemOpen
  \bibfield  {author} {\bibinfo {author} {\bibfnamefont {Ming-Yang}\
  \bibnamefont {Cui}}, \bibinfo {author} {\bibfnamefont {Qiang}\ \bibnamefont
  {Yuan}}, \bibinfo {author} {\bibfnamefont {Yue-Lin~Sming}\ \bibnamefont
  {Tsai}}, \ and\ \bibinfo {author} {\bibfnamefont {Yi-Zhong}\ \bibnamefont
  {Fan}},\ }\bibfield  {title} {\enquote {\bibinfo {title} {{Possible dark
  matter annihilation signal in the AMS-02 antiproton data}},}\ }\href
  {\doibase 10.1103/PhysRevLett.118.191101} {\bibfield  {journal} {\bibinfo
  {journal} {Phys. Rev. Lett.}\ }\textbf {\bibinfo {volume} {118}},\ \bibinfo
  {pages} {191101} (\bibinfo {year} {2017})},\ \Eprint
  {http://arxiv.org/abs/1610.03840} {arXiv:1610.03840 [astro-ph.HE]}
  \BibitemShut {NoStop}%
\bibitem [{\citenamefont {Cuoco}\ \emph
  {et~al.}(2017{\natexlab{b}})\citenamefont {Cuoco}, \citenamefont {Kr\"amer},\
  and\ \citenamefont {Korsmeier}}]{Cuoco:2016eej}%
  \BibitemOpen
  \bibfield  {author} {\bibinfo {author} {\bibfnamefont {Alessandro}\
  \bibnamefont {Cuoco}}, \bibinfo {author} {\bibfnamefont {Michael}\
  \bibnamefont {Kr\"amer}}, \ and\ \bibinfo {author} {\bibfnamefont {Michael}\
  \bibnamefont {Korsmeier}},\ }\bibfield  {title} {\enquote {\bibinfo {title}
  {{Novel Dark Matter Constraints from Antiprotons in Light of AMS-02}},}\
  }\href {\doibase 10.1103/PhysRevLett.118.191102} {\bibfield  {journal}
  {\bibinfo  {journal} {Phys. Rev. Lett.}\ }\textbf {\bibinfo {volume} {118}},\
  \bibinfo {pages} {191102} (\bibinfo {year} {2017}{\natexlab{b}})},\ \Eprint
  {http://arxiv.org/abs/1610.03071} {arXiv:1610.03071 [astro-ph.HE]}
  \BibitemShut {NoStop}%
\bibitem [{\citenamefont {Cholis}\ \emph {et~al.}(2019)\citenamefont {Cholis},
  \citenamefont {Linden},\ and\ \citenamefont {Hooper}}]{Cholis:2019ejx}%
  \BibitemOpen
  \bibfield  {author} {\bibinfo {author} {\bibfnamefont {Ilias}\ \bibnamefont
  {Cholis}}, \bibinfo {author} {\bibfnamefont {Tim}\ \bibnamefont {Linden}}, \
  and\ \bibinfo {author} {\bibfnamefont {Dan}\ \bibnamefont {Hooper}},\
  }\bibfield  {title} {\enquote {\bibinfo {title} {{A Robust Excess in the
  Cosmic-Ray Antiproton Spectrum: Implications for Annihilating Dark
  Matter}},}\ }\href {\doibase 10.1103/PhysRevD.99.103026} {\bibfield
  {journal} {\bibinfo  {journal} {Phys. Rev. D}\ }\textbf {\bibinfo {volume}
  {99}},\ \bibinfo {pages} {103026} (\bibinfo {year} {2019})},\ \Eprint
  {http://arxiv.org/abs/1903.02549} {arXiv:1903.02549 [astro-ph.HE]}
  \BibitemShut {NoStop}%
\bibitem [{\citenamefont {Cuoco}\ \emph {et~al.}(2019)\citenamefont {Cuoco},
  \citenamefont {Heisig}, \citenamefont {Klamt}, \citenamefont {Korsmeier},\
  and\ \citenamefont {Kr\"amer}}]{Cuoco:2019kuu}%
  \BibitemOpen
  \bibfield  {author} {\bibinfo {author} {\bibfnamefont {Alessandro}\
  \bibnamefont {Cuoco}}, \bibinfo {author} {\bibfnamefont {Jan}\ \bibnamefont
  {Heisig}}, \bibinfo {author} {\bibfnamefont {Lukas}\ \bibnamefont {Klamt}},
  \bibinfo {author} {\bibfnamefont {Michael}\ \bibnamefont {Korsmeier}}, \ and\
  \bibinfo {author} {\bibfnamefont {Michael}\ \bibnamefont {Kr\"amer}},\
  }\bibfield  {title} {\enquote {\bibinfo {title} {{Scrutinizing the evidence
  for dark matter in cosmic-ray antiprotons}},}\ }\href {\doibase
  10.1103/PhysRevD.99.103014} {\bibfield  {journal} {\bibinfo  {journal} {Phys.
  Rev. D}\ }\textbf {\bibinfo {volume} {99}},\ \bibinfo {pages} {103014}
  (\bibinfo {year} {2019})},\ \Eprint {http://arxiv.org/abs/1903.01472}
  {arXiv:1903.01472 [astro-ph.HE]} \BibitemShut {NoStop}%
\bibitem [{\citenamefont {Dor\'e}\ \emph {et~al.}(2014)\citenamefont {Dor\'e}
  \emph {et~al.}}]{SPHEREx:2014bgr}%
  \BibitemOpen
  \bibfield  {author} {\bibinfo {author} {\bibfnamefont {Olivier}\ \bibnamefont
  {Dor\'e}} \emph {et~al.} (\bibinfo {collaboration} {SPHEREx}),\ }\bibfield
  {title} {\enquote {\bibinfo {title} {{Cosmology with the SPHEREX All-Sky
  Spectral Survey}},}\ }\href@noop {} {\  (\bibinfo {year} {2014})},\ \Eprint
  {http://arxiv.org/abs/1412.4872} {arXiv:1412.4872 [astro-ph.CO]} \BibitemShut
  {NoStop}%
\bibitem [{\citenamefont {Spergel}\ \emph {et~al.}(2015)\citenamefont {Spergel}
  \emph {et~al.}}]{Spergel:2015sza}%
  \BibitemOpen
  \bibfield  {author} {\bibinfo {author} {\bibfnamefont {D.}~\bibnamefont
  {Spergel}} \emph {et~al.},\ }\bibfield  {title} {\enquote {\bibinfo {title}
  {{Wide-Field InfrarRed Survey Telescope-Astrophysics Focused Telescope Assets
  WFIRST-AFTA 2015 Report}},}\ }\href@noop {} {\  (\bibinfo {year} {2015})},\
  \Eprint {http://arxiv.org/abs/1503.03757} {arXiv:1503.03757 [astro-ph.IM]}
  \BibitemShut {NoStop}%
\bibitem [{\citenamefont {Amendola}\ \emph {et~al.}(2018)\citenamefont
  {Amendola} \emph {et~al.}}]{Amendola:2016saw}%
  \BibitemOpen
  \bibfield  {author} {\bibinfo {author} {\bibfnamefont {Luca}\ \bibnamefont
  {Amendola}} \emph {et~al.},\ }\bibfield  {title} {\enquote {\bibinfo {title}
  {{Cosmology and fundamental physics with the Euclid satellite}},}\ }\href
  {\doibase 10.1007/s41114-017-0010-3} {\bibfield  {journal} {\bibinfo
  {journal} {Living Rev. Rel.}\ }\textbf {\bibinfo {volume} {21}},\ \bibinfo
  {pages} {2} (\bibinfo {year} {2018})},\ \Eprint
  {http://arxiv.org/abs/1606.00180} {arXiv:1606.00180 [astro-ph.CO]}
  \BibitemShut {NoStop}%
\bibitem [{MUS()}]{MUST_web}%
  \BibitemOpen
  \href@noop {} {}\bibinfo {howpublished} {\url{
  https://must.astro.tsinghua.edu.cn/en}}\BibitemShut {NoStop}%
\end{thebibliography}%

\end{document}